\documentclass[a4paper,11pt]{article}
\pdfoutput=1 

\usepackage{jcappub} 

\usepackage{multirow}      

\usepackage{blkarray, bigstrut}
\usepackage{xparse}
                     
\usepackage{graphicx}
\usepackage{dcolumn}
\usepackage{hyperref}
\hypersetup{colorlinks}
\usepackage[T1]{fontenc} 
\usepackage[normalem]{ulem}
\usepackage[utf8]{inputenc}

\usepackage{bm}
\let\vec\bm

\newcommand{\hmpc}{\,h^{-1}\text{Mpc}}
\newcommand{\hmpci}{\,h \, \text{Mpc}^{-1}}
\newcommand{\hgpc}{\,h^{-1}\text{Gpc}}

\newcommand{\mI}{\mathcal{I}}

\newcommand{\Dk}[1]{\frac{d^3#1}{(2\pi)^3}}

\newcommand{\ve}[1]{{\text{\bf #1}}} 
\newcommand{\vk}{\vec k}
\newcommand{\vp}{\vec p}
\newcommand{\vq}{\vec q}
\newcommand{\vx}{\vec x}
\newcommand{\vv}{\vec v}

\newcommand{\vhn}{\hat{\vec n}}
\newcommand{\va}{\ve a}

\newcommand{\mA}{\mathcal{A}}
\newcommand{\mB}{\mathcal{B}}

\newcommand{\tm}{m}

\newcommand{\ikk}{\underset{\vk_{12}= \vk}{\int}}

\newcommand{\ip}{\int_{\vp}} 

\newcommand{\dD}{\delta_\text{D}}
\newcommand{\Ps}{\vec \Psi}

\newcommand{\fk}{\texttt{fk}}

\usepackage{float} 

\usepackage[svgnames,table]{xcolor}
\usepackage{booktabs}

\newcommand{\revised}[1]{#1}

\title{\boldmath Fast computation of non-linear power spectrum in cosmologies with massive neutrinos} 

\author[a, b]{Hernán E. Noriega,}
\emailAdd{henoriega@estudiantes.fisica.unam.mx}

\author[b, c, 1]{Alejandro Aviles,\note{Corresponding author.}}
\emailAdd{avilescervantes@gmail.com}

\author[d]{Sebastien Fromenteau,}
\emailAdd{sfroment@icf.unam.mx}

\author[a]{Mariana Vargas-Maga\~na}
\emailAdd{mmaganav@fisica.unam.mx}

\affiliation[a]{Instituto de Física, Universidad Nacional Autónoma de México, Apdo. Postal 20-364, 01000, D.F, México.}

\affiliation[b]{Departamento de F\'isica, Instituto Nacional de Investigaciones Nucleares,
Apartado Postal 18-1027, Col. Escand\'on, Ciudad de M\'exico,11801, M\'exico.}

\affiliation[c]{Consejo Nacional de Ciencia y Tecnolog\'ia, Av. Insurgentes Sur 1582,
Colonia Cr\'edito Constructor, Del. Benito Ju\'arez, 03940, Ciudad de M\'exico, M\'exico.}

\affiliation[d]{Instituto de Ciencias F\'isicas, Universidad Nacional
Autónoma de México,  62210, Cuernavaca, Morelos.}


\abstract{
We  compute 1-loop corrections to the redshift space galaxy power spectrum in cosmologies containing additional scales, and hence kernels different from Einstein-de Sitter (EdS). Specifically, our method is tailored for cosmologies in the presence of massive neutrinos and some modified gravity models; in this article we concentrate on the former case. The perturbative kernels have contributions that we notice appear either from the logarithmic growth rate $f(k,t)$, which is scale-dependent because of the neutrino free-streaming, or from the failure of the commonly used approximation $f^2=\Omega_m$. The latter contributions make the computation of loop corrections quite slow, precluding {\it full-shape} analyses for parameter estimation. However, we identify that the dominant pieces of the kernels come from the growth factor, allowing us to simplify the kernels but retaining the characteristic free-streaming scale introduced by the neutrinos' mass.  Moreover, with this simplification one can exploit FFTLog methods to speed up the computations even more.  We validate our analytical modeling and numerical method with halo catalogs extracted from the \textsc{Quijote} simulations finding good agreement with the, {\it a priori}, known cosmological parameters.  We make public our Python code \texttt{FOLPS$\nu$} to compute the redshift space power spectrum in a fraction of second. Code available at \href{https://github.com/henoriega/FOLPS-nu}{https://github.com/henoriega/FOLPS-nu}.
}

\begin{document} 
\maketitle
\flushbottom


\begin{section}{Introduction}\label{sec:Introduction}

The neutrinos are perhaps the most elusive particles in the Universe, they couple very weakly to the rest of the standard model particles, and they always interact or are created in definite flavor states. However, the observed oscillations of atmospheric and solar neutrinos indicate that their energy and flavor states are not the same, but they mix, being this only possible if they have non-zero mass \cite{1968JETP...26..984P}. These observations also put lower bounds on the sum of their masses, being $0.06\, \text{eV}$ for the normal hierarchy and $0.11\, \text{eV}$ for the inverted one \cite{Esteban:2018azc}. On the other hand, from the energy spectrum of beta electrons emitted in the decay of tritium into helium-3, the KATRIN experiment has put an upper bound of $0.8 \,\text{eV}$ at 0.90 confidence level (c.l.) in the mass of the lightest energy state \cite{Aker:2019uuj,Aker:2021gma}, meaning an upper bound of $2.4\,\text{eV}$ in the sum of the three masses. Hence, neutrinos have excelling small masses. Despite this, they are so abundant that they contribute considerably to the cosmic energy budget. They also have specific, well known signatures on the large scale properties of the Universe which are highly dependent on their masses; see \cite{Lesgourgues:2006nd,Wong:2011ip} for reviews on neutrino physics in Cosmology. Primeval neutrinos decoupled early on from the cosmic plasma while they were still relativistic, as they continued to be until well inside the matter dominated epoch when they have slowed down sufficiently to become non-relativistic. Since then, primordial neutrinos have behaved as an additional dark matter component. However, they are a non-cold component since they still have large velocity dispersion, preventing them from clustering below a certain scale, named the free-streaming scale. Since neutrinos do participate in the clustering at large scale, they leave a well understood signature in the power spectrum, suppressing it above the free-streaming wave-number $k_\text{FS}$ \cite{Hu:1997vi,Hu:1997mj}. For realistic masses, this is located at about $0.1\, \hmpc$ nowadays, that is, at the onset of the non-linear scales.

Since gravity responds to the total energy momentum tensor, cosmological probes are particularly sensitive to the sum of the neutrino masses, while the mass splitting and hierarchy can be safely neglected for current and near future experiments \cite{Archidiacono:2020dvx}. Cosmic Microwave Background (CMB) temperature and polarization anisotropies from {\it Planck} 2018 legacy data put stringent constraints in the neutrino mass of $\sum_i m_{\nu, i} \lesssim 0.26 \,\text{eV}$ (0.95 c.l.), which is reduced to $0.12\,\text{eV}$ when combined with DR12 BOSS, MGS and 6dFGS Baryon Acoustic Oscillations (BAO) data  and CMB-lensing \cite{Akrami:2018vks}; and even more, down to $0.09\,\text{eV}$ when the SDSS eBOSS DR14 Lyman-alpha forest flux power spectrum data is added \cite{Palanque-Delabrouille:2019iyz}. Other recent analysis as minimal extensions to the base $\Lambda$CDM have yielded similar bounds, e.g \cite{dePutter:2012sh,Vagnozzi:2017ovm,Vagnozzi:2018jhn,DiValentino:2021hoh,Giusarma:2018jei}.\footnote{No minimal extensions as dynamical dark energy of modified gravity can lead to very different constraints on the neutrino masses; e.g. \cite{Vagnozzi:2018jhn,Garcia-Arroyo:2022vvy}} That is, the cosmological bounds on the neutrino mass absolute scale are very impressive and excel those from particle physics experiments by about one order of magnitude. Clearly, this success comes with the cost of being indirect measurements and hence very sensitive to our ability to estimate the other cosmological parameters, particularly the Hubble rate $H_0$ and the amplitude of fluctuations $\sigma_8$. Future galaxy surveys such as DESI\footnote{\href{https://www.desi.lbl.gov/}{desi.lbl.gov}} \cite{Aghamousa:2016zmz,DESI:2022xcl}, EUCLID\footnote{\href{https://sci.esa.int/web/euclid}{sci.esa.int/web/euclid}} \cite{laureijs2011euclid} and the LSST at the Vera C. Rubin Observatory\footnote{\href{https://www.lsst.org/}{lsst.org/}} \cite{Ivezic:2008fe} will be deeper in redshift and wider in subtended angle, so that it is expected that the neutrino masses will be measured with unprecedented precision in the near future. This is mainly because of the reduction of statistical errors in the data. Accordingly, methods to reduce the systematic errors in the modeling, well below the statistical errors, are very valuable. This is the subject of this paper.

Recently, the methods of Effective Field Theory (EFT) for large scale structure formation \cite{Baumann:2010tm,Carrasco:2012cv,Porto:2013qua,Vlah:2015sea,Angulo:2015eqa} built on Perturbation Theory (PT) \cite{Bernardeau:2001qr} have reached sufficient maturity to be used to compare the galaxy power spectrum (as well as other statistics) of real survey data directly to the analytical modeling of non-linearities \cite{DAmico:2019fhj,Ivanov:2019pdj}; see also \cite{Colas:2019ret,Wadekar:2020hax,Chudaykin:2020aoj,Philcox:2021kcw,Philcox:2022frc,Tanseri:2022zfe,Nishimichi:2020tvu,Chen:2020zjt,Tsedrik:2022cri,Carrilho:2022mon,Nunes:2022bhn}. These {\it full-shape} analyses capture at the same time the BAO and the redshift space distortions (RSD) features, reaching similar constrictive power as Planck experiment for some cosmological parameters.\footnote{Meanwhile, more traditional methods that compress cosmological information are in continue development, and recent implementations have similar success than full-shape analyses \cite{Brieden:2021edu,Brieden:2022lsd}.} Furthermore, with the aid of FFTLog methods \cite{Hamilton:1999uv}, the computation of loop corrections has been speed up \cite{McEwen:2016fjn,Fang:2016wcf,Schmittfull:2016jsw,Schmittfull:2016yqx,Simonovic:2017mhp} in order to use efficient Markov Chain Monte Carlo (MCMC) algorithms to draw the posterior distribution of parameters. All the existing methods have used Einstein-de Sitter (EdS) kernels so far, which is an excellent approximation as long as the neutrino mass is small ($\sim 0.1\,\text{eV}$), as expected from {\it Planck} data. However, the use of uninformative priors together with EdS kernels can introduce systematic errors when performing full-shape analyses of data, since one has to explore up to high mass values. Perhaps the main obstacle, and objection, against the use of more proper kernels for modeling cosmologies that include massive neutrinos is the computational time, which for full kernels is prohibitive for parameter estimation purposes. The objective of this work is to show that under simple approximations, it is possible to retain the main characteristics of the proper kernels without sacrificing valuable and expensive computational time.

Before entering into the more technical details in the following sections, here we present a schematic, but general form of our framework and describe how we overcome the difficulties. 

There are by now, many PT in the presence of massive neutrinos with different complexity levels and types of approximations when treating non-linearities \cite{Saito:2008bp,Wong:2008ws,Saito:2009ah,Shoji:2009gg,Lesgourgues:2009am,Upadhye:2013ndm,Dupuy:2013jaa,Blas:2014hya,Fuhrer:2014zka,Levi:2016tlf,Wright:2017dkw,Senatore:2017hyk,Garny:2020ilv}, in this work we adopt the one developed in \cite{Aviles:2017aor,Aviles:2020cax,Aviles:2020wme,Aviles:2021que}. 
The velocity and density fields of the combined cold dark matter and baryons fluid $cb$ in the presence of massive neutrinos are non-local related even at large scales by \cite{Aviles:2020cax} 
\begin{equation}\label{Eq:linear_theta}
\theta^{(1)}_{cb}(\vk) = \frac{f(k)}{f_0}\delta^{(1)}_{cb}(\vk) 
\end{equation}
with $f_0(t) = f(k\rightarrow 0,t)$ the cosmic growth rate at very large scales, or simply the growth rate $f=f_0$ in models without massive neutrinos. Strictly, the above expression holds to linear order in perturbation theory. However, this property is inherited to higher orders due to the advection of fields, enforcing the appearance of the free-streaming scale not only through the linear power spectrum but also via the non-linear kernels. 

The redshift space power spectrum can be written as \cite{Aviles:2020wme}
\begin{equation}\label{Pkmu}
P(k,\mu) = \sum_{m=0}^\infty \sum_{n=0}^m \mu^{2n} f_0^m I^m_n(k)
\end{equation} 
where $\mu = \hat{\vk} \cdot \vhn$ is the cosine angle between the wave-vector $\vk$ and the line-of-sight direction $\vhn$. Each function $I^m_n$ is obtained via a ``loop'' integral over internal momenta $\vp$,
\begin{equation}\label{Iofk}
I(k) = \int d^3 p \, \mathcal{I}(\vk,\vp)
\end{equation}
of a function $\mathcal{I}(\vp,\vk)$. These $\mathcal{I}$  functions are invariant against rotations, and so they only depend on the length of the sides of the triangle formed with the momenta $\vp$, $\vk$ and $\vk-\vp$. 
In the absence of massive neutrinos, the functions $\mI$ are known analytically, up to the linear power spectra, and the 1-loop integrals in \eqref{Iofk} can be performed with high precision and very rapidly using FFTLog methods, since one can write 
\begin{equation}\label{IofkExp}
\mathcal{I}(\vk,\vp) = \sum_{a,b,c} \, C_{abc} \, p^{z_a}|\vk-\vp|^{z_b} k^{z_c} 
\end{equation}
for a collection of complex powers $z_a$, $z_b$ and $z_c$ and coefficients $C_{abc}$.

However, in the presence of massive neutrinos the functions $\mathcal{I}(\vk,\vp)$ have no analytical form, but instead should be obtained from a set of differential equations at each point of the 3-dimensional grid in \eqref{Iofk}, that is, at each configuration $k$, $p$ and $|\vk-\vp|$. This, of course, makes the computation of the loop integrals quite slow, precluding efficient Markov Chain exploration in order to draw cosmological parameter constraints via a full-shape analysis.

The perturbative kernels developed in \cite{Aviles:2020cax,Aviles:2021que} differ from EdS by two kind of contributions: The first one is due to the growth rate discussed above. The second is pure non-linear coming from the failure of the approximation $f^2 = \Omega_m$ that should be obtained by solving differential equations via Green functions in a similar manner than in $\Lambda$CDM, with the difference that in our case these corrections are also scale-dependent and so, the solution to these equations should be obtained at each configuration of internal and external wave-vectors in loop integrals, and not only one time as in the massless neutrino case. It turns out that the former contribution, introducing the free-streaming scale, is much more important than the latter \cite{Aviles:2021que}, that plays a role similar to the normalization factors that exist between $\Lambda$CDM and EdS kernels. If we keep only the growth rates in the kernels, there is no necessity of solving differential equations and one can write the functions $\mI$ as in eq.~\eqref{IofkExp}. This approach has two important advantages over the use of full kernels for efficient evaluation: there is no need to solve differential equations and one can use FFTLog methods.

In this paper we present our analytical modeling based on the above mentioned approximation, and a numerical method with its implementation into the \texttt{FOLPS$\nu$} Python code.\footnote{Publicly available at \href{https://github.com/henoriega/FOLPS-nu}{https://github.com/henoriega/FOLPS-nu}.} We validate the formalism by comparing against the redshift space power spectrum multipoles of halos obtained from the \textsc{Quijote} simulations suite \cite{Villaescusa-Navarro:2019bje}.\footnote{\href{https://github.com/franciscovillaescusa/Quijote-simulations}{https://github.com/franciscovillaescusa/Quijote-simulations}.} We obtain very satisfactory results, recovering the {\it standard} cosmological parameters and the neutrino mass.   
We do some unrealistic simplifications in our analyses since the reach of this paper is to validate our methodology: first we use halos, instead of galaxies, which together with the use of 100 simulation boxes each with volume $(1 \, h^{-1}\text{Gpc})^3$, allow us to have a small covariance and so, any mismatch in recovering the cosmological parameters of the simulations is probably more due to errors on our perturbative approach and not due to statistical errors, or the imperfection on the galaxy-halo connection modeling that would be present if we use e.g. HODs instead of halos. Second, by using a large value for the sum of the neutrinos' three mass states, $\sum_i m_{\nu,i} = 0.4 \, \text{eV}$, which correspond to the largest mass available in the \textsc{Quijote} suite, we test the theory in the regimes where the use of non-EdS kernels are more relevant. Although, we also test against the mass neutrinos' case with $\sum_i m_{\nu,i} = 0.1 \, \text{eV}$.

The rest of the paper is organized as follows: In   \S\ref{sect:neutrinos}, we introduce the Neutrino free-streaming scale, our key approximations and the linear theory.   \S\ref{sec:PT} reviews the PT theory developed in \cite{Aviles:2020cax,Aviles:2020wme, Aviles:2021que} and introduces the \texttt{fk}-kernels. In   \S\ref{sec:FFTLog} we present the FTTLog method we use which is complemented in appendix \ref{appB}. The validation of our model is presented in   \S\ref{sec:modelvalidation} where we test our modeling to the \textsc{Quijote} suite of simulations; complementary plots are shown in appendix \ref{app:plots}. In   \S\ref{sec:5} we give a brief review and some useful information about our code \texttt{FOLPS}$\nu$. Finally, in   \S\ref{sect:conclusions} we present our conclusions.

\end{section}

\begin{section}{Neutrino free-streaming scale and linear evolution} \label{sect:neutrinos}

At very early times, primordial neutrinos are coupled to the primeval plasma, with a
common temperature $T_\nu=T$. At this stage, neutrinos are relativistic following a Fermi-Dirac distribution. As the Universe expands, it cools down as $T \propto 1/a$ until the weak interaction rate $\Gamma_{\nu} \sim G^2_\text{F} T^5$  falls below  the expansion rate $H\sim G^{1/2}_\text{N} T^2$, where $G_\text{F}$ and $G_\text{N}$ are the Fermi and Newton constants, and neutrinos decouple from the cosmic plasma when $\Gamma_{\nu} \sim H$. The decoupling temperature is $T_\text{dec} \sim 1 \, \text{MeV}$. Since particle physics experiments put a limit on the neutrino masses $m_\nu \lesssim \mathcal{O}(1\,\text{eV})\ll T_\text{dec}$, the neutrinos are still relativistic at decoupling.
After that, the neutrinos do not interact anymore and follow the same Fermi-Dirac distribution, with $T_\nu \propto 1/a$ no longer a ``true'' temperature, but 
a parameter of a distribution that has been freezed out.
Neutrinos become non-relativistic when the mean energy per particle falls below their mass $m_{\nu,\text{i}}$. This occurs at redshift $1+z_{\text{nr, i}} \approx 1890 \,m_{\nu,\text{i}} \, \text{eV}^{-1}$ for each species. Since then, the neutrinos behave as dark matter, and nowadays their abundance is
\begin{equation}
\Omega_{\nu} =
\frac{M_\nu}{93.14 \, h^2\,\text{eV}}, \qquad \qquad
M_\nu \equiv \sum_\text{i} m_{\nu,\text{i}}.
\end{equation}
where we defined capital ``$M_\nu$'' as the sum of the masses of the three species.
Despite massive neutrinos become non-relativistic at very early times, their velocity dispersion is large  \cite{Shoji:2010hm,Lesgourgues:2006nd}
\begin{equation}
  \sigma_\nu(z)  \simeq \sqrt{\frac{15 \zeta(5)}{\zeta(3)}}  \frac{ T_\nu}{m_{\nu}} \simeq 1811\, (1+z) \left( \frac{0.1 \,\text{eV}}{m_\nu}\right) \,\text{km/s}.
\end{equation}

This velocity dispersion acts as a pressure support counteracting the action of gravity. As in a Jeans-like mechanism, the neutrinos cannot be confined and contribute to the formation of structure below a certain scale, named the {\it free-streaming scale}. This corresponds to a wave-number \cite{Shoji:2010hm}
\begin{equation} \label{eqkFS}
 k_\text{FS}(z) = \sqrt{ \frac{4 \pi G \bar{\rho} a^2}{\frac{5}{9}\sigma^2_\nu}} \simeq 0.0908\frac{H(z)/H_0}{(1+z)^2}  \left( \frac{m_\nu}{0.1 \,\text{eV}} \right) \,h\,\text{Mpc}^{-1}. 
\end{equation}
This expression gives us an idea of the involved scales, but is never used in this work. In fact, there are several definitions for the free-streaming scale in the literature, all of them having similar numerical values. Notice that the above equation is valid when neutrinos are non-relativistic, while before that, their free-streaming was equal to the Hubble horizon. Hence, the free-streaming has a minimum at the non-relativistic transition redshift, 
and all clustering on length scales below it becomes suppressed by the presence of massive neutrinos  \cite{Lesgourgues:2006nd}. For realistic neutrino masses this scale is about $1\, h^{-1} \text{Gpc}$, much larger than $1/k_\text{FS}(z)$ at recent times.
Further, eq.~\eqref{eqkFS} is valid for a single neutrino species, corresponding to the three different masses. From now on, we will consider equal mass neutrinos, so $M_\nu = 3\times m_{\nu,i}$ and there is only one single free-streaming scale; although, our method does not depend on this choice. Three equal masses is also commonly assumed in $N$-body simulations. Below the free-streaming scale $k \ll k_\text{FS}$ neutrinos behave as cold dark matter contributing to the formation of structures, while above it  $k \gg k_\text{FS}$ they do not cluster anymore, producing a suppression of the power spectrum and correlation function for scales below than the free-streaming.


At late times, inside the matter dominated epoch, the gravitational potential is related to the matter fields through the Poisson equation
\begin{equation}
   - \frac{k^2}{a^2} \Phi(\vk,t) = \frac{3}{2} \Omega_m(t) H^2(t) \big( f_{cb} \delta_{cb}(\vk,t) + f_\nu \delta_{\nu}(\vk,t) \big),
\end{equation}
where $f_{X}= \Omega_X/\Omega_m$ is the relative abundance of the matter component $X$ to the total matter $m$. Since the only clustering components are assumed to be neutrinos, cold dark matter and baryons, then $f_\nu = 1-f_{cb}$. We are primarily interested in the evolution of the $cb$ field, since to a very good approximation galaxies are tracers of it, and not of the total matter field \cite{Ichiki:2011ue,Villaescusa-Navarro:2013pva,Castorina:2013wga,Vagnozzi:2018pwo,Banerjee:2019omr}. Hence, our approach takes the \revised{non-linear} neutrino density field as an intermediate quantity that we approximate as \cite{Aviles:2021que}
\begin{equation} \label{deltanu}
\delta_{\nu}(\vk,t) = \frac{T_\nu(k,t)}{T_{cb}(k,t)} \delta_{cb}(\vk,t),
\end{equation}
with $T_{\nu,cb}$ the  \revised{linear} transfer functions of the neutrinos and the $cb$ field. 
That is, we do not assume that the neutrino field is linear, which would violate momentum conservation \cite{Blas:2014hya}, but instead that it is proportional to the non-linear $cb$ density field. Our approximation was shown to be very accurate for the total matter power spectrum up to non-linear scales using simulations (see figure 3 of \cite{Aviles:2021que}). The same, or very similar, approximations were considered in some works \cite{Lesgourgues:2009am,Upadhye:2013ndm,Levi:2016tlf}, but their perturbative theories differ than our in almost any other aspect. 

Using eq.~\eqref{deltanu}, the Poisson equation becomes
\begin{equation}
    - \frac{k^2}{a^2} \Phi(\vk,t) = A(k,t) \delta_{cb}(\vk,t)
\end{equation}
with
\begin{equation} \label{Aofk}
  A(k,t) \equiv \frac{3}{2} \Omega_m(t) H^2(t) \left( f_{cb} + f_\nu \frac{T_\nu(k,t)}{T_{cb}(k,t)}\right). 
\end{equation}

At large scales, the ratio of transfer functions is unity and $A \rightarrow 4 \pi G \bar{\rho}_m$, while the neutrino transfer function falls sharply about the free-streaming scale and $A  \rightarrow 4 \pi G \bar{\rho}_{cb}$ at small scales.

Under our approximation, the $cb$ field evolves according to the fluid equations,
\begin{align}
\partial_t\delta_{cb}(\vx,t)  + \frac{1}{a}\partial_i \big[(1+\delta_{cb})v^i_{cb} \big] &= 0, \label{FEcontpre}\\  
\partial_t v^i_{cb}(\vx,t)+ \frac{1}{a} v^j_{cb}\partial_j v^i_{cb}  + H v^i_{cb}  + \frac{1}{a} \partial^i \Phi & =0 \label{FEEulerpre},  
\end{align}
where $\vv_{cb}$ is the peculiar velocity of the $cb$ field. Linearizing the above equations for a longitudinal velocity field, we can solve for the linear order density field
\begin{equation} \label{deltacb_decomposition}
     \delta_{cb}^{(1)}(k,t) = D_+(k,t) \delta_{cb}^{(1)}(k,t_0), 
\end{equation}
where $D_+$ is the scale-dependent growth function given by
\begin{equation} \label{DplusEq}
    \frac{d^2 D_+(k,t)}{dt^2} + 2 H  \frac{d D_+(k,t)}{dt} - A(k,t) D_+(k,t) = 0
\end{equation}
with the initial conditions given in \cite{Hu:1997mj} to pick up the growing solution.  We choose the normalization  $D_+(\revised{k\rightarrow 0},t_0)=1$ with $t_0$ the present time. Notice that although the decomposition of eq.~\eqref{deltacb_decomposition} looks arbitrary, with this normalization it becomes unique and coincides with the massless neutrinos case at large scales.

 \begin{figure}
 	\begin{center}
 	\includegraphics[width=5.0 in]{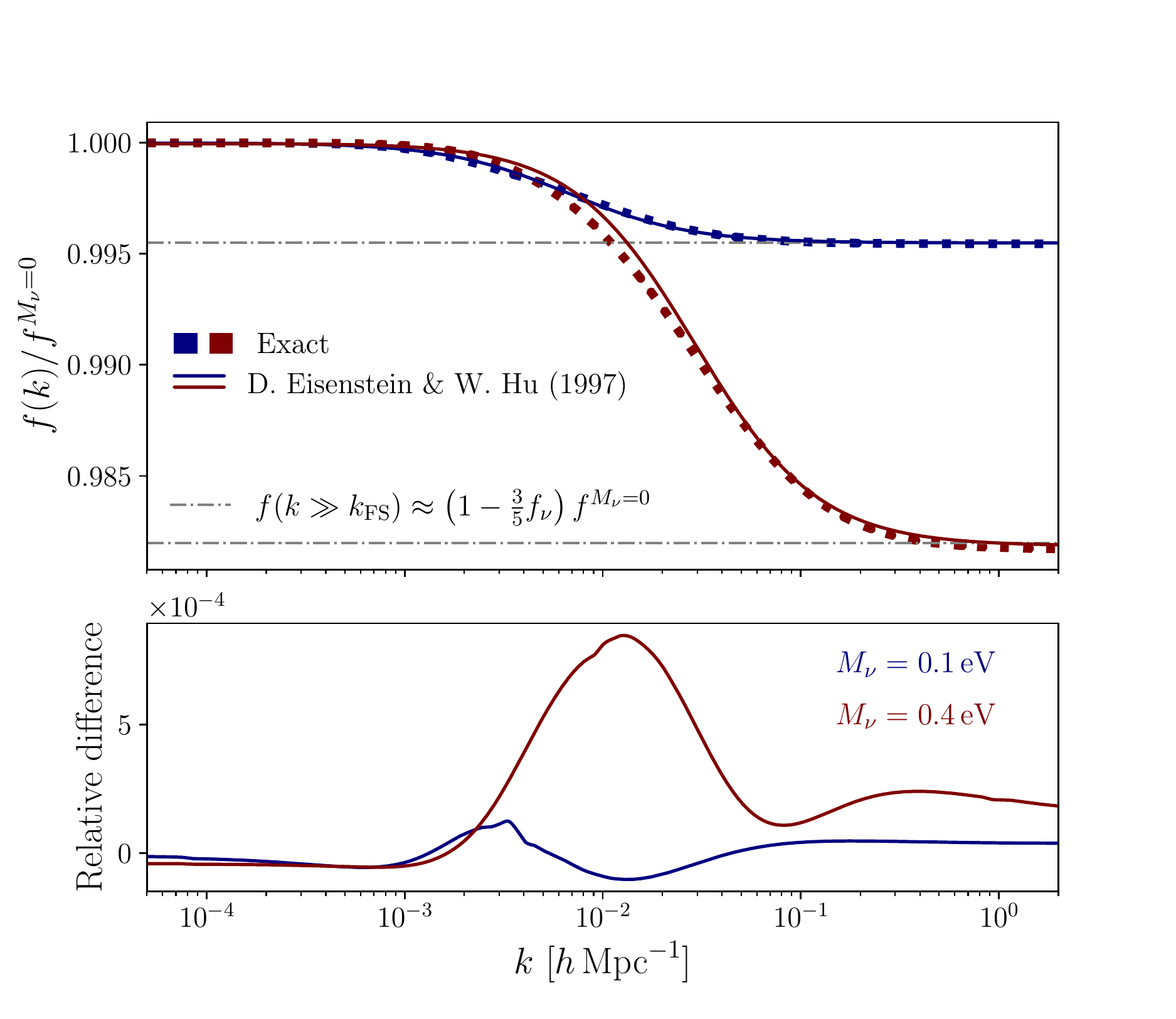}
 	\caption{Growth rate $f(k)$ divided by their large scale limits $f_0=f(k\rightarrow 0) \equiv f^{M_\nu=0}$ evaluated at redshift $z=0.5$, for two cosmologies including massive neutrinos with $M_\nu=0.1\,\text{eV}$ (blue lines) and $M_\nu=0.4\,\text{eV}$ (red lines). We use eqs.~\eqref{DplusEq} and \eqref{fk} (dotted lines) which we call it ``exact'', and the algebraic approximation of Hu \& Eisenstein (1998) given in \cite{Hu:1997vi} (solid lines).   The horizontal dot-dashed gray lines show the small scales limit.
 	\label{fig:fk_over_f0}}
 	\end{center}
 \end{figure}

The linear (logarithmic) growth rate is given by 
\begin{equation} \label{fk}
    f(k,t) = \frac{d \ln D_+(k,t)}{d \ln a(t)},
\end{equation}
which is again scale-dependent. We define its value at very large scales as  \cite{Valogiannis:2019nfz}
\begin{equation}
    f_0(t) \equiv f(\revised{k\rightarrow 0} ,t) = f^{M_\nu=0}(t),
\end{equation}
which coincides with the scale-independent growth rate $f(t)$ of a cosmology with the same amount of total matter, but on which the neutrinos are massless. In figure \ref{fig:fk_over_f0} we show a plot for $f(k,t)$ computed directly using eqs.~\eqref{DplusEq} and \eqref{fk} in dotted lines and using the approximate formulae given in \cite{Hu:1997vi} in solid lines. The discrepancy among both methods is smaller than the $0.1\%$ even for the largest mass of $M_\nu=0.4\, \text{eV}$. The gray horizontal lines show the limit of small scales, $f(k \gg k_\text{FS}) \rightarrow (1-\frac{3}{5} f_\nu) f_0$, which corresponds to the maximum suppression, that is, when the contributions from neutrinos is completely negligible. The Hu-Eisenstein approximation is used in our code because otherwise one has to compute the neutrinos and $cb$ transfer functions for each wave-number and for each cosmology, which is very time-consuming.

As is common, we define the longitudinal piece of the velocity field as
\begin{equation}
    \theta_{cb}(\vx,t) = -\frac{\partial_i v^i_{cb}}{a H f_0 }(\vx,t).
\end{equation}
With this, we obtain the relation \eqref{Eq:linear_theta} between linear velocity and density fields.

As a recapitulation, our method relies on the approximation for the neutrino density field given by eq.~\eqref{deltanu}. Further, we are assuming that the neutrinos late times velocity dispersion tensor has no impact on the gravitational potentials of the metric, and so they become equal $\Psi=\Phi$. It is known that this is no precise and actually the velocity dispersion leads to the very definition of free-streaming scale \cite{Aviles:2015osc} and the generation of vorticity \cite{Cusin:2016zvu}. But, by considering eq.~\eqref{deltanu}, we are actually restricting our theory to the $cb$ field in the presence of massive neutrinos, and we do not deal directly with the non-linear evolution of the neutrinos themselves. This can be seen as a disadvantage, since one cannot construct a consistent non-linear total matter power spectrum, and in this respect other approaches, particularly the one of \cite{Senatore:2017hyk}, are more complete. But as we mentioned, the galaxy field is a biased tracer of the $cb$ fluid and not of the total matter fluid, and this is the object of our primary interest in this work. However, for some applications one deals with the total real space matter power spectrum, as in weak lensing. In such a case, the approximation of eq.~\eqref{deltanu} implies
\begin{equation}
    P_m(k,t) = P_{cb}(k,t) \left[ (1-f_\nu)^2 + 2 (1-f_\nu) f_\nu \frac{T_\nu(k,t)}{T_{cb}(k,t)}  + f_\nu^2 \left( \frac{T_\nu(k,t)}{T_{cb}(k,t)} \right)^2\right],
\end{equation}
which still gives very good results \cite{Aviles:2020cax}.
\end{section}

\begin{section}{Perturbative model in massive neutrino cosmologies}\label{sec:PT}

The peculiar \revised{velocities} of galaxies $\vec v$ produce a Doppler effect along the line-of-sight direction $\vhn$ (we assume the distant observer approximation where we fix $\vhn$), breaking the real space isotropy, such that the power spectrum is given by \cite{Scoccimarro:2004tg,Vlah:2018ygt}
\begin{equation} \label{pkScocc}
 (2\pi)^3 \dD(\vk)  + P_s(\vk) = \int d^3x \, e^{-i\vk \cdot \vx} \Big[ 1 + \mathcal{M}(\vec J=\vk, \vx )\Big],
\end{equation}
in terms of the density-weighted velocity moments generating function
\begin{equation}
    \mathcal{M}(\vec J, \vx ) = \left\langle \big(1+\delta(\vx_1)\big)\big(1+\delta(\vx_2)\big) e^{- i \vec J \cdot \Delta \vec u  }  \right\rangle,
\end{equation}
with $\vx = \vx_2-\vx_1$ and $\Delta \vec u= \vec u(\vx_2) - \vec u(\vx_1)$. The ``velocity'' $\vec u$ and the peculiar velocities are related by 
\begin{equation}
    \vec u(\vx) = \vhn  \frac{\vec v (\vx) \cdot \vhn }{a H}.
\end{equation}

The $m$-th density weighted velocity field moment of the generation function is a $m$-rank tensor defined as
\begin{align}
 \Xi^{m}_{ i_1 \cdots i_\tm}(\vx) &\equiv i^\tm \frac{\partial^\tm}{\partial J_{i_1}\cdots \partial J_{i_m}} \big[ 1+\mathcal{M}(\vec J,\vx) \big] \Big|_{\vec J = 0}
 = \langle \big(1+\delta_1\big)\big(1+\delta_2\big)\Delta u_{i_1}\cdots\Delta u_{i_\tm} \rangle,
\end{align}
with $\delta_1 =\delta(\vx_1)$ and $\delta_2 =\delta(\vx_2)$.

By expanding in this way eq.~\eqref{pkScocc}, we get the power spectrum at the moment expansion approach,
\begin{align} \label{pkME}
   (2\pi)^3 \dD(\vk) + P_s(\vk)
      &= \sum_{\tm=0}^\infty \frac{(-i)^\tm}{\tm!} k_{i_1}\dots k_{i_\tm}  \tilde{\Xi}_{i_{1}\cdots i_{\tm}}^{\tm}(\vk),
\end{align}
where the $\tilde{\Xi}^{\tm}_{i_{1}\cdots i_\tm}(\vk)$  are the Fourier moments of the generating function 
---the Fourier transforms of their configuration space counterparts, $\Xi^{\tm}_{i_{1}\cdots i_{n}}(\vx)$:

\begin{equation}
  \tilde{\Xi}_{i_{1}\cdots i_{\tm}}^{\tm}(\vk) = \int d^3x \, e^{-i\vk\cdot \vx} \,\Xi^{\tm}_{ i_1 \cdots i_\tm}(\vx). 
\end{equation}

For rotational invariant $S(\vk,\vp)$ function one has  \cite{Philcox:2020srd,Aviles:2020wme}
\begin{align} \label{pdotnG}
 \int \Dk{p} (\hat{\vp} \cdot \vhn)^n S(\vk,\vp)  &= 
 \sum_{m=0}^n \mu^m \int \Dk{p} G_{nm}(\hat{\vk}\cdot\hat{\vp})  S(\vk,\vp) \nonumber\\
 &= \frac{k^3}{4\pi^2} \sum_{m=0}^n \mu^m  \int_0^\infty dr \, r^2 \int_{-1}^1 dx \, G_{nm}(x)  S(k,r,x), 
\end{align}
with $x=\hat{\vk}\cdot\hat{\vp}$ and $r= p/k$, and
\begin{align}\label{Gnm}
G_{nm}(x) &=   \sum_{\ell=0}^n \frac{(1+(-1)^{\ell+n}) (2 \ell+1)}{2(1+\ell+n)}  \binom{\ell}{m} \binom{2 \ell}{\ell}  \binom{\frac{\ell+m-1}{2}}{\ell} \nonumber\\
&\quad \times \,
{}_3F_2\left(\frac{1-\ell}{2},-\frac{\ell}{2},\frac{1}{2} (-1-\ell-n);\frac{1}{2}-\ell,\frac{1}{2} (1-\ell-n); 1 \right)
\mathcal{L}_\ell(x),
\end{align}
with $\mathcal{L}_\ell$ the Legendre polynomial of degree $\ell$, and ${}_3F_2(a_1,a_2,a_3;b_1,b_2; z)$ the extended hypergeometric function of the type $p=3$ and $q=2$.

Using eq.\eqref{pdotnG},
and the properties of $n$-point correlators of homogeneous and isotropic fields it can be shown that
\begin{equation}
  \frac{(-i)^\tm}{\tm!} k_{i_1}\dots k_{i_\tm}  \tilde{\Xi}_{i_{1}\cdots i_{\tm}}^{\tm}(\vk) = \sum_{n=0}^m \mu^{2n} f_0^m I^m_n(k),  
\end{equation}
for only momentum magnitude dependent functions $I^m_n(k)$,
recovering eq.\eqref{Pkmu}. The case that we will use, up to 1-loop corrections, has been shown in \cite{Aviles:2020wme}.

\begin{subsection}{Perturbation Theory with full kernels}

As was mentioned in  \S\ref{sect:neutrinos}, there is some evidence that galaxies and other cosmic objects are tracers of the combined dark matter and baryonic fields, and not of the total matter field that also includes the neutrinos. The underlying reason is not completely clear but can be understood from peak theory since primeval density overdensities, from which halos will arise, are formed early on, before the neutrinos are non-relativistic; and moreover, these proto-halos do not capture the neutrinos because of the free-streaming. Up to mildly non-linear scales, the combined CDM and baryonic fields behave essentially as a single fluid, named $cb$. The scale at which this stops to occur is also not clear and is highly dependent on how the baryonic physics are modeled \cite{Chisari:2019tus,2020MNRAS.495.4800A}; however, the consensus is that at large scales, those reached by perturbation theory and EFTofLSS, baryonic effects can be neglected to a very good approximation \cite{Schneider:2018pfw,Chisari:2019tus,2020MNRAS.495.4800A}. Therefore, we are interested in the loop corrections of the $cb$ power spectrum, and on top of that we model the tracers with a flexible biasing scheme.  

There are several perturbation theory schemes in the literature, we follow here the one developed recently in \cite{Aviles:2020cax,Aviles:2021que}. The first step was to compute the Lagrangian Perturbation Theory (LPT) kernels $L^{(n)}_i$, which are easier to obtain than their Eulerian, Standard Perturbation Theory (SPT) counterparts when additional scales are involved. Thereafter, these kernels can be mapped to the Eulerian frame, to obtain the $F_n$ and $G_n$.  In the following, we summarize the results of \cite{Aviles:2020cax,Aviles:2021que}.

The $cb$ fluid elements follow trajectories given by
\begin{equation}
    \vx = \vq + \Ps(\vq,t), 
\end{equation}
where $\vq$ is the initial position and $\Ps$ is the Lagrangian displacement, which is obtained through the geodesic equation
\begin{equation}
    \ddot{\Ps}(\vq,t) + 2 H \dot{\Ps}(\vq,t) = - \nabla \Phi(\vx,t)\big|_{\vx= \vq + \Ps(\vq,t)}.
\end{equation}
The Lagrangian displacement is expanded in a Taylor series in Fourier space, $\Ps = \sum_n \Ps^{(n)}$. At order $n$\footnote{For simplicity, we adopt the notations
\begin{equation}
 \underset{\vk_{1\cdots n}= \vk}{\int} 
 = \int \Dk{\vk_1}\cdots \Dk{\vk_n}(2\pi)^3 \dD(\vk_{1\cdots n}-\vk)\quad \mathrm{ and }\quad  \vk_{1\cdots n} = \vk_1 + \cdots + \vk_n.
\end{equation}
}
\begin{equation}\label{LDkernels}
 \Psi_i^{(n)}(\vk,t) = \frac{i}{n!} \underset{\vk_{1\cdots n}= \vk}{\int} L^{(n)}_i(\vk_1,\cdots,\vk_n;t)\delta_{cb}^{(1)}(\vk_1,t)\cdots \delta_{cb}^{(1)}(\vk_n,t). 
\end{equation}
In \cite{Aviles:2020cax} the $L^{(n)}_i$ kernels are obtained, and we reproduce them here. The linear order kernel is
\begin{equation}
     L_i^{(1)}(\vk)=  \frac{k_i}{k^2},
\end{equation}
which is identical when neutrinos are massless. To second and third orders,
\begin{align}
 L^{(2)}_i(\vk_1,\vk_2,t) &=  \frac{3}{7} \frac{k^i_{12}}{k^2_{12}} \left( \mathcal{A} - \mathcal{B} \, \frac{(\vk_1\cdot\vk_2)^2}{k_1^2 k_2^2} \right), \label{LKL2}\\
  L^{(3)}_i(\vk_1,\vk_2,\vk_3,t) &=  \frac{k^i_{123}}{k^2_{123}}
     \Bigg\{ \frac{5}{7} \left( \mathcal{A}^{(3)} -\mB^{(3)}
               \frac{(\vk_2 \cdot \vk_3)^2}{k^2_2 k^3_2} \right) 
              \left( 1-  \frac{(\vk_1 \cdot \vk_{23})^2}{k_1^2 k_{23}^2} \right)  \nonumber\\
     &   -\frac{1}{3} \left( \, \mathcal{C}^{(3)}  -3 \mathcal{D}^{(3)} \frac{(\vk_2 \cdot \vk_3)^2}{k^2_2 k^2_3} 
      + 2 \mathcal{E}^{(3)} \frac{(\vk_1 \cdot \vk_2)(\vk_2 \cdot \vk_3)(\vk_3 \cdot \vk_1)}{k_1^2 k^2_2 k^2_3} \, \right)
      \Bigg\}. \label{LKL3}
 \end{align} 
Functions $\mA$ and $\mB$ in the second order kernels depend on the wave-vectors $\vk_1$ and $\vk_2$ and on time $t$, while all the functions in the third order Lagrangian kernel $\mathcal{A}^{(3)}, \dots, \mathcal{E}^{(3)}$ depend on the same arguments as $L^{(3)}$, that is on $\vk_1$, $\vk_2$ and $\vk_3$ \revised{and on time $t$}. In the case of $\Lambda$CDM with massless neutrinos all these functions are only time dependent and reduce to unity for EdS evolution. 
These functions completely determine the kernels and should be obtained by solving a system of linear differential equations at each wave-vector configuration and for each set of cosmological parameters $\Omega_{cb}$, $\Omega_\nu$ and $h$. The expressions for them are not displayed here, not only because they are large but also because they are not used by the method presented in this work, they can be found in \cite{Aviles:2020cax}.   

We use the functions $C^{(n)} \Gamma_n(\vk_1,\cdots,\vk_n,t)$ and $C^{(n)} \Gamma_n^f(\vk_1,\cdots,\vk_n,t)$ as the kernels of the fields $-i\vk \cdot \Ps^{(n)}$ and $-i (n H f_0)^{-1} \vk \cdot \dot{\Ps}^{(n)}$, respectively \cite{Valogiannis:2019nfz}. The numbers $C^{(n)}$ are chosen for algebraical convenience and have the values $C^{(1)}=C^{(3)}=1$ and $C^{(2)}=3/7$. As usual, we assume the Lagrangian displacement is longitudinal at early times, and as such the only relevant kernels for 1-loop, 2-point statistics are the $\Gamma_n$ functions. In terms of the standard Lagrangian kernels $L^{(n)}$, one can write  
\begin{align} \label{Gamman}
 C^{(n)} \Gamma_n(\vk_1,\cdots,\vk_n,t) &= k^i_{1\cdots n} L_i^{(n)}(\vk_1,\cdots,\vk_n,t).
\end{align}
and 
\begin{equation}\label{Gammafn}
 \Gamma^f_n(\vk_1,\cdots,\vk_n,t) =  \Gamma_n(\vk_1,\cdots,\vk_n,t) \frac{f(k_1)+\cdots+f(k_n)}{n f_0} + \frac{1}{n f_0 H} \dot{\Gamma}_n(\vk_1,\cdots,\vk_n,t) .
\end{equation}
For the case of EdS evolution, $\Gamma_n = \Gamma_n^f$, and one obtains the widely used relation  $\dot{\Ps}^{(n)} = n H f \Ps^{(n)}$ \cite{Matsubara:2007wj,Carlson:2012bu}.  

The $cb$ density fluctuation becomes \cite{Aviles:2021que}
\begin{align}\label{dcb_LF}
\delta_{cb}(\vk) 
&= \sum_{m=1}^\infty \frac{(-i)^m}{m!} k_{i_1}\cdots  k_{i_m} \underset{\vp_{1 \cdots m}= \vk}{\int} \Psi_{i_1}(\vp_1)\cdots \Psi_{i_m}(\vp_m).
\end{align}
which is obtained from the well-known relation $\delta(\vk) = \int d^3q\; e^{-i\vk\cdot \vq}\Big[ e^{-i \vk \cdot \Ps(\vq)} - 1 \Big] $.
Similarly, considering the peculiar velocity $\vv_{cb} = a \dot{\Ps}$, the $\theta_{cb}$ field is obtained through \cite{Aviles:2021que,Aviles:2018saf} 
\begin{align} \label{tfpsid}
\theta_{cb}(\vk) 
              &= -\frac{1}{H f_0 } \sum_{m=0}^{\infty} \frac{(-i)^m}{m!}\int d^3 q\; e^{-i\vk \cdot \vq} \big(\vk \cdot \Ps\big)^m  \nonumber\\
              &\quad \times \Big[   \, \dot{\Psi}_{i,i} +  (\delta_{ij}\delta_{ab} - \delta_{ia}\delta_{jb})  \Psi_{a,b}\dot{\Psi}_{i,j}    
               +  \frac{1}{2}\epsilon_{ikp}\epsilon_{jqr}    \Psi_{k,q} \Psi_{p,r}\dot{\Psi}_{i,j} \Big], 
\end{align}
with the Lagrangian displacements evaluated at the Lagrangian coordinate $\vq$, \revised{and $\epsilon_{ijk}$ is the Levi-Civita symbol and $\delta_{ij}$ the Kronocker delta}.

On the other hand, the kernels $F_n$ and $G_n$ are defined via the following equations for the $n$-th order density fluctuation and velocity
\begin{align}
 \delta_{cb}^{(n)}(\vk,t) &= \underset{\vk_{1\cdots n}= \vk}{\int} F_n(\vk_1,\cdots,\vk_n;t)\delta_{cb}^{(1)}(\vk_1,t)\cdots \delta_{cb}^{(1)}(\vk_n,t), \label{dcbExp}\\
 \theta_{cb}^{(n)}(\vk,t) &= \underset{\vk_{1\cdots n}= \vk}{\int} G_n(\vk_1,\cdots,\vk_n;t)\delta_{cb}^{(1)}(\vk_1,t)\cdots \delta_{cb}^{(1)}(\vk_n,t). \label{tcbExp}
\end{align}
Comparing order by order eq.~\eqref{dcbExp} with \eqref{dcb_LF} and eq.~\eqref{tcbExp} with \eqref{tfpsid} one obtains the Eulerian kernels in terms of the Lagrangian ones. For our purposes, we are interested in the specific configurations of wave-vectors that enter in 1-loop integrals of 2-point statistics: for second order kernels, these are $\vk_1=\vp$ and $\vk_2=\vk - \vp$, while for third order $\vk_1= \vk$, $\vk_2= -\vp$ and $\vk_3= \vp$. We obtain
\begin{equation}
 F_1(k) = 1, \qquad   G_1(k) = \frac{f(k)}{f_0},
\end{equation}
\begin{align}
 F_2(\vp,\vk-\vp) &= \frac{3}{14} \Gamma_2(\vp,\vk-\vp) + \frac{1}{2}\frac{(\vk\cdot \vp) (\vk \cdot (\vk-\vp))}{p^2 |\vk-\vp|^2},  \label{LPTtoF2}\\ 
G_2(\vp,\vk-\vp)  &= \frac{3}{7} \Gamma^f_2(\vp,\vk-\vp) + \frac{(\vp \cdot (\vk-\vp))^2}{p^2 |\vk-\vp|^2} \frac{f(|\vk-\vp|)+f(p)}{2f_0} \nonumber\\
&\quad + \frac{1}{2}\frac{\vp \cdot (\vk-\vp)}{p \,|\vk-\vp|}  \left( \frac{|\vk-\vp|}{p}\frac{f(|\vk-\vp|)}{f_0}+\frac{p}{|\vk-\vp|}\frac{f(p)}{f_0} \right), \label{LPTtoG2} 
\end{align}
and,
\begin{align}
 F_3(\vk,-\vp,\vp)  &= \frac{1}{6}  \Gamma_3(\vk,-\vp,\vp)  +  \frac{1}{7} \frac{\vk\cdot(\vk-\vp) \, \vk \cdot \vp}{p^2|\vk-\vp|^2} \Gamma_2(\vk,-\vp)  - \frac{1}{6} \frac{(\vk \cdot \vp)^2}{p^4}, \label{LPTtoF3}\\
    G_3(\vk,-\vp,\vp) &=  
   \frac{1}{2} \Gamma^{f}_3(\vk,-\vp,\vp)   +  \frac{2}{7} \frac{\vk \cdot \vp}{p^2} \Gamma^{f}_2(\vk,-\vp)   + \frac{1}{7}  \frac{f(p)}{f_0}  \Gamma_2(\vk,-\vp) \frac{\vk \cdot (\vk-\vp)}{|\vk-\vp|^2} \nonumber\\
&\quad     -  \frac{1}{7}   \left[ 2 \Gamma_2^f(\vk,-\vp) + \Gamma_2(\vk,-\vp) \frac{f(p)}{f_0} \right]\left[1 -\frac{(\vp \cdot (\vk-\vp))^2}{p^2|\vk-\vp|^2}  \right] \nonumber\\
&\quad - \frac{1}{6} \frac{(\vk \cdot \vp)^2}{p^4} \frac{f(k)}{f_0} . \label{LPTtoG3}
\end{align}
where the $\Gamma_3$ kernels used in the above equations are constructed by first symmetrizing $L^{(3)}_i(\vk_1,\vk_2,\vk_3)$ in eq.~\eqref{LKL3} over its three arguments and thereafter evaluating at $\vk_1= \vk$, $\vk_2= -\vp =- \vk_3$. For $\Omega_{cb}(a)=1$, one has $f(k)=f_0$ and all the scale- and time-dependent functions $\mA, ... $ are unity, reducing the above expressions to the standard EdS kernels.

The integrals for the loop corrections have exactly the same form as in the standard massless neutrino case, but with the use of the corrected kernels. The $\delta\delta$, $\delta\theta$ and $\theta\theta$ power spectra are
\begin{equation}\label{Pab}
P^{\text{1-loop}}_{ab}(k) = P^L_{ab}(k) +P_{ab}^{22}(k) + P_{ab}^{13}(k),  
\end{equation}
where $a$ and $b$ refer to $\delta$ or $\theta$ fields, and linear power spectra $P^L_{ab}(k)$, 
 \begin{equation} \label{Plinear}
 P^L_{\delta\delta}(k) \equiv P_L(k) , \quad
 P^L_{\delta\theta}(k) =  \frac{f(k)}{f_0} P_L(k), \quad
 P^L_{\theta\theta}(k) =  \left(\frac{f(k)}{f_0} \right)^2 P_L(k),
 \end{equation}
with $P_L$ the linear $cb$ real space power spectrum, and leading non-linear contributions  given by\footnote{Hereafter we use the notation 
\begin{equation}
    \ip = \int \Dk{p}.
\end{equation}}
\begin{align}
P_{\delta\delta}^{22}(k) &=   2 \ip \big[F_2(\vp,\vk-\vp)\big]^2 P_L(p) P_L(|\vk -\vp|), \label{P22dd}\\
P_{\delta\theta}^{22}(k) &=   2 \ip F_2(\vp,\vk-\vp)G_2(\vp,\vk-\vp) P_L(p) P_L(|\vk -\vp|), \\
P_{\theta\theta}^{22}(k) &=   2 \ip \big[G_2(\vp,\vk-\vp)\big]^2 P_L(p) P_L(|\vk -\vp|), \label{P22vv} \\
P_{\delta\delta}^{13}(k) &=   6  P_L(k) \ip F_3(\vk,-\vp, \vp) P_L(p),\\
P_{\delta\theta}^{13}(k) &=   3  P_L(k) \ip \big[F_3(\vk,-\vp, \vp)\frac{f(k)}{f_0} + G_3(\vk,-\vp, \vp) \big] P_L(p),\\
P_{\theta\theta}^{13}(k) &=   6  \frac{f(k)}{f_0} P_L(k) \ip G_3(\vk,-\vp, \vp) P_L(p). \label{P13vv}
\end{align}
These integrals can be performed straightforwardly with standard methods in perturbation theory. However, the computational time becomes quite large because one has to solve \revised{a set of differential equations} for the functions $\mA, \mB, \mA^{(3)}, \mB^{(3)}$ 
\revised{at each volume element in the 2-dimensional quadratures, since these do not have analytical expressions for
arbitrary momenta, making the computations considerably slower than in the massless neutrinos case.}
This makes the use of these kernels not viable for parameters estimation, where the integrals must be evaluated thousands of times.

We will come back later in \S\ref{subsec:EFT} to the rest of the necessary functions entering eq.~\eqref{Pkmu} and the biasing scheme. But before that we are showing the approximation we use to overcome the difficulties explained in the above paragraph.

\end{subsection}

\begin{subsection}{\fk-kernels}\label{subsect:fkkernels}

The \fk-kernels method consists on fixing the functions $\mA,\cdots$ equal to their $\Lambda$CDM or EdS values, but keeping the linear growth rates $f(k)$. More precisely, \revised{one constructs} the \fk-kernels by taking the $\Gamma_n$ kernels [eq.~\eqref{Gamman}] as their EdS or $\Lambda$CDM counterparts, which are known analytically (we use EdS in this paper) or computed by solving a system of differential equations only once (for $\Lambda$CDM),\footnote{For $\Lambda$CDM, $\mA=\mB$, $\mA^{(3)}= \mB^{(3)}$ and $\mathcal{C}^{(3)}=\mathcal{D}^{(3)}=\mathcal{E}^{(3)}$, and are all functions of time only. Thus, they can be solved just once and thereafter one performs the loop integrals.} and in the $\Gamma_n^f$ kernels of eq.~\eqref{Gammafn} one keeps the $f(k)$ rates. Thereafter, one uses these $\Gamma_n$ into the SPT kernels $F_n$ and $G_n$ expressions of eqs.~\eqref{LPTtoF2}-\eqref{LPTtoG3}.

In this way, the perturbative kernels inherit the free-streaming scale imprinted by the large neutrinos' velocity dispersion, such that also the advection of fields is well modeled at non-linear orders. For example, the difference of linear velocity fields $\theta^{(1)}(\vx + \Ps) - \theta^{(1)}(\vx)$ is a second order effect that yields the contribution
\begin{equation}\label{intheta2}
 \ikk  \frac{\hat{\vk}_1 \cdot \hat{\vk}_2}{2} \left( \frac{k_1}{k_2}\frac{f(k_1)}{f_0} + \frac{k_2}{k_1}\frac{f(k_2)}{f_0}\right) \delta_{cb}^{(1)}(\vk_1)\delta_{cb}^{(1)}(\vk_2)\in \theta^{(2)}(\vk),
\end{equation}
to the velocity field,  where the rates $f(k)/f_0$ appear because of the non-trivial linear relation between density and velocity fields in eq.~\eqref{Eq:linear_theta}.

In ref.~\cite{Aviles:2021que} it is shown that this approximation is quite accurate and the growth rates constitute the most important additional pieces for kernels of the $cb$ fields; see figure 9 of that paper. 

A second advantage of the use of \fk-kernels is that it is possible to write them as the sum of power laws
\begin{equation}\label{plaws}
   p^{2 n_1}|\vk-\vp|^{2 n_2} k^{-2(n_1 + n_2)}
\end{equation}
multiplied by growth rates evaluated either at wave-numbers $p$, $|\vk-\vp|$ or $k$. This allows us to use FFTLog methods \cite{Hamilton:1999uv} to speed up even more the calculations \cite{McEwen:2016fjn,Fang:2016wcf,Schmittfull:2016jsw,Schmittfull:2016yqx,Simonovic:2017mhp}.

For example, the second order kernels are $F_2^\text{\fk} = F_2^\text{EdS}$ and
\begin{align}
G_2^\text{\fk}(\vp,\vk-\vp) &=  \left(\frac{k^4}{14 |\vk-\vp|^2 p^2}+\frac{3 k^2}{28 |\vk-\vp|^2}-\frac{k^2}{7 p^2}-\frac{5 p^2}{28|\vk-\vp|^2}+\frac{|\vk-\vp|^2}{14 p^2}+\frac{3}{28} \right)  \frac{f(p)}{f_0} \nonumber\\
&\quad +  \big( \, p \leftrightarrow |\vk-\vp| \, \big),
\end{align}
and clearly for $f(p)= f(|\vk-\vp|) = f_0$, one has $G_2^\text{\fk}(\vp,\vk-\vp) = G_2^\text{EdS}$. 
We will come back to the numerical method to speed up the loop integrals in \S\ref{sec:FFTLog}

\end{subsection}

\begin{subsection}{EFT galaxy power spectrum multipoles}\label{subsec:EFT}

We use the EFT galaxy power spectrum as given by \cite{Kaiser:1984sw,Scoccimarro:2004tg,Taruya:2010mx,Aviles:2020wme,Aviles:2021que,Perko:2016puo,Chen:2020zjt,Vlah:2018ygt} 
\begin{align}\label{PS_EFT}
P^\text{EFT}_s(k,\mu) &= P_{\delta\delta}(k) + 2 f_0 \mu^2 P_{\delta\theta}(k) + f_0^2 \mu^4 P_{\theta\theta}(k) + A^\text{TNS}(k,\mu) + D(k,\mu) \nonumber\\
&\quad + (\alpha_0 + \alpha_2 \mu^2 + \alpha_4 \mu^4 +\alpha_6 \mu^6) k^2 P_L(k) + \tilde{c} (f_0 \sigma_v k \mu)^4 P_s^K(k,\mu) \nonumber\\
&\quad + P_\text{shot} \big[\alpha^{shot}_0 + \alpha^{shot}_{2} (k\mu)^2 \big].
\end{align}
The first line of the above equation is obtained by using the density-weighted velocity momentum expansion \eqref{pkME} up to 1-loop in perturbation theory: $P_{\delta\delta}$,  $P_{\delta\theta}$ and  $P_{\theta\theta}$ are the tracers 1-loop real space power spectra for the velocity and density fields, as given below in \S \ref{sect:biasedtracers}. The functions $A^\text{TNS}(k,\mu)$ and $D(k,\mu)$ are \cite{Aviles:2020wme,Taruya:2010mx}
\begin{align}
A^\text{TNS}(k,\mu) &= 2 k \mu f_0 \ip  \frac{\vp\cdot \vhn}{p^2} B_\sigma(\vp,-\vk,\vk-\vp) \, , \label{ATNS} \\
D(k,\mu) &= (k\mu f_0)^2  \ip \Big\{ F(\vp)F(\vk-\vp)  \nonumber\\
      &\quad+   \frac{(\vp\cdot\vhn)^2}{p^4}P_{\theta\theta}^L(p) \big[ P^K_s(|\vk-\vp|, \mu_{\vk-\vp}) -P^K_s(k, \mu) \big] \Big\}, \label{DTNS}
\end{align}
with $\mu_{\vk-\vp}$ the cosine angle between the wave-vector $\vk-\vp$ and the line-of-sight direction $\vhn$. Here, the bispectrum $B_\sigma$ and function $F$ are
\begin{align}\label{defBsigma}
&(2 \pi)^3 \dD(\vk_1 + \vk_2 + \vk_3) B_\sigma(\vk_1,\vk_2,\vk_3) = \nonumber\\
&\qquad \qquad \Big\langle \theta(\vk_1)\left[\delta(\vk_2) + f_0 \frac{(\vk_2\cdot\vhn)^2}{k_2^2}\theta(\vk_2) \right] \left[\delta(\vk_3) + f_0 \frac{(\vk_3\cdot\vhn)^2}{k_3^2}\theta(\vk_3) \right]\Big\rangle,
\end{align}
and
\begin{align}
F(\vp) &= \frac{\vp\cdot\vhn}{p^2}\Big[ P_{\delta\theta}(p) + f_0 \frac{(\vp\cdot\vhn)^2}{p^2} P_{\theta\theta}(p) \Big].
\end{align}
While
\begin{equation}
    P^K_s(k, \mu) = \big( 1 + \mu^2 f(k) \big)^2 P_L(k)
\end{equation}
is the linear Kaiser power spectrum \cite{Kaiser:1984sw}. In Appendix \ref{sect:AandD} we will write the functions $A^\text{TNS}(k,\mu)$ and $D(k,\mu)$ in the form of eq.~\eqref{Pkmu} which is more suitable for numerical analyses.

In the second line of eq.~\eqref{PS_EFT}, we add EFT parameters $\alpha_0$,   $\alpha_2$,  $\alpha_4$ and $\alpha_6$ that model the backreaction of small scales over large scales and the non-linear map between real and redshift spaces. Further, along the line-of-sight direction, 2-point statistics are dominated by Fingers of God as a non-linear coupling between the velocity and density fields, with a characteristic scale given by the velocity dispersion $\sigma_v$, yielding the next-to-leading order counterterm $\tilde{c}$  \cite{Ivanov:2019pdj}.

The third line in eq.~\eqref{PS_EFT} includes the noise parameters, uncorrelated with long wave-length fluctuations. $P_\text{shot}$ is considered a constant that can be set equal to the Poisson process shot noise $P_\text{Poisson}= 1/\bar{n}_X$ with $\bar{n}_X$ the number density of tracers, or to any other constant whose value is not relevant since it is completely degenerate with $\alpha_0^{shot}$. We have added a tilt proportional to $(k\mu)^2$ \cite{McDonald:2009dh,Desjacques:2016bnm,Perko:2016puo,Chen:2020fxs,Nishimichi:2020tvu,Schmittfull:2020trd,Chen:2020zjt}. The shot noise departs from a pure white noise (when $\alpha^{shot}_{2} \neq 0$) because stochasticity is not localized at a single point \cite{McDonald:2009dh,Desjacques:2016bnm} and because of the stochastic nature of peculiar velocities at small scales \cite{Chen:2020fxs}.

Regarding the set of EFT and noise parameters introduced above, not all are necessary at the same time. As we already said, the value of $P_\text{shot}$ is not relevant and only alters the numerical values of  $\alpha^{shot}_{0}$ and $\alpha^{shot}_{2}$, so it can take any value, e.g. the Poissonian shot noise. On the other hand, one usually leads with only the first two or three non-vanishing multipoles of the power spectrum, hence not all the EFT parameters are necessary: for example, below we will compute only the monopole and the quadrupole, thus the counterterms $\alpha_4$ and $\alpha_6$ would be redundant and not considered. Finally, the introduced functional dependence by $\tilde{c}$ is $k^4 P_L(k)$, which is approximately $\propto k^2$ at high-$k$, and hence degenerate with $\alpha_2^{shot}$. And so, the common is to choose between the use of either $\tilde{c}$ or  $\alpha_2^{shot}$. In this work, we use the latter as a nuisance parameter.

\subsubsection{Biased tracers}\label{sect:biasedtracers}
We adopt the bias prescription of McDonald \cite{McDonald:2006mx,McDonald:2009dh,Saito:2014qha}, which has been tweaked to work for theories with additional scales \cite{Aviles:2020cax}. Notice the introduction of the neutrino mass impedes to have a complete bias expansion at any given PT order; however, curvature and higher order biases can be used to partially tame this inconsistency \cite{Desjacques:2016bnm}. Ultimately, we obtain good numerical results with the same number of bias parameters as in $\Lambda$CDM. That is, we consider the bias parameters $b_1$, $b_2$, $b_{s^2}$ and $b_{3nl}$. We do not add the curvature bias $b_{\nabla^2\delta}$, because it is absorbed by the Effective Field Theory parameters (introduced below) since they are degenerate. The isotropic power spectra for velocity and density fields become
\begin{align}
 P_{\delta\delta}(k) &= b_1^2 P^\text{1-loop}_{\delta\delta}(k)  + 2 b_1 b_2 P_{b_1b_2}(k) + 2 b_1 b_{s^2} P_{b_1b_{s^2}}(k) + b_2^2 P_{b_2^2}(k)  \nonumber\\
                     &\quad + 2 b_2 b_{s^2} P_{b_2 b_{s^2}}(k) + b_{s^2}^2 P_{b_{s^2}^2}(k) + 2 b_1 b_{3nl} \sigma^2_3(k) P_L(k),  \label{PddTNL}\\
 P_{\delta\theta}(k) &=  b_1 P^\text{1-loop}_{\delta\theta}(k) + b_2 P_{b_2,\theta}(k)  + b_{s^2} P_{b_{s^2},\theta}(k) + b_{3nl} \sigma^2_3(k)  P^L_{cb,\delta\theta}(k), \label{PdtTNL}\\
 P_{\theta\theta}(k) &=   P^\text{1-loop}_{\theta\theta}(k),  \label{PttTNL}                 
\end{align}
where the quantities $P^{\text{1-loop}}_{\delta \delta}(k)$, $P^{\text{1-loop}}_{\delta \theta}(k)$ and $P^{\text{1-loop}}_{\theta \theta}(k)$ represent the 1-loop power spectra obtained from eqs.~(\ref{Pab}) and~(\ref{P22dd})–(\ref{P13vv}). The other contributions are defined as \cite{Saito:2014qha}
\begin{align}
 P_{b_1b_2}(k)     &= \ip F_2(\vp,\vk-\vp) P_L(p)P_L(|\vk-\vp|), \label{Eq:Pb_1b_2} 
 \\
 P_{b_1b_{s^2}}(k) &= \ip F_2(\vp,\vk-\vp) S_2(\vp,\vk-\vp) P_L(p)P_L(|\vk-\vp|),  
 \\
 P_{b_2^2}(k)      &= \frac{1}{2} \ip P_L(p) \big[P_L(|\vk-\vp|) - P_L(p) \big], \label{Pb22} 
 \\
 P_{b_2 b_{s^2}}(k)&= \frac{1}{2} \ip P_L(p) \left[P_L(|\vk-\vp|)S_2(\vp,\vk-\vp) - \frac{2}{3}P_L(p) \right], \label{Pb2bs2} 
 \\
 P_{b_{s^2}^2}(k)&= \frac{1}{2} \ip P_L(p) \left[P_L(|\vk-\vp|)[S_2(\vp,\vk-\vp)]^2 - \frac{4}{9}P_L(p) \right], \label{Pb2s2} 
 \\
P_{b_2,\theta}(k) &=\ip G_2(\vp,\vk-\vp) P_L(p)P_L(|\vk-\vp|), \label{Eq:P_b_2_t} 
\\
P_{b_{s^2},\theta}(k) &=\ip G_2(\vp,\vk-\vp) S_2(\vp,\vk-\vp) P_L(p)P_L(|\vk-\vp|), \label{Eq:P_b_s2_t} 
\end{align}
with
\begin{equation}
 S_2(\vk_1,\vk_2) = \frac{(\vk_1\cdot\vk_2)^2}{k_1^2 k_2^2} -\frac{1}{3}.
\end{equation} 
Additionally, the function $\sigma^2_3(k)$ is given by
\begin{equation} \label{sigma23EdS}
 \sigma^{2}_{3}(k) = \frac{105}{16} \ip P_L(p) \left[ S_2(\vp,\vk-\vp)\left(\frac{2}{7}S_2(-\vp,\vk)  -\frac{4}{21} \right) + \frac{8}{63} \right].
\end{equation}

Alternatively, one can use the bias expansion with $b_{\mathcal{G}_2}$ and  $b_{\Gamma_3}$ \cite{Assassi:2014fva}, instead of $b_{s^2}$ and $b_{3nl}$ \cite{McDonald:2009dh}.

\subsubsection{IR-resummations}\label{sect:IRresummations}

Coherent flows of the matter field stream over a scale settled by the variance of Lagrangian displacements $\sigma^2_\Psi$. Because the scale is comparable in size to the BAO peak width, overdense regions in the Universe become partially depleted, while underdense regions partially populated, broadening the acoustic peak \cite{Eisenstein:2006nk,Crocce:2007dt,Tassev:2013rta}. This effect is well described by LPT, since bulk flows are captured by the displacement fields even at linear order \cite{Matsubara:2007wj,Carlson:2012bu,Vlah:2015sea,Chen:2020fxs,Chen:2020zjt}. In contrast, in SPT the convergence is very slow \cite{Lewandowski:2018ywf}, and therefore non-perturbative methods, called IR-resummations, must be used \cite{Senatore:2014via}. We follow the prescription of \cite{Baldauf:2015xfa,Ivanov:2018gjr,Chudaykin:2020aoj}, that splits the linear power spectrum in a piece that does not contain the BAO (the non-wiggle power spectrum, $P_{nw}$) and a wiggle piece $P_w$, such that the real space linear power spectrum can be written as $P_L = P_{nw} + P_w$. As a result of this splitting, the 1-loop IR-resummed EFT redshift space power spectrum becomes \cite{Ivanov:2018gjr}
\begin{align}\label{PsIR}
P_s^\text{IR}(k,\mu) &= 
 e^{-k^2 \Sigma^2_\text{tot}(k,\mu)} P_s^\text{EFT}(k,\mu) +  \big(1-e^{-k^2 \Sigma^2_\text{tot}(k,\mu)} \big) P_{s,nw}^\text{EFT}(k,\mu) \nonumber\\
 &\quad +  e^{-k^2 \Sigma^2_\text{tot}(k,\mu)} P_w(k) k^2 \Sigma^2_\text{tot}(k,\mu), 
\end{align}
where the wiggle piece $P_s^\text{EFT}(k,\mu)$ is the 1-loop power spectrum computed using eq.~\eqref{PS_EFT}. The non-wiggle part, $P_{s,nw}^\text{EFT}(k,\mu)$ is also computed with eq.~\eqref{PS_EFT} but using as input the non-wiggle linear power spectrum $P_{nw}$. The function $\Sigma^2_\text{tot}$ is given by
\begin{equation}\label{Sigma2T}
\Sigma^2_\text{tot}(k,\mu) = \big[1+f(k) \mu^2 \big( 2 + f(k) \big) \big]\Sigma^2 + f^2(k) \mu^2 (\mu^2-1) \delta\Sigma^2,    
\end{equation}
with
\begin{align}
\Sigma^2 &= \frac{1}{6 \pi^2}\int_0^{k_s} dp \,P_{nw}(p) \left[ 1 - j_0\left(p \,\ell_\text{BAO}\right) + 2 j_2 \left(p \,\ell_\text{BAO}\right)\right],\\
\delta\Sigma^2 &= \frac{1}{2 \pi^2}\int_0^{k_s} dp \,P_{nw}(p)  j_2 \left(p \,\ell_\text{BAO}\right),
\end{align}
where $\ell_\text{BAO}\simeq 105 \hmpc$ is the BAO peak scale and $j_0$ and  $j_2$ are the spherical Bessel functions of degree 0 and 2. The scale $k_s$ splits the long and short modes, whose choice is somewhat arbitrary. We use the value $k_s =0.4 \hmpci$. 

\subsubsection{Multipoles}

The IR-resummed EFT power spectrum of eq.~\eqref{PsIR} is the one we compare the simulations to. More precisely, we take its monopole, quadrupole and hexadecapole multipoles from
\begin{equation}\label{Pells}
P_\ell(k) = \frac{2 \ell + 1}{2} \int_{-1}^{1} d\mu \; P_s^\text{IR}(k,\mu) \mathcal{L}_{\ell}(\mu),    
\end{equation}
where $\mathcal{L}_{\ell}$ are the Legendre polynomial of degree $\ell$.

\end{subsection}

\end{section}

\section{FFTLog formalism}\label{sec:FFTLog}

FFTLog algorithms give us the 1-dimensional Fast Fourier Transform or Fast Hankel Transform of a function evaluated over a set of log-spaced intervals \cite{1978JCoPh..29...35T,Hamilton:1999uv}. Its usefulness in cosmology arise since one has to manipulate functions with relevant features spanning several orders of magnitude \cite{Hamilton:1999uv}.\footnote{\href{https://jila.colorado.edu/~ajsh/FFTLog/index.html}{https://jila.colorado.edu/$\sim$ajsh/FFTLog/}} For example, it serves to obtain the real space matter 2-point correlation function from the matter power spectrum faster, with higher accuracy, and notably reduced ringing and aliasing than when using direct integration. More recently, these methods have been reintroduced to help to bring the problem of integrating loop expressions onto the problem of matrix multiplication, which can be handled very fast with the use of proper numerical methods. 

The starting point is to decompose the linear real space matter power spectrum in to a series of power laws
\begin{equation}\label{Eq:FFTLog}
\bar{P}_{L}(k) = \sum_{m=-N/2}^{N/2} c_m \, k^{\nu+i\eta_m}, 
\end{equation}
where $N$ is the number of sampling points logarithmic spaced over the interval $[k_\text{min},\dots,k_\text{max}]$. The Fourier coefficients  $c_m$ and exponents $\eta_m$ are given by
\begin{align}
c_m &= \frac{W_m}{N} k_\text{min}^{-(\nu + i\eta_m)} \sum_{l=0}^{N-1}P_{L}(k_l)\left( \frac{k_l}{k_\text{min}} \right)^{-\nu}  e^{-2\pi iml/N}, \label{Eq:c_m}\\
\eta_{m} &= \frac{N-1}{N}\frac{2\pi m}{\log(k_\text{max}/k_\text{min})}, \label{Eq:eta_m}
\end{align}
with $W_m=1$ except $W_{\pm N/2} =1/2$. The parameter $\nu$ is an arbitrary real number known as the FFTLog bias. It serves as a leverage to obtain better convergence of the loop integrals. That is, the $c_m$ are not the FFT of the set $P_L(k_m)$, but of $P_L(k_m) k_m^{-\nu}$. Notice $\bar{P}_{L}$ in eq.~\eqref{Eq:FFTLog} denotes the approximation of the linear power spectrum, while the $P_{L}$ appearing in eq.~\eqref{Eq:c_m} is the real one as obtained from an Einstein-Boltzmann code. For brevity, we will not continue to use this ``bar'' notation.

In the presence of massive neutrinos one must distinguish between the linear power spectra $\delta\delta$, $\delta\theta$ and $\theta\theta$, since these are not equal due to eq.~\eqref{Eq:linear_theta}. Hence, similar expansions to eq.~\eqref{Eq:FFTLog} are constructed,
\begin{align}
P_{\delta\theta}^{L}(k) & =\frac{f(k)}{f_0}P_L(k) = \sum_{m=-N/2}^{N/2} c_m^f \, k^{\nu+i\eta_m}, \label{Eq:FFTLogdt} \\
P_{\theta\theta}^{L}(k) &= \left( \frac{f(k)}{f_0}\right)^2P_L(k) = \sum_{m=-N/2}^{N/2} c_m^{ff} \, k^{\nu+i\eta_m}, \label{Eq:FFTLogtt}
\end{align}
where the coefficients $c_m^f$ and $c_m^{ff}$ are obtained analogously to $c_m$ in eq.~\eqref{Eq:c_m} but with the substitution of $P_L$ by $P^L_{\delta\theta}$ and $P^L_{\theta\theta}$, respectively.

\begin{subsection}{Velocity power spectrum}

All the $I^m_n(k)$ functions are either of the ``$P_{22}$-{\it type}'' or ``$P_{13}$-{\it type}''. The former type means that one has to integrate over a kernel that can be written as a power series  of $p^2$, $k^2$ and $|\vk-\vp|^2$, multiplied by growth rates and by one linear power spectrum evaluated at  $p$ and other evaluated at $|\vk-\vp|$. For the $P_{13}$-{\it type} case is similar but with the power spectra evaluated one at the internal momentum $p$ and the other at external momentum $k$. 

In this subsection, we consider the velocity power spectrum $P^\text{loop}_{\theta\theta}(k)$ as an example. The rest of the functions $I^m_n$ are computed in the same form, but the algebra is lengthy, so we show the results in appendix \ref{appB}.

We start with the ``$P_{22}$-{\it type}'' piece given by eq.~(\ref{P22vv}), where
\begin{align}\label{Eq:P22tt}
    P^{22}_{\theta \theta}(k) & = 2 \ip \big[ G_2 (\vp, \vk-\vp)\big]^2 P_L(p) P_L(|\vk -\vp|) \nonumber \\ & 
    = 2 \ip K^{f_{\vk-\vp}f_{\vk-\vp} }_{\theta \theta }(\vp, \vk-\vp) \frac{f^2(|\vk-\vp|)}{f^2_0} P_L(p)P_L(|\vk-\vp|) \nonumber \\ & \quad
    +2 \ip K^{f_{\vp}f_{\vp}}_{\theta \theta }(\vp, \vk-\vp) \frac{f^2(p)}{f^2_0} P_L(p)P_L(|\vk-\vp|)
    \nonumber \\ & \quad
    +2  \ip K^{f_{\vk-\vp} f_{\vp}}_{\theta \theta }(\vp, \vk-\vp)  \frac{f(|\vk-\vp|) f(p)}{f^2_0}P_L(p)P_L(|\vk-\vp|),
\end{align}
with kernels
%
\begin{align}
   K^{f_{\vp}f_{\vp}}_{\theta \theta }(\vp, \vk-\vp) & =  \frac{k^8}{196 |\vk-\vp|^4 p^4}+\frac{3 k^6}{196 |\vk-\vp|^4 p^2}-\frac{k^6}{49 |\vk-\vp|^2 p^4}-\frac{3 k^4}{196 |\vk-\vp|^2 p^2} \nonumber \\ & \quad
  -\frac{11 k^4}{784 |\vk-\vp|^4}+\frac{3 k^4}{98 p^4}-\frac{15 k^2 p^2}{392 |\vk-\vp|^4}   -\frac{k^2 |\vk-\vp|^2}{49 p^4}+\frac{29 k^2}{392 |\vk-\vp|^2} \nonumber \\ & \quad
-\frac{3 k^2}{196 p^2}-\frac{15 p^2}{392 |\vk-\vp|^2}+\frac{25 p^4}{784 |\vk-\vp|^4}+\frac{3 |\vk-\vp|^2}{196 p^2} + \frac{|\vk-\vp|^4}{196 p^4} \nonumber\\ 
 & \quad -\frac{11}{784},    \label{Eq:K_fpfp} \\
 K^{f_{\vk-\vp} f_{\vp}}_{\theta \theta }(\vp, \vk-\vp) & =     \frac{k^8}{196 |\vk-\vp|^4 p^4}-\frac{k^6}{196 |\vk-\vp|^4 p^2}+\frac{37 k^4}{784 |\vk-\vp|^2 p^2} 
  -\frac{9 k^4}{196 |\vk-\vp|^4} \nonumber \\ & \quad
  +\frac{13 k^2 p^2}{196 |\vk-\vp|^4}    -\frac{13 k^2}{196 |\vk-\vp|^2}  -\frac{9 p^2}{392 |\vk-\vp|^2}    -\frac{5 p^4}{196 |\vk-\vp|^4} +\frac{19}{392} \nonumber\\ & \quad
  +  \big( \, p \leftrightarrow |\vk-\vp| \, \big), \label{Eq:K_fkmpfp} \\ 
    K^{f_{\vk-\vp}f_{\vk-\vp} }_{\theta \theta }(\vp, \vk-\vp) &= K^{f_{\vp}f_{\vp}}_{\theta \theta }(\vk-\vp,\vp). \label{Kfkmpfkmp}
\end{align}
In the second equality of eq.~(\ref{Eq:P22tt}) we have split $\big[ G_2 (\vp, \vk-\vp)\big]^2$ in three pieces, each of them proportional to the growth rate evaluated at different wave-numbers. Interchanging the integration variable $\vp \rightarrow \vk-\vp$ of the first of these three terms, we can rewrite eq.~(\ref{Eq:P22tt}) as
\begin{align}
    P^{22}_{\theta \theta}(k) &  = 4 \ip  K^{f_{\vp}f_{\vp}}_{\theta \theta }(\vp, \vk-\vp)  P^L_{\theta \theta}(p)P_L(|\vk-\vp|)   \nonumber\\
    &\quad
    + 2  \ip K^{f_{\vk-\vp} f_{\vp}}_{\theta \theta }(\vp, \vk-\vp)  P^L_{\delta \theta}(p)P^{L}_{\delta \theta}(|\vk-\vp|) ,
\end{align}
with
\begin{equation}
 K^{f_{\vp}f_{\vp}}_{\theta \theta }(\vp, \vk-\vp) \equiv    \displaystyle\sum\limits_{n_1, n_2 = -2}^2 f^{f_{\vp}f_{\vp}}_{22, \theta \theta} (n_1, n_2) k^{-2(n_1+n_2)} p^{2n_1} |\vk-\vp|^{2n_2},
\end{equation}
\begin{equation}
f^{f_{\vp}f_{\vp}}_{22, \theta \theta} (n_1, n_2) = 
 \begin{blockarray}{cccccc} 
 \text{\scriptsize $n_2=-2$} &  \text{\scriptsize $-1$} & \text{\scriptsize $0$} & \text{\scriptsize $1$} & \text{\scriptsize $2$} \\
\begin{block}{(ccccc)c}        
 \frac{1}{196} & -\frac{1}{49} & \frac{3}{98} & -\frac{1}{49} & \frac{1}{196}  & \text{\scriptsize $n_1=-2$}\\[5pt]
 \frac{3}{196} & -\frac{3}{196} & -\frac{3}{196} & \frac{3}{196} & 0 & \text{\scriptsize $-1$}\\[5pt]
 -\frac{11}{784} & \frac{29}{392} & -\frac{11}{784} & 0 & 0 & \text{\scriptsize $0$}\\[5pt]
 -\frac{15}{392} & -\frac{15}{392} & 0 & 0 & 0  & \text{\scriptsize $1$}\\[5pt]
 \frac{25}{784} & 0 & 0 & 0 & 0 & \text{\scriptsize $2$}\\[5pt]
\end{block}
\end{blockarray}
,
\end{equation}
and kernel
\begin{equation}
K^{f_{\vk-\vp} f_{\vp}}_{\theta \theta }(\vp, \vk-\vp) \equiv    \displaystyle\sum\limits_{n_1, n_2 = -2}^2 f^{f_{\vk-\vp} f_{\vp}}_{22, \theta \theta} (n_1, n_2) k^{-2(n_1+n_2)} p^{2n_1} |\vk-\vp|^{2n_2},
\end{equation}
with matrix
\begin{equation}\label{eq3.1.46}
 f^{f_{\vk-\vp}f_{\vp}}_{22, \theta \theta} (n_1, n_2) = \begin{blockarray}{cccccc} 
 \text{\scriptsize $n_2=-2$} &  \text{\scriptsize $-1$} & \text{\scriptsize $0$} & \text{\scriptsize $1$} & \text{\scriptsize $2$} \\
\begin{block}{(ccccc)c}        
  \frac{1}{98} & -\frac{1}{196} & -\frac{9}{196} & \frac{13}{196} & -\frac{5}{196}  & \text{\scriptsize $n_1=-2$}\\[5pt]
 -\frac{1}{196} & \frac{37}{392} & -\frac{13}{196} & -\frac{9}{392} & 0 & \text{\scriptsize $-1$}\\[5pt]
-\frac{9}{196} & -\frac{13}{196} & \frac{19}{196} & 0 & 0 & \text{\scriptsize $0$}\\[5pt]
\frac{13}{196} & -\frac{9}{392} & 0 & 0 & 0 & \text{\scriptsize $1$}\\[5pt]
 -\frac{5}{196} & 0 & 0 & 0 & 0 & \text{\scriptsize $2$}\\[5pt]
\end{block}
\end{blockarray}
,
\end{equation}
where we have written the expansions~(\ref{Eq:K_fpfp}) and~(\ref{Eq:K_fkmpfp}) compactly since all their summands have the form $k^{-2(n_1+n_2)}p^{2n_1}|\vk-\vp|^{2n_2}$ with $n_1, n_2 \in \{-2,-1,0,1,2\}$. 

Using the FFTLog decompositions of eq.~\eqref{Eq:FFTLog}, \eqref{Eq:FFTLogdt} and \eqref{Eq:FFTLogtt}, and the above defined matrices we approximate
\begin{align}
P^{22}_{\delta\theta}(k) &= 4 \sum_{m_1,m_2} c^{ff}_{m_1} c_{m_2} \sum_{n_1,n_2=-2}^2 f^{f_{\vp}f_{\vp}}_{22, \theta \theta} (n_1,n_2) k^{-2(n_1+n_2)} \ip \frac{1}{p^{2\nu_1-2 n_1} |\vk-\vp|^{2\nu_2-2 n_2}} \nonumber\\
& +  2 \sum_{m_1,m_2} c_{m_1}^f c_{m_2}^f \sum_{n_1,n_2=-2}^2 f^{f_{\vk-\vp}f_{\vp}}_{22, \theta \theta} (n_1,n_2) k^{-2(n_1+n_2)} \ip \frac{1}{p^{2\nu_1-2 n_1} |\vk-\vp|^{2\nu_2-2 n_2}}, 
\end{align}
where the sums over $m_1$ and $m_2$ run from $-N/2$ to $N/2$, and for simplicity we defined
\begin{equation} \label{nu1nu2defs}
\nu_1 = -\frac{1}{2}( \nu + i \eta_{m_1}) , \qquad \nu_2 = -\frac{1}{2}( \nu + i \eta_{m_2}).   
\end{equation}
To further simplify the previous expression, we define
\begin{align}
M_{22,\theta\theta}^{f_{\vp}f_{\vp}}(\nu_1,\nu_2) &=  4\sum_{n_1,n_2=-2}^2  f^{f_{\vp}f_{\vp}}_{22,\theta\theta}(n_1,n_2) I (\nu_1-n_1,\nu_2-n_2), \label{M22fp} \\
M_{22,\theta\theta}^{f_{\vk-\vp}f_{\vp}}(\nu_1,\nu_2) &=  2\sum_{n_1,n_2=-2}^2 f^{f_{\vk-\vp}f_{\vp}}_{22,\theta\theta}(n_1,n_2) I (\nu_1-n_1,\nu_2-n_2), \label{M22fkmp}
\end{align}
where the $I(\nu_1,\nu_2)$ function is defined as  \cite{Smirnov:1991jn,Scoccimarro:1996jy}
\begin{align} \label{Eq:Integral}
I(z_1, z_2) &\equiv k^{-3+2 z_{12}} \ip \frac{1}{p^{2z_1}|\vk -\vp|^{2z_2}} = \frac{1}{8 \pi^{3/2}} \frac{\Gamma\left(\frac{3}{2}-z_1\right)\Gamma\left(\frac{3}{2}-z_2\right)\Gamma\left(z_{12} -\frac{3}{2}\right)}{\Gamma{(z_1)}\Gamma{(z_2)}\Gamma{(3-z_{12})}},
\end{align}
for complex numbers $z_1$, $z_2$ and $z_{12} \equiv z_1+ z_2$, and $\Gamma(z)$ is the Gamma function.

Then, using the ``$M$'' matrices we can write the approximation for $P^{22}_{\theta \theta}(k)$ as 
\begin{align}
    P^{22}_{\theta \theta} (k) & = k^3 \displaystyle\sum\limits_{m_1, m_2} c^{ff}_{m_1} k^{-2\nu_1} M^{f_{\vp}f_{\vp}}_{22, \theta \theta}(\nu_1, \nu_2) \, c_{m_2} k^{-2\nu_2}   
    \nonumber \\ & \quad
    + k^3 \displaystyle\sum\limits_{m_1, m_2} c^{f}_{m_1} k^{-2\nu_1} M^{f_{\vk-\vp} f_{\vp}}_{22, \theta \theta}(\nu_1, \nu_2) \, c^f_{m_2} k^{-2\nu_2},
\end{align}
with
\begin{align}
M^{f_{\vp}f_{\vp}}_{22, \theta \theta}(\nu_1, \nu_2)  &= 
\Big[98 \nu_1^3 \nu_2+7 \nu_1^2 \big(2 \nu_2 (7 \nu_2-8)+1 \big) -\nu_1 \big(2 \nu_2 (7 \nu_2+17)+53 \big)-12 (1-2 \nu_2)^2\Big]
    \nonumber\\
&\quad \times \frac{2\nu_{12}-3}{98 \nu_1 (\nu_1+1) \nu_2 (\nu_2+1) (2 \nu_2-1)}
    I(\nu_1, \nu_2), \\
 M^{f_{\vk-\vp} f_{\vp}}_{22, \theta \theta}(\nu_1, \nu_2) &= \Big[7 \nu_1^2 (7 \nu_2+3)+\nu_1 \big(7 \nu_2 (7 \nu_2-1)-10\big)+\nu_2 (21 \nu_2-10)-37\Big]
     \nonumber\\
&\quad \times \frac{2\nu_{12} -3 }{98 \nu_1 (\nu_1+1) \nu_2 (\nu_2+1)} I(\nu_1, \nu_2).
\end{align}

Before continuing with the rest of the terms in $P^\text{loop}_{\theta \theta}$ let us discuss a little bit on what we have done. The formalism developed above allows us to transform the evaluation of the loop integrals into the contraction of cosmology-independent matrices $M$ and cosmology-dependent terms $c_m$ (and $c_m^f$ and $c_m^{ff}$ when massive neutrinos are present), speeding up the loop calculations compared to the usual, direct integration, because one is able to perform matrix multiplications with fast algorithms which are very common in standard software library for numerical linear algebra. Moreover, since the matrices do not depend on cosmology, they can be pre-computed and stored, while the cosmology-dependent terms can be found quickly through Fast Fourier Transform (FFT). Naively, one may think of that for EdS the number of computations were reduced drastically, about half, since in this case $c_m=c_m^f =c_m^{ff}$. 
However, this is not correct since some of the same computations are shared for the different $I^m_n$ functions and we take advantage of this fact as much as possible. Hence, we have to perform a total of 26 $P_{22}$-{\it type} matrix multiplications with our method, instead of 24 if EdS is used, and all of the multiplications take about the same amount of time. For $P_{13}$-{\it type} the numbers of multiplications is reduced to 7 from 11. Since most of the time the code is performing $P_{22}$-{\it type} matrix multiplications, the time spent with the use of EdS kernels is only marginally smaller than with our method.

Finally, we develop the approximation to $P^{13}_{\theta \theta}(k)$,
\begin{align}
 P^{13}_{\theta \theta}(k)  & =   6  \frac{f(k)}{f_0} P_L(k) \ip G_3(\vk,-\vp, \vp) P_L(p)
 \nonumber \\ & 
 =  6 \left(\frac{f(k)}{f_0}\right)^2P_L(k) \ip K^{f_{\vk}}_{\theta \theta}(\vk, -\vp, \vp) P_L(p) 
 + 6 \frac{f(k)}{f_0}P_L(k) \ip K^{f_{\vp}}_{\theta \theta}(\vk, -\vp, \vp) P^L_{\delta \theta}(p).
\end{align}
Observe that we have decomposed $G_3(\vk,-\vp,\vp)$ in two pieces, one of them proportional to $ f(k)/f_0$ and the other proportional to $ f(p)/f_0$. The former involves the kernel
\begin{align}
K^{f_{\vk}}_{\theta \theta}(\vk, -\vp, \vp) &=  -\frac{k^6}{252 |\vk-\vp|^2 p^4}+\frac{k^4}{63 |\vk-\vp|^2 p^2} -\frac{11 k^4}{252 p^4}+\frac{19 k^2 |\vk-\vp|^2}{168 p^4}-\frac{k^2}{42 |\vk-\vp|^2}
\nonumber \\ &\quad 
-\frac{41 k^2}{504 p^2}-\frac{p^2}{504 k^2}+\frac{p^2}{63 |\vk-\vp|^2} -\frac{p^4}{252 k^2 |\vk-\vp|^2} +\frac{13 |\vk-\vp|^2}{126 p^2} +\frac{5 |\vk-\vp|^2}{168 k^2}
\nonumber \\ &\quad 
-\frac{5 |\vk-\vp|^4}{63 p^4}-\frac{19 |\vk-\vp|^4}{504 k^2 p^2}+\frac{|\vk-\vp|^6}{72 k^2 p^4}  -\frac{5}{126}
\nonumber \\ & 
\equiv \displaystyle\sum\limits_{n_1= -2}^{2} \displaystyle\sum\limits_{n_2 = -1}^{3} f^{f_{\vk}}_{13, \theta \theta}(n_1, n_2) k^{-2(n_1+n_2)}p^{2n_1} |\vk-\vp|^{2n_2},
\end{align}
with matrix
\begin{equation}
f^{f_{\vk}}_{13, \theta \theta}(n_1, n_2) = 
 \begin{blockarray}{cccccc} 
 \text{\scriptsize $n_2=-1$} &  \text{\scriptsize $0$} & \text{\scriptsize $1$} & \text{\scriptsize $2$} & \text{\scriptsize $3$} \\
\begin{block}{(ccccc)c}        
 -\frac{1}{252} & -\frac{11}{252} & \frac{19}{168} & -\frac{5}{63} & \frac{1}{72} & \text{\scriptsize $n_1=-2$}\\[5pt]
  \frac{1}{63} & -\frac{41}{504} & \frac{13}{126} & -\frac{19}{504} & 0 & \text{\scriptsize $-1$}\\[5pt]
 -\frac{1}{42} & -\frac{5}{126} & \frac{5}{168} & 0 & 0 &
 \text{\scriptsize $0$}\\[5pt]
 \frac{1}{63} & -\frac{1}{504} & 0 & 0 & 0 & \text{\scriptsize $1$}\\[5pt]
 -\frac{1}{252} & 0 & 0 & 0 & 0 & \text{\scriptsize $2$}\\[5pt]
\end{block}
\end{blockarray}
,
\end{equation}
while the latter involves the kernel
\begin{align}
 K^{f_{\vp}}_{\theta \theta}(\vk, -\vp, \vp)  & =     -\frac{k^6}{126 |\vk-\vp|^2 p^4}+\frac{k^4}{72 |\vk-\vp|^2 p^2}+\frac{k^4}{72 p^4}+\frac{k^2 |\vk-\vp|^2}{168 p^4} +\frac{k^2}{168 |\vk-\vp|^2}
\nonumber \\ & \quad
+\frac{k^2}{252 p^2}-\frac{5 p^2}{126 k^2} -\frac{11 p^2}{504 |\vk-\vp|^2}+\frac{5 p^4}{504 k^2 |\vk-\vp|^2}+\frac{11 |\vk-\vp|^2}{504 p^2} +\frac{5 |\vk-\vp|^2}{84 k^2}
\nonumber \\ &  \quad
-\frac{11 |\vk-\vp|^4}{504 p^4}-\frac{5 |\vk-\vp|^4}{126 k^2 p^2}+ \frac{5 |\vk-\vp|^6}{504 k^2 p^4}+\frac{11}{504}
\nonumber \\ & 
\equiv \displaystyle\sum\limits_{n_1= -2}^{2} \displaystyle\sum\limits_{n_2 = -1}^{3} f^{f_{\vp}}_{13, \theta \theta}(n_1, n_2) k^{-2(n_1+n_2)}p^{2n_1} |\vk-\vp|^{2n_2},
\end{align}
with matrix
\begin{equation}\
f^{f_{\vp}}_{13, \theta \theta}(n_1, n_2) = \begin{blockarray}{cccccc} 
 \text{\scriptsize $n_2=-1$} &  \text{\scriptsize $0$} & \text{\scriptsize $1$} & \text{\scriptsize $2$} & \text{\scriptsize $3$} \\
\begin{block}{(ccccc)c}        
 -\frac{1}{126} & \frac{1}{72} & \frac{1}{168} & -\frac{11}{504} & \frac{5}{504} & \text{\scriptsize $n_1=-2$}\\[5pt]
 \frac{1}{72} & \frac{1}{252} & \frac{11}{504} & -\frac{5}{126} & 0 & \text{\scriptsize $-1$}\\[5pt]
 \frac{1}{168} & \frac{11}{504} & \frac{5}{84} & 0 & 0 &
 \text{\scriptsize $0$}\\[5pt]
 -\frac{11}{504} & -\frac{5}{126} & 0 & 0 & 0 & \text{\scriptsize $1$}\\[5pt]
\frac{5}{504} & 0 & 0 & 0 & 0 & \text{\scriptsize $2$}\\[5pt]
\end{block}
\end{blockarray}
.
\end{equation}
Then, the approximation for $P^{13}_{cb, \theta \theta}(k)$ gives
\begin{align}
    P^{13}_{\theta \theta}(k) = k^{3} \frac{f(k)}{f_0}  P_L(k) \left(  \frac{f(k)}{f_0} \displaystyle\sum\limits_{m_1} c_{m_1} k^{-2 \nu_1}M^{f_{\vk}}_{13, \theta \theta} (\nu_1) +  \displaystyle\sum\limits_{m_1} c^{f}_{m_1}k^{-2\nu_1}M^{f_{\vp}}_{13, \theta \theta} (\nu_1) \right),
\end{align}
where
\begin{align}
M^{f_{\vk}}_{13, \theta \theta} (\nu_1) &= -\frac{\tan (\nu_1\pi )}{14 \pi  (\nu_1+1)\nu_1(\nu_1-1)(\nu_1-2)(\nu_1-3)},
    \\
M^{f_{\vp}}_{13, \theta \theta} (\nu_1) &= \frac{9\nu_1-7}{4}
     \frac{\tan(\nu_1\pi)}{28\pi (\nu_1+1)\nu_1(\nu_1-1)(\nu_1-2)(\nu_1-3)}.
\end{align}
In Appendix \ref{appB}, we present the FFTLog approximations to obtain all the functions $I^m_n(k)$ in eq.~\eqref{Pkmu}.

\end{subsection}

\subsection{UV and IR corrections}

Notice the function $I(z_1, z_2)$ vanishes if one of the arguments is zero or a negative integer. For example, consider $z_2 = 0$. In this case, eq.~(\ref{Eq:Integral}) becomes $\ip p^{-2z_1} = 0$. On the other hand, it is well-known that $\ip P_{L}(p)$ diverges. It seems contradictory because the linear power spectrum can be decomposed into power laws, and as a consequence of applying eq.~(\ref{Eq:Integral}) we would find zero as the final result. The latter suggests that $I(z_1, z_2)$ gives inappropriate results when the integral is divergent. This is because $I(z_1, z_2)$ calculates only the finite part of the loop integral. Therefore, if the integral we are interested in has an ultraviolet (UV) or an infrared (IR) divergence, to get the correct answer, one simply has to add the UV/IR contribution by hand \cite{Simonovic:2017mhp, Lewandowski:2018ywf}.

Now, let us discuss the convergence properties of the loop integrals~(\ref{P22dd})\,–\,(\ref{P13vv}). The behavior of \fk-kernels under the UV $(k \ll p)$ and IR $ (p \ll k)$ regimes are the same as $\Lambda$CDM, with the difference of some multiplications by $f_0$ and $f(k\rightarrow \infty)$ factors.
Then, when considering the FFTLog decomposition, eq.~(\ref{Eq:FFTLog}), the convergence of the loop integrals is determined by the bias parameter $\nu$, as much as in the massless neutrino case. For example, if we take the UV/IR limit and consider $P_L(k) \sim k^\nu$, one can find that the $P^{22}_{ab}$ contributions are UV convergent for $\nu < 1/2$ and IR convergent for $-1 < \nu$. Then for $-1<\nu<1/2$ the $P^{22}_{ab}$ integrals are convergent, and consequently, the use of eq.~(\ref{Eq:Integral}) return the same results as with the traditional direct computation. Nevertheless, when we choose values for the bias $\nu$ outside the convergence range, the $P^{22}_{ab}$ integrals become UV or IR divergent. Then, eq.~\eqref{Eq:Integral} does not guarantee the correct answer because the divergent pieces are not captured by dimensional regularization. Therefore, to obtain the correct answer, the corresponding UV or IR piece must be added to the result obtained through the FFTLog formalism \cite{Simonovic:2017mhp, Lewandowski:2018ywf}.

\begin{center}
\begin{table}
\setlength{\tabcolsep}{1.10em} 
\renewcommand{\arraystretch}{1.0}
\begin{tabular}{lll}
\hline
\multicolumn{3}{c}{\bf{UV and IR corrections}} \\
\hline
  & UV  & IR  \\ \hline 
\multirow{10}{*}{$P^{22}$ } 
\\
& $ \tfrac{1}{2} < \nu < \tfrac{3}{2}$  & $-3 < \nu  < -1$ \\ \\
&  $P^{22, \text{UV}}_{\delta \delta} (k) =  \frac{9k^4}{196 \pi^2}  \int_{0}^{\infty} dp \, \frac{P^2_L(p)}{p^2}$  & $P^{22, \text{IR}}_{\delta \delta} (k) =  P_L(k) k^2 \sigma_\Psi^2$  \\ \\
& $P^{22, \text{UV}}_{\delta \theta} (k) = - \frac{3k^4}{196 \pi^2}  \int_{0}^{\infty} dp \, \frac{f(p)}{f_0} \frac{P^2_L(p)}{p^2} $  & $P^{22, \text{IR}}_{\delta \theta} (k) =  P^L_{\delta \theta}(k) k^2 \sigma_\Psi^2$ \\ \\
 & $P^{22, \text{UV}}_{\theta \theta} (k) =  \frac{k^4}{196 \pi^2}  \int_{0}^{\infty} dp \, \left(\frac{f(p)}{f_0}\right)^2 \frac{P^2_L(p)}{p^2} $  & $P^{22, \text{IR}}_{\theta \theta} (k) =  P^L_{\theta \theta}(k) k^2 \sigma_\Psi^2$ \\ 
 \\
 \hline
\multirow{10}{*}{$P^{13}$ } 
\\
& $-1 < \nu  < 1$  & $ -3 < \nu  < -1$ \\ \\
&  $P^{13, \text{UV}}_{\delta \delta}(k) = - \frac{61}{105} P_L(k) k^2 \sigma^2_{\Psi}$  & $P^{13, \text{IR}}_{\delta \delta} (k) = - P_L(k) k^2 \sigma_\Psi^2 $ \\ \\
& $P^{13, \text{UV}}_{\delta \theta} (k) =  - \left(  \frac{23}{21}  \frac{f(k)}{f_0} \sigma^2_\Psi + \frac{2}{21} \sigma^2_v \right) k^2 P_L(k) $  & $P^{13, \text{IR}}_{\delta \theta} (k) = - P^L_{\delta \theta}(k) k^2 \sigma_\Psi^2$ \\ \\
 & $ P^{13, \text{UV}}_{\theta \theta} (k) = -\left( \frac{169}{105} \frac{f(k)}{f_0} \sigma^2_\Psi  +\frac{4}{21} \sigma^2_v \right)  k^2 P^L_{\delta \theta}(k)$  & $P^{13, \text{IR}}_{\theta \theta} (k) = - P^L_{\theta \theta}(k) k^2 \sigma_\Psi^2$ \\ 
 \\
 \hline
\end{tabular}
\caption{\label{Table:divergences} Leading UV/IR contributions for different values of bias $\nu$}
\end{table}
\end{center}

A similar analysis for contributions $P^{13}_{ab}$ finds these loop integrals are divergent for $\nu > -1$ and $\nu < -1$, i.e. the integrals never converge. Thus, based on the $\nu$ value, we have to add the UV or IR divergence, as appropriate. For example, for $\nu > -1$ the $P^{13}_{ab}$ loop integrals are UV divergent. In this limit the leading contribution to $P^{13}_{\theta \theta}$ is
\begin{equation}\label{eqP13UV}
    P^{13, \text{UV}}_{\theta \theta} (k) = -\left( \frac{169}{105} \frac{f(k)}{f_0} \sigma^2_\Psi  +\frac{4}{21} \sigma^2_v \right)  k^2 P^L_{\delta \theta}(k),
\end{equation}
with
\begin{equation}\label{Eq:sigmas_psi_v}
    \sigma^2_{\Psi} \equiv \frac{1}{6 \pi^2}  \int_{0}^{\infty} dp \, P_L(p), \qquad \sigma^2_{v} \equiv \frac{1}{6 \pi^2}  \int_{0}^{\infty} dp \, P^L_{\delta \theta}(p).
\end{equation}
Notice that we have kept only the leading UV contribution, which is valid in the range $-1 < \nu < 1$. For biases $\nu > 1$, one has to consider subleading UV corrections. 

Similarly, if we consider a bias $\nu < -1$, the IR corrections are required. For these biases, the leading IR contribution of $P^{13}_{\theta \theta}$ is 
\begin{equation}\label{eqP13IR}
    P^{13, \text{IR}}_{\theta \theta} (k) = - P^L_{\theta \theta}(k) k^2 \sigma_\Psi^2,
\end{equation}
which holds for $-3 < \nu < -1$. For smaller values of the bias, the subleading IR contributions are necessary.

Returning to $P^{22}_{ab}$, we find that for the velocity spectra, the leading UV and IR contributions are
\begin{align}
P^{22, \text{UV}}_{\theta \theta} (k)  &=  \frac{k^4}{196 \pi^2}  \int_{0}^{\infty} dp \, \left(\frac{f(p)}{f_0}\right)^2 \frac{P^2_L(p)}{p^2},
    \qquad (\tfrac{1}{2} < \nu < \tfrac{3}{2})\\
P^{22, \text{IR}}_{\theta \theta} (k) &=  P^L_{\theta \theta}(k) k^2 \sigma_\Psi^2, 
    \qquad \qquad \qquad \qquad \quad (-3 < \nu < -1)
\end{align}
In table \ref{Table:divergences} we summarize all the leading UV and IR contributions with the corresponding range where they become necessary. Notice that $P^{22}_{ab}$ and $P^{13}_{ab}$ are individually IR divergent for $\nu < -1$. Furthermore, notice in the range $-3 < \nu < -1$ the IR-contributions from $P^{22}_{ab} $ and $P^{13}_{ab} $ cancel out, so they do not must to be added, as much as when neutrinos are massless \cite{Pajer:2013jj}. 


\section{Model validation}
\label{sec:modelvalidation}

We are now in position to validate our analytical model and numerical recipes by comparing against the \textsc{Quijote} simulations. \textsc{Quijote} is a suite of 44,100 full N-body simulations whose fiducial cosmology is $\{ \Omega_m =0.3175, \Omega_b=0.049, h=0.6711, n_s=0.9624, \sigma_8=0.834, M_\nu=0 \}$, and contains additional neutrino cosmologies with total mass $M_\nu=0.1$, $0.2$ and $0.4$ eV equally distributed among the three neutrino species. In this work, we use mainly $M_\nu= 0.4\, \text{eV}$ since this is the most massive case, and hence the effects of having kernels beyond EdS are more important. But also, we compare against the more likely value of $M_\nu= 0.1\, \text{eV}$. We use $N_T=100$ realizations for each cosmology, containing $512^3$ cold dark matter ($cdm$) particles and $512^3$ neutrino particles. We consider halos with masses in the range $13.1 < \log (M/ h^{-1} M_\odot) < 13.5$ identified with a Friends-of-Friends algorithm \cite{Davis:1985rj} over the $cdm$ particles, with a linking length parameter $b=0.2$. 
\revised{These are the same halos utilized in \cite{Aviles:2020cax,Aviles:2021que}, which facilitates the comparisons to previous works}. Each realization has a volume of  $1 \, (\hgpc)^3$, hence the total volume is $100\,(\hgpc)^3$, allowing us to have very small statistical errors and hence any departure from recovering the correct parameters from the simulations are expected to come from effects that are not present or not well modeled in our theory, e.g. 2-loop corrections, halo identifications, biasing and effective field theory counterterms, and more important for us, to test the \texttt{fk}-kernels method, since even small inaccuracies in the modeling can lead to very large biased constraints on the parameters.

 \begin{figure}
 	\begin{center}
 	\includegraphics[width=6.0 in]{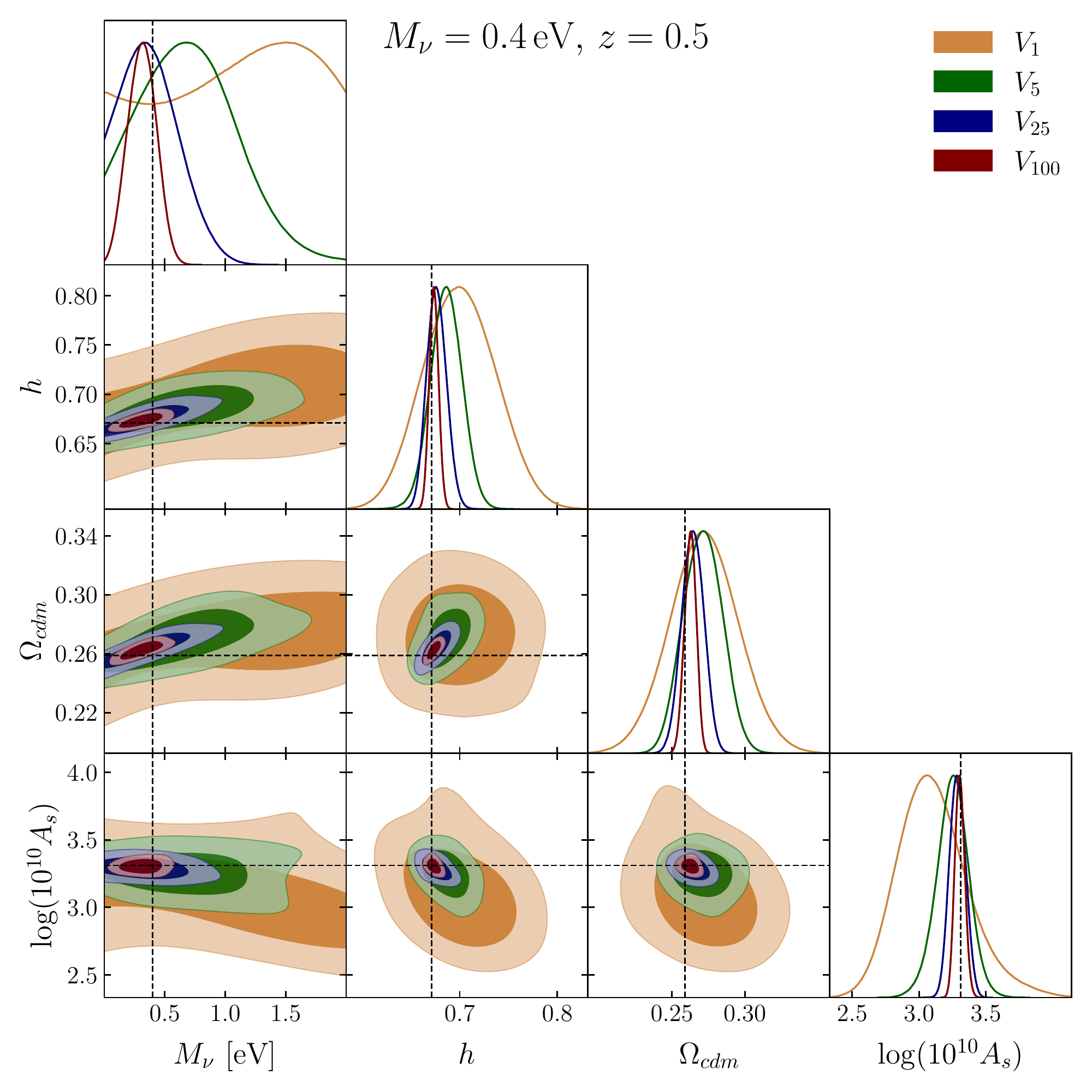}
 	\caption{Contour plots for the posterior distributions at 68 and 95\% confidence level computed with different covariance matrices, corresponding to $V_1$, $V_5$, $V_{25}$ and  $V_{100}$ effective volume data sets. This is the case of mass $M_\nu=0.4 \,\text{eV}$ at redshift $z=0.5$. The fittings are performed using the monopole and quadrupole of the power spectrum up to $k_{\text{max}} = 0.2 \, \hmpci$. Vertical and horizontal dashed lines show the values of the simulations. The triangular plot with all the parameters, including nuisances, is shown in figure \ref{fig:Base_TriangularPlot_All}. 
 	\label{fig:Base_TriangularPlot}}
 	\end{center}
 \end{figure}



\begin{table}
\centering
\setlength{\tabcolsep}{0.80em} 
\renewcommand{\arraystretch}{1.4}
\begin{tabular}{llll}
\rowcolors{2}{White}{LightGray!30}
\begin{tabular}{*5c}\toprule
  \textbf{Parameter} &  $\vec V_1$    &  $\vec V_5$   &  $\vec V_{25}$  & $\vec V_{100}$ \\\midrule
 $h$                 &  $0.700\pm 0.035$       & $0.686\pm 0.015$       & $0.6763\pm 0.0096$     &   $0.6736\pm 0.0046$ \\
 $\Omega_{cdm}$      &  $0.272\pm 0.023$       & $0.271\pm 0.013$       & $0.2643\pm 0.0073$     & $0.2626^{+0.0042}_{-0.0038}$ \\
 $M_{\nu}$ [eV]          &  ---        & $0.71^{+0.33}_{-0.47}$ & $0.39^{+0.16}_{-0.30}$ & $0.31\pm 0.11$ \\
 $\log(10^{10}A_s)$  &  $3.10^{+0.19}_{-0.28}$ & $3.25\pm 0.12$         & $3.288\pm 0.057$       & $3.305\pm 0.035$ \\
 $b_1$               &  $1.83^{+0.28}_{-0.25}$ & $1.70\pm 0.13$         & $1.645\pm 0.066 $      & $1.626\pm 0.038$ \\
 $b_2$               &  $-0.3\pm 1.8$          & $-0.4\pm 1.1$          & $-0.46^{+0.56}_{-0.68}$& $-0.53\pm 0.34$ \\
 $\alpha_0$ [$ h^{-2} \, \text{Mpc}^2$]         &  $12\pm 81$             & $15\pm 35$             & $3^{+17}_{-20}$        & $1.8^{+8.9}_{-10}$ \\
 $\alpha_2$ [$ h^{-2} \, \text{Mpc}^2$]         &  $0^{+31}_{-62}$        & $-26^{+19}_{-22}$      & $-28.9\pm 7.2$         & $-29.9\pm 3.9$ \\
 $\alpha^{shot}_0-1$ &  $-0.91^{+1.0}_{-0.80}$ & $-1.01\pm 0.51$        & $-0.92\pm 0.25$        & $-0.90^{+0.16}_{-0.13}$ \\
 $\alpha^{shot}_2$  [$ h^{-2} \, \text{Mpc}^2$] &  $-8.5^{+5.9}_{-5.2}$   & $-7.7\pm 2.7$          & $-8.1\pm 1.1 $         & $-8.07\pm 0.54$ \\\bottomrule
\end{tabular}
\end{tabular}
\caption{\label{Table:bestfit} 1-dimensional constraints for the different covariance matrices, corresponding to $V_1$, $V_5$, $V_{25}$ and  $V_{100}$ effective volume data sets. We show the mean of the posterior distributions with the 68\% confidence level intervals. We consider the case of massive neutrinos with $M_\nu=0.4 \,\text{eV}$ at redshift $z=0.5$. The fittings are performed using the monopole and quadrupole of the power spectrum up to $k_{\text{max}} = 0.2 \, \hmpci$. The {\it em dash} ``---'' in the neutrino mass for $V_1$ means no conclusive results because the posterior distribution saturates the prior of $[0,2]\,\text{eV}$. The corresponding triangular plot is shown in figure \ref{fig:Base_TriangularPlot}.
 }
\end{table}

We construct four data sets to compare to. All of them having the average of the 100 realizations as central points, but with different covariance matrices
\begin{equation} \label{CovMat}
    C_{N_r} = \frac{1}{N_r} C_1, \qquad \text{with}     \qquad N_r=1,\,5,\,25,\,\text{and}\,100,
\end{equation}
with $C_1$ the covariance of just one realization, but using the average of the $N_T=100$ realizations:
\begin{equation} \label{CovMat1}
    C_1(k_i,k_j) = \left\langle \Big(\bar{P}(k_i) - \hat{P}(k_i) \Big) \Big( \bar{P}(k_j) - \hat{P}(k_j) \Big)\right\rangle
\end{equation}
where $\hat{P}(k_i)$ is the value of the power spectrum of a single realization at bin $k_i$ and $\bar{P}(k_i) = \langle \hat{P}(k_i) \rangle$  is the mean over the ensemble of realizations at bin $k_i$.
It may be more precise the use of $N_r \leq N_T$ random subsamples  of realizations and fit to their average  \cite{Fumagalli:2022plg}, instead of fitting to the average of the $N_T$ realizations as we do. However, we have made tests down to $N_r=5$ and found no significant difference between the two approaches. Ultimately, our interest is to have a consistent way to compare between different effective volumes to validate our modeling. Another way to understand our approach is under the assumption of a Gaussian covariance matrix, scaling with the total volume. 

We choose four different volumes, 
corresponding to {\emph 1)} the volume of one realization $V_{1} = 1\hgpc $ ($N_r=1$),  {\emph 2)} the volume of the $N_T= 100$ realizations $V_{100} = 100 \hgpc$  ($N_r=N_T=100$),  {\emph 3)} we also consider a ``BOSS-like'' volume of $V_{5} = 5 \hgpc$  ($N_r=5$), and  {\emph 4)} a ``DESI-like'' volume  $V_{25} = 25 \hgpc$  ($N_r=25$).

 \begin{figure}
 	\begin{center}
 	\includegraphics[width=3.0 in]{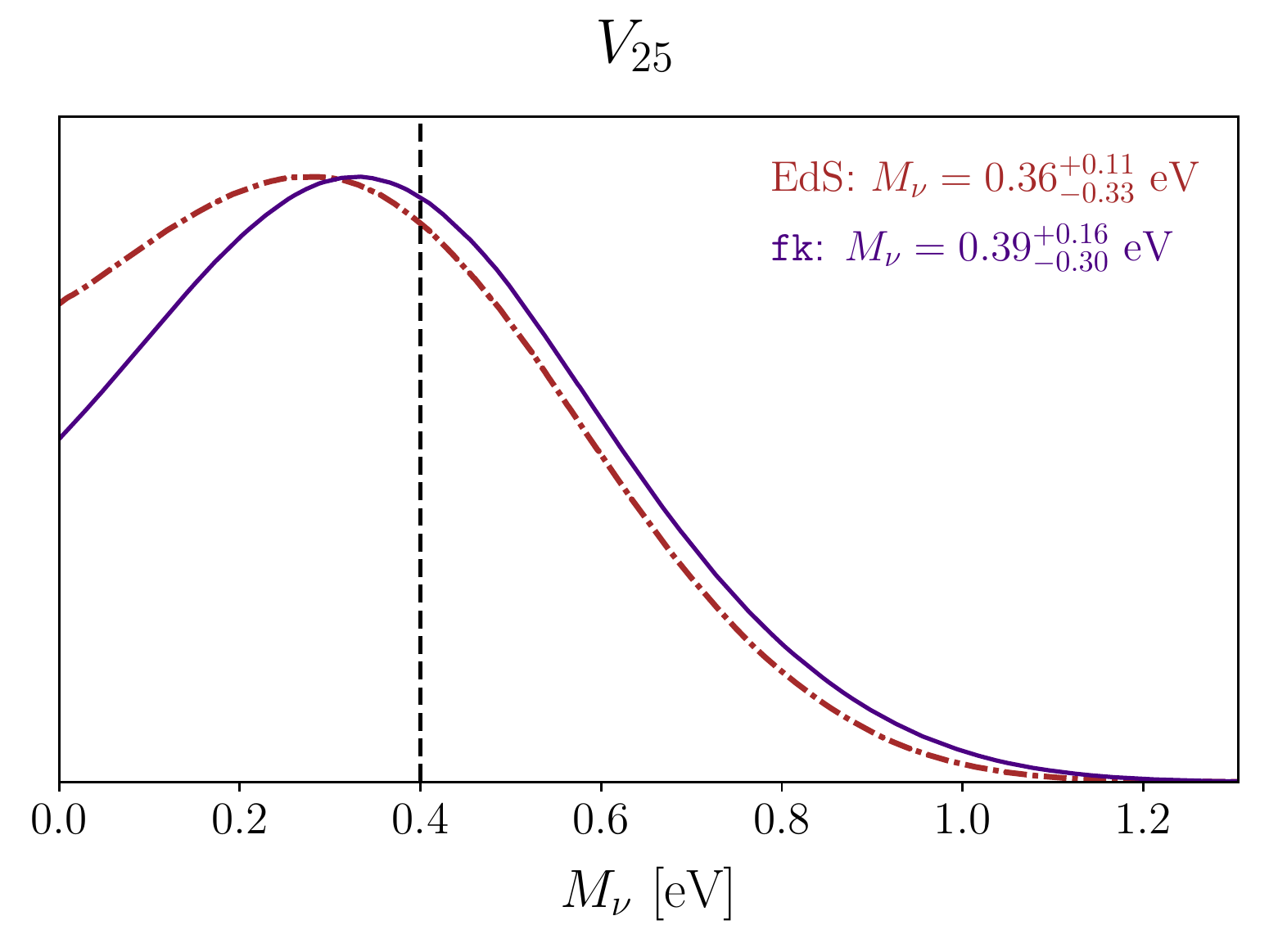}
 	\includegraphics[width=3.0 in]{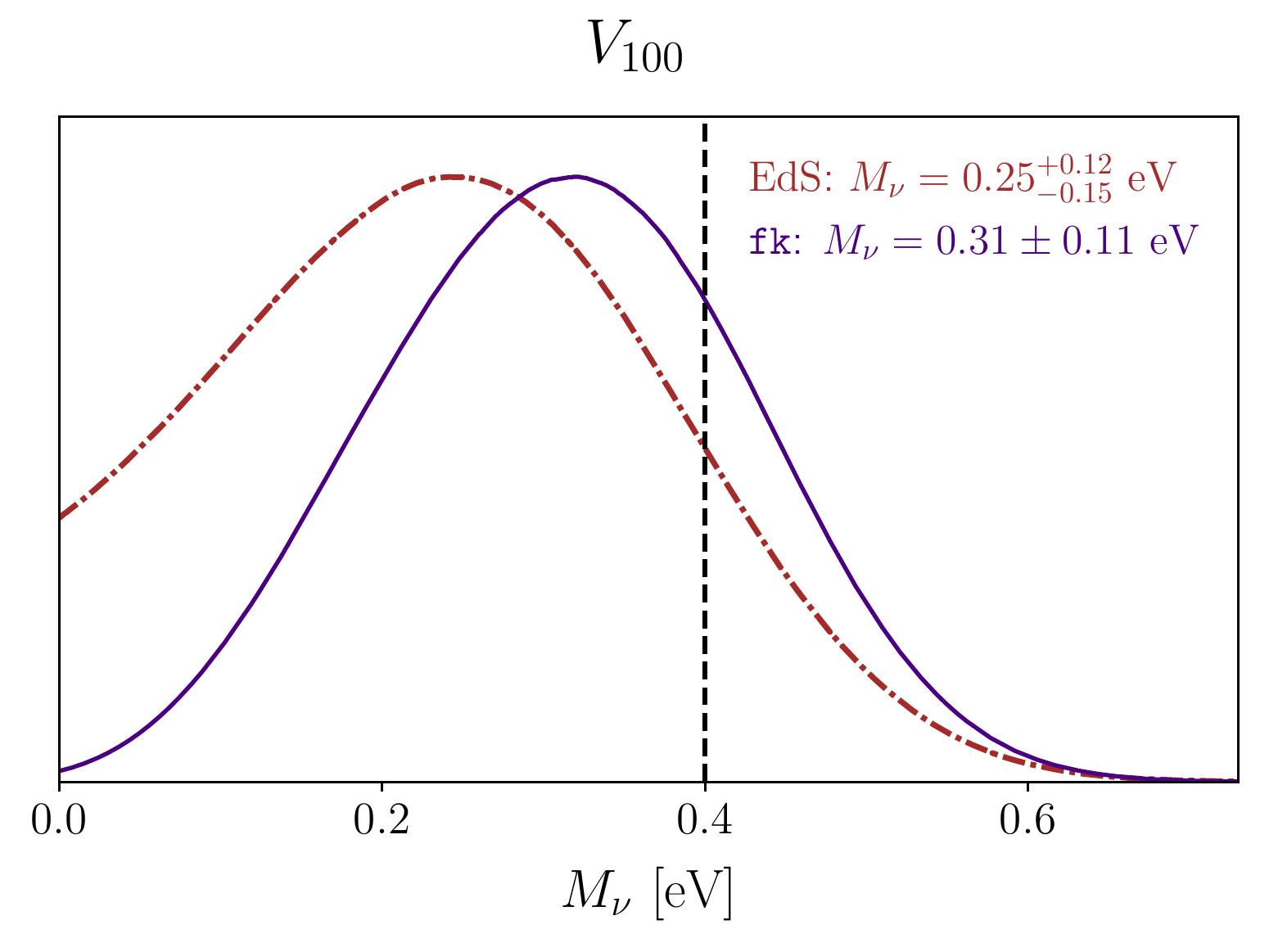}
 	\caption{Comparison on the constraints of the sum of the neutrino masses with the use of \fk- and EdS-kernels for the volumes $V_{25}$ (left panel) and $V_{100}$ (right panel). 
 	\label{fig:1dimfkvsEdS}}
 	\end{center}
 \end{figure}

Our baseline fitting consist on letting free the cosmological parameters $\Omega_{cdm}$, $h$, $\log(10^{10} A_s)$ and $M_\nu$, where $\Omega_{cb}=\Omega_{cdm} + \Omega_b$ with the baryon abundance fixed to the \textsc{Quijote} fiducial value $\omega_b=0.02207$. 
We consider also the bias parameters $b_1$, $b_2$, the EFT counterterms $\alpha_0$, $\alpha_2$, and the shot noise $\alpha_0^{shot}$ and $\alpha_2^{shot}$. The tidal and non-local bias are fixed by co-evolution theory 
\cite{Chan:2012jj,Baldauf:2012hs,Saito:2014qha}, implying that
\begin{equation}\label{coevbiases}
 b_{s^2}= -\frac{4}{7}(b_1-1),\qquad b_{3nl}=\frac{32}{315}(b_1-1).
\end{equation}
These expressions neglect an early time, Lagrangian tidal bias and are obtained using EdS evolution. However, we find that they yield good results also in the presence of massive neutrinos when the biased field is $cb$, so we use them to reduce the number of free parameters of the theory. Later we will relax this assumption and let free the biases $b_{s^2}$ and $b_{3nl}$, as well as the baryon abundance $\omega_b$. We consider uninformative priors on the cosmological and nuisances\revised{, with the exception of the neutrino mass prior, which is uniform over the interval $[0,2]\,\text{eV}$}. The shot noise constant is fixed to $P_{shot} = 1/\bar{n}_x = 4719.7 \, \revised{ h^{-3} \,\text{Mpc}^{3}}$, hence any departure from the Poissonian process is encoded in the values of $\alpha_0^{shot}$ and the tilt $\alpha_2^{shot}$. The linear $cb$ power spectrum is obtained from the Einstein-Boltzmann code \texttt{CLASS}\footnote{\href{https://lesgourg.github.io/class_public/class.html}{https://lesgourg.github.io/class\_public/class.html}} \cite{Blas:2011rf} which serves as an input of \texttt{FOLPS$\nu$} giving us the galaxy power spectrum multipoles that we compare against the simulations using a Gaussian likelihood,
$L \propto \exp(-\chi^2/2)$,  \revised{where $\chi^2$ is defined as $\chi^2\equiv D^T C_{N_r}^{-1}D$ and $D$ is the residual of the data vector and the model and $C_{N_r}$ is the covariance matrix in eq.~\eqref{CovMat}}.

 \begin{figure}
 	\begin{center}
 	\includegraphics[width=6.0 in]{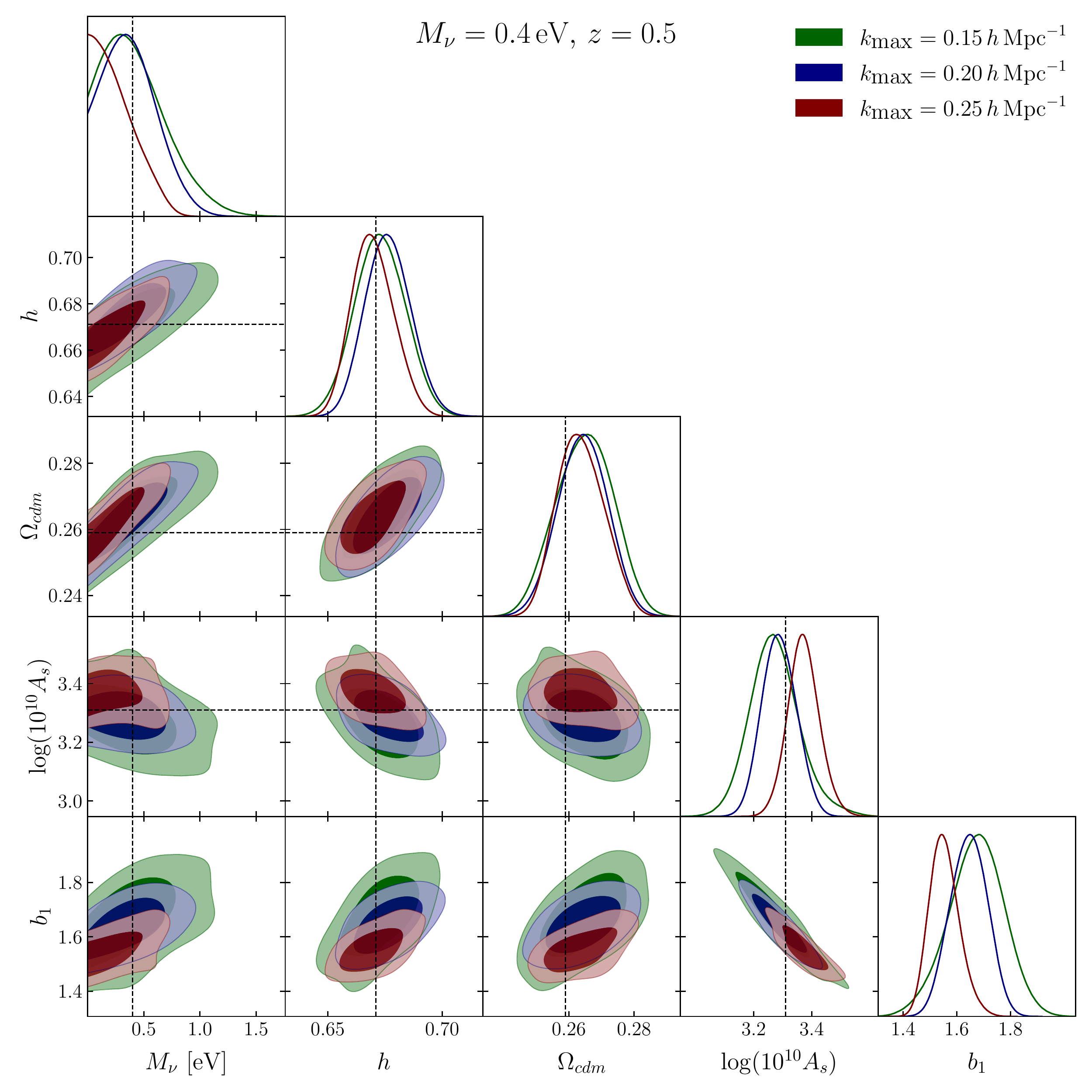}
 	\caption{Effects on the posterior distributions of the cosmological parameters of considering different values of the maximum wave-number $k_{\text{max}} = 0.15,\,0.2,\,0.25 \hmpci$.  We use the DESI-like volume $V_{25}$. Vertical and horizontal dashed lines show the values of the simulations. \revised{The triangular plot with all the parameters, including nuisances, is shown in figure \ref{fig:kmax_all}.}
 	\label{fig:kmax}}
 	\end{center}
 \end{figure}

 \begin{figure}
 	\begin{center}
 	\includegraphics[width=6.0 in]{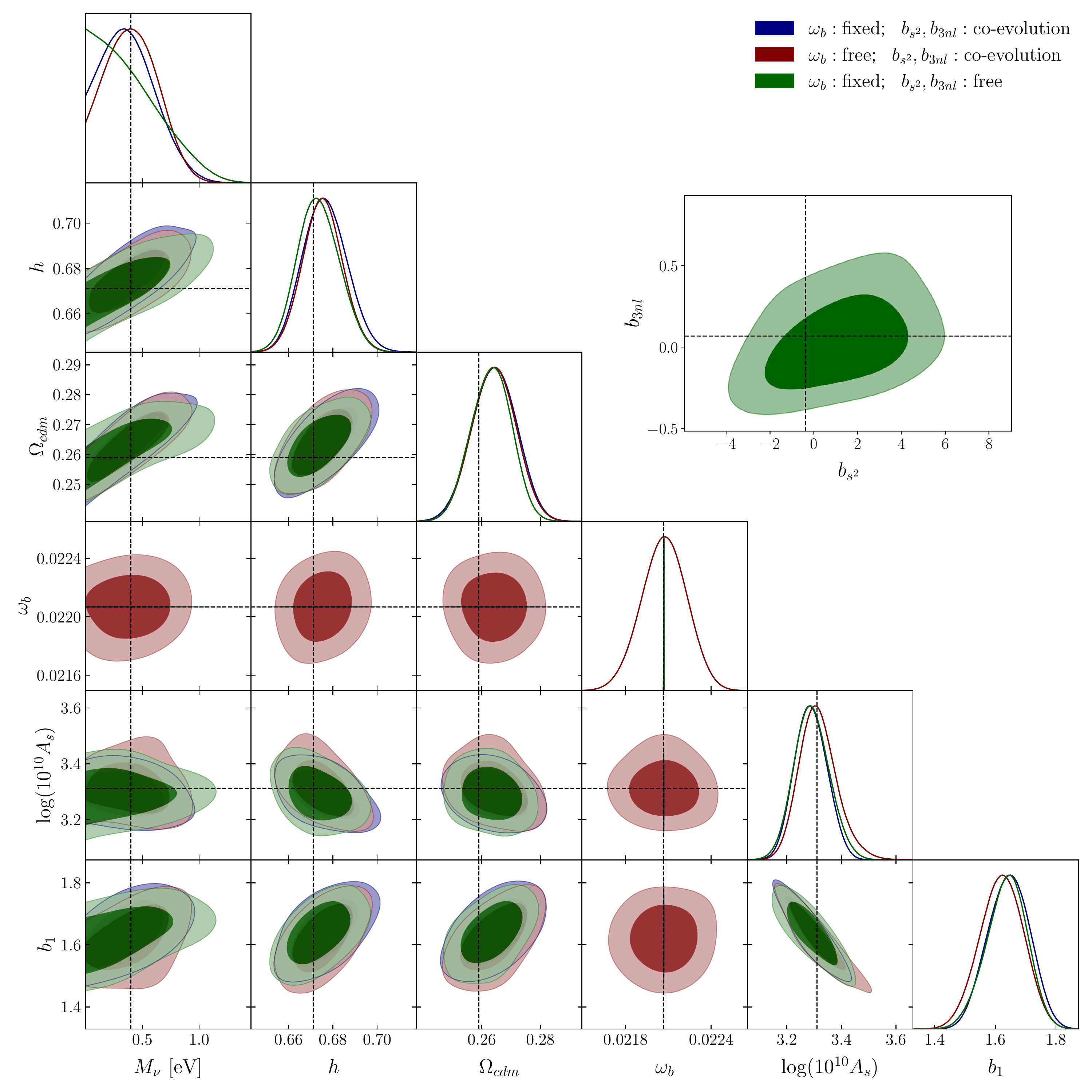}
 	\caption{Effects on the posterior distributions of the cosmological parameters of letting free the biases $b_{s^2}$ and $b_{3nl}$ instead of fixing them to co-evolution theory \cite{Saito:2014qha}. Also, we vary the baryon abundance with a Gaussian prior  $\omega_b = 0.02207 \pm 0.00015$. We use the DESI-like volume $V_{25}$. Vertical and horizontal dashed lines show the values of the simulations. 
 	\label{fig:Allparams}}
 	\end{center}
 \end{figure}

To sample the parameter space, we run the code \texttt{emcee}\footnote{\href{https://emcee.readthedocs.io/}{https://emcee.readthedocs.io/}} \cite{ForemanMackey:2012ig} which utilizes the affine-invariant ensemble sampler method for MCMC \cite{2010CAMCS...5...65G}.  Finally, we obtain the plots and confidence intervals using the \texttt{GetDist}  \texttt{Python} package \cite{Lewis:2019xzd}.

We compare against the monopole and quadrupole of the power spectrum up to a wave-number $k_{\text{max}}= 0.20 \hmpci$ for redshift $z=0.5$. 
In table \ref{Table:bestfit} we show the mean values and 0.68 c.l. intervals \revised{of the 1-dimensional marginalized} posterior distributions, where we considered the four volumes $V_1$, $V_5$, $V_{25}$, $V_{100}$. 
A triangular plot for these fits is shown in figure \ref{fig:Base_TriangularPlot} where we only show the cosmological parameters (the plot for all the parameters, including nuisances, is given in Appendix \ref{app:plots}, figure \ref{fig:Base_TriangularPlot_All}). 
\revised{We notice that for the smallest volume, $V_1$, we are not able to recover the mass of the neutrinos, but the posterior saturates the prior of $[0, 2]\, \text{eV}$, as it is represented with an em dash in table \ref{Table:bestfit}}. However, for the other effective volume cases we indeed recover all the cosmological parameters inside the 0.68 confidence intervals, it is easily noticed that even for the largest volume, where we expect that the modeling systematics dominate the error, we are able to obtain back all the parameters. 

 \begin{figure}
 	\begin{center}
 	\includegraphics[width=5.5 in]{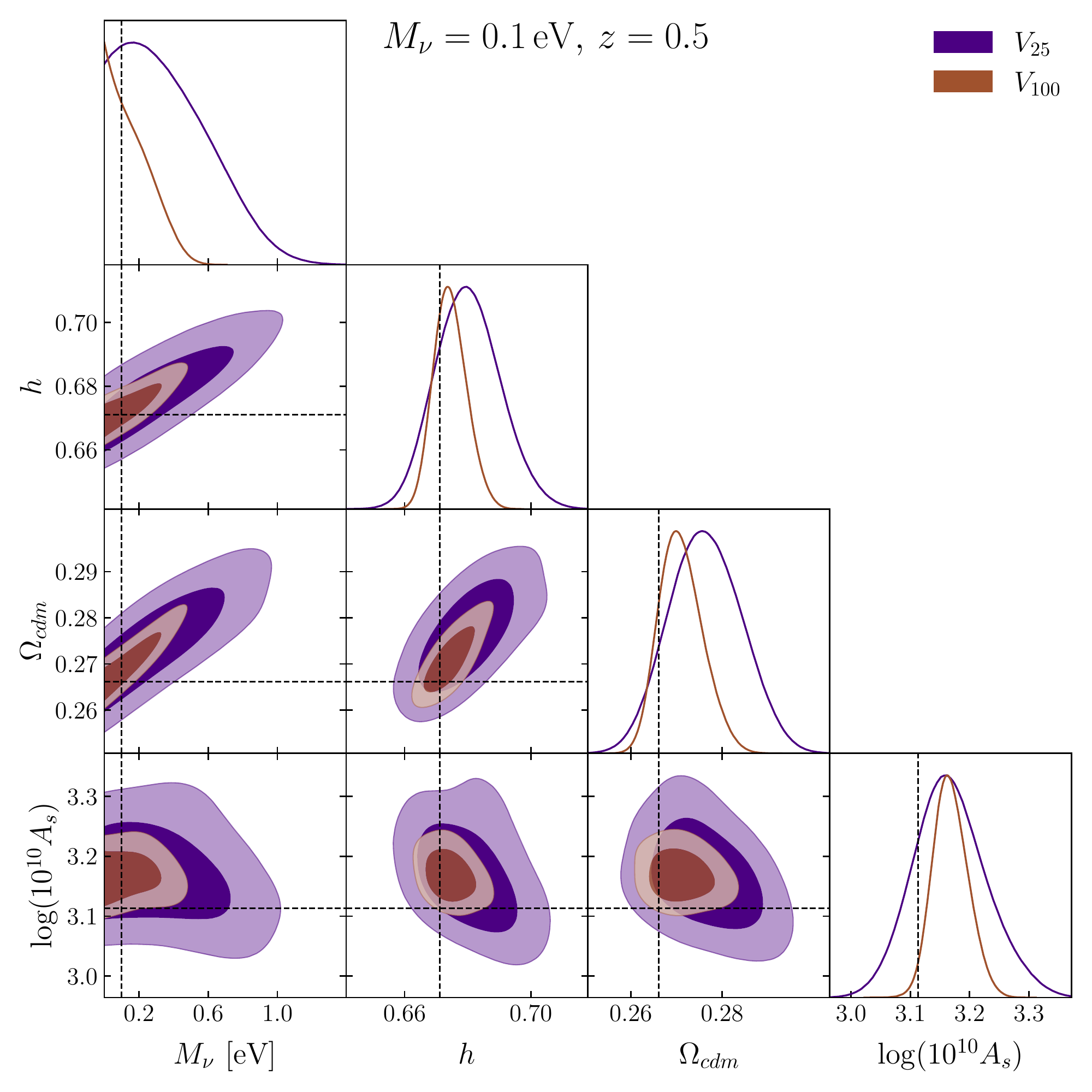}
 	\caption{Contour plots for the posterior distributions at 68 and 95\% confidence level computed with different covariance matrices, corresponding to $V_{25}$ and  $V_{100}$ effective volume data sets. This is the case of mass $M_\nu=0.1 \,\text{eV}$ at redshift $z=0.5$. The fittings are performed using the monopole and quadrupole of the power spectrum up to $k_{\text{max}} = 0.2 \, \hmpci$. Vertical and horizontal dashed lines show the values of the simulations. 
 	\label{fig:mass01z05}}
 	\end{center}
 \end{figure}

We perform the same exercise using EdS kernels, by means of simply setting $f(k) \rightarrow f_0$ in our \texttt{FOLPS$\nu$} code, and although the fittings are good as well, when our method of \fk-kernels is used the results for $M_\nu$ are modestly better. This is shown in figure \ref{fig:1dimfkvsEdS} where we plot the marginalized 1-dimensional posterior for $M_\nu$ for volumes $V_{25}$ and $V_{100}$. The estimation for the sum of the neutrino masses is within the 68\% limits for both cases when using the \texttt{fk}-kernels, but it fails when using EdS with volume $V_{100}$, although is recovered within the 95\% limits. The corresponding triangular plot is shown in Appendix \ref{app:plots}, figures \ref{fig:fkvsEdS} and \ref{fig:fkvsEdS25}, where all the rest of the cosmological parameters are recovered within the 0.68 c.l. intervals both with EdS and with \texttt{fk} kernels.

\begin{table}
\centering
\setlength{\tabcolsep}{0.80em} 
\renewcommand{\arraystretch}{1.4}
\begin{tabular}{llll}
\rowcolors{2}{White}{LightGray!30}
\begin{tabular}{*3c}\toprule
  \textbf{Parameter}                                        &  $\vec V_{25}$            & $\vec V_{100}$ \\\midrule
 $h$                                                        & $0.6800\pm 0.0099$        & $0.6744^{+0.0048}_{-0.0056}$\\
 $\Omega_{cdm}$                                             & $0.2763\pm 0.0077$        &  $0.2709^{+0.0042}_{-0.0052}$\\
 $M_{\nu}$  [eV]                                  & $<0.493$                  &   $<0.219$\\
 $\log(10^{10}A_s)$                                         & $3.166^{+0.057}_{-0.065}$ &   $3.168^{+0.027}_{-0.032}$\\
 $b_1$                                                      & $1.696\pm 0.073$          &   $1.667\pm 0.041$\\
 $b_2$                                                      & $-0.48^{+0.63}_{-0.52}$   &   $-0.28\pm 0.34$\\
 $\alpha_0$ [$ h^{-2} \, \text{Mpc}^2$]           & $12^{+18}_{-23}$          &   $-0.9^{+8.2}_{-9.7} $\\
 $\alpha_2$ [$ h^{-2} \, \text{Mpc}^2$]          & $-25.5\pm 7.7$            &   $-28.0\pm 3.4$\\
 $\alpha^{shot}_0-1$                                        & $-1.00\pm 0.22$           &   $-0.95\pm 0.15$\\
 $\alpha^{shot}_2$ [$ h^{-2} \, \text{Mpc}^2$]    & $-8.0\pm 1.1$             &   $-7.98\pm 0.53 $\\\bottomrule
\end{tabular}
\end{tabular}
\caption{\label{Table:bestfit_m01z05} 1-dimensional constraints for the different covariance matrices, corresponding to $V_{25}$ and  $V_{100}$ effective volume data sets. We show the mean of the posterior distributions with the 68\% confidence level intervals. We consider the case of massive neutrinos with $M_\nu=0.1 \,\text{eV}$ at redshift $z=0.5$. The fittings are performed using the monopole and quadrupole of the power spectrum up to $k_{\text{max}} = 0.2 \, \hmpci$. The corresponding triangular plot is shown in figure \ref{fig:mass01z05}.}
\end{table}

 \begin{figure}
 	\begin{center}
 	\includegraphics[width=5.5 in]{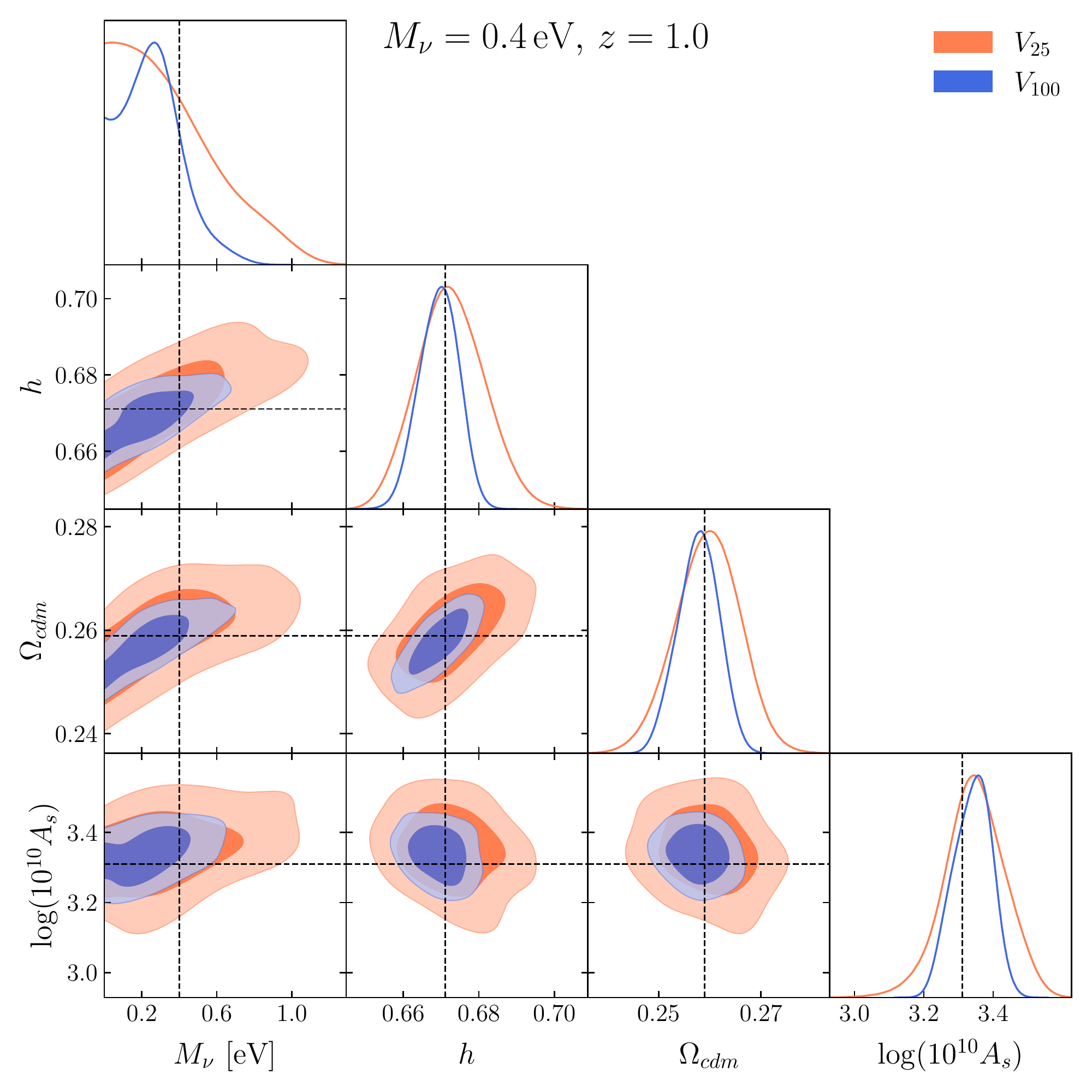}
 	\caption{\revised{Contour plots for the posterior distributions at 0.68 and 0.95 c.l. computed with different covariance matrices, corresponding to $V_{25}$ and  $V_{100}$ effective volume data sets. This is the case of mass $M_\nu=0.4 \,\text{eV}$ at redshift $z=1.0$. The fittings are performed using the monopole and quadrupole of the power spectrum up to $k_{\text{max}} = 0.2 \, \hmpci$. Vertical and horizontal dashed lines show the values of the simulations.} 
 	\label{fig:mass04z1}}
 	\end{center}
 \end{figure}

\begin{table}
\centering
\setlength{\tabcolsep}{0.80em} 
\renewcommand{\arraystretch}{1.4}
\begin{tabular}{llll}
\rowcolors{2}{White}{LightGray!30}
\begin{tabular}{*3c}\toprule
  \textbf{Parameter}                                &  $\vec V_{25}$             & $\vec V_{100}$ \\\midrule
 $h$                                                & $0.6726\pm 0.0089$         & $0.6694^{+0.0056}_{-0.0049}$   \\
 $\Omega_{cdm}$                                     & $0.2596\pm 0.0062$         & $0.2578^{+0.0043}_{-0.0038}$   \\
 $M_{\nu}$  [eV]                         & $< 0.460$                  & $0.26^{+0.12}_{-0.19}$     \\
 $\omega_{b}$                                       & $0.02208^{+0.00014}_{-0.00020}$ & $0.02209^{+0.00018}_{-0.00016}$ \\
 $\log(10^{10}A_s)$                                 & $3.347^{+0.090}_{-0.081}$       & $3.340^{+0.061}_{-0.051}$   \\
 $b_1$                                              & $2.27^{+0.12}_{-0.14}$          & $2.270^{+0.070}_{-0.084}$   \\
 $b_2$                                              & $0.4^{+1.7}_{-1.5}$             & $-0.8^{+1.6}_{-1.4}$   \\
 $b_{s^2}$                                          & $2.4^{+3.3}_{-2.8}$             & $-1.0^{+2.2}_{-3.5}$   \\
 $b_{3nl}$                                          & $0.23^{+0.38}_{-0.46}$          & $0.14^{+0.20}_{-0.25}$   \\
 $\alpha_0$ [$ h^{-2} \, \text{Mpc}^2$]   & $-10.5^{+1.6}_{-1.8}$           & $-7.6^{+4.5}_{-5.1}$   \\
 $\alpha_2$ [$ h^{-2} \, \text{Mpc}^2$]   & $-14.3\pm 2.7$                  & $-15.2^{+8.9}_{-11}$   \\
 $\alpha^{shot}_0-1$                                & $-0.71^{+0.29}_{-0.34}$         & $-0.78^{+0.15}_{-0.20}$   \\
 $\alpha^{shot}_2$ [$ h^{-2} \, \text{Mpc}^2$]    & $-5.25^{+0.84}_{-1.1}$  & $-5.10\pm 0.83$   \\\bottomrule
\end{tabular}
\end{tabular}
\caption{\label{Table:bestfit_m04z01} 1-dimensional constraints for the different covariance matrices, corresponding to $V_{25}$ and  $V_{100}$ effective volume data sets. We show the mean of the posterior distributions with the 68\% confidence level intervals. We consider the case of massive neutrinos with $M_\nu=0.4 \,\text{eV}$ at redshift $z=1.0$. The fittings are performed using the monopole and quadrupole of the power spectrum up to $k_{\text{max}} = 0.2 \, \hmpci$. The corresponding triangular plot is shown in figure \ref{fig:mass04z1}.}
\end{table}

So far, we have fixed the maximum wave-number value to $k_{\text{max}} =0.2 \hmpci$ since we found it optimal under our testings. In figure  \ref{fig:kmax} we show contour and 1-dimensional posterior distributions for three different values, $k_{\text{max}} = 0.15,\,0.2$ and $0.25 \hmpci$, using the DESI-like volume $V_{25}$. The case of  $k_{\text{max}} =0.25 \hmpci$ underestimate\revised{s} the neutrino mass $M_\nu$ considerably and overestimate\revised{s} the amplitude of fluctuations $A_s$, or alternatively underestimate the linear local bias which was measured for this halo sample to be $b_1= 1.665$ using the power spectrum \cite{Aviles:2021que} and the correlation function \cite{Aviles:2020cax}.\footnote{\revised{In those works, the cosmological parameters are fixed to the Quijote fiducial values, while the nuisances are fitted to the simulations.} } On the other hand, using either $k_{\text{max}} =0.15$ or $0.20 \hmpci$ we recover the cosmological parameters within the 68\% limits, but  $k_{\text{max}} = 0.20 \hmpci$ does it with smaller standard deviations. Hence, the justification of our choice of $k_{\text{max}}$.  \revised{The failure to recover the correct parameters when using large values for $k_\text{max}$ has been observed in previous works, for example in \cite{Aviles:2020cax} where only the nuisances were left free. In that work it was clear that as larger was $k_\text{max}$, the more inaccurate was the inferred value for $b_1$. Since this parameter is largely degenerate with the primordial fluctuation amplitude $A_s$, it is natural to expect a wrong estimation of the cosmological parameters as well. We attribute this to the fact that at small scales the form of the analytical power spectrum is largely dominated by the shot-noise and the EFT counterterms, which on the other hand become tightly constrained by the data large $k$ modes (see the red contours in the complementary figure \ref{fig:kmax_all}), leaving little room to the cosmological parameters to find their ``true'' values.}

Now, we present the results when letting free the biases $b_{s^2}$ and $b_{3nl}$ instead of fixing them to the co-evolution relations \eqref{coevbiases}. We show the triangular plots of posterior distribution using the DESI-like volume $V_{25}$ in figure \ref{fig:Allparams} with (blue  coloured) and without (green coloured) co-evolution, noticing no significant change in the estimation of parameters. Also, in this figure we show the results when we let free the baryon abundance with a Gaussian prior centered at the fiducial value of the simulations and standard deviation from {\it Planck} data 2018 $\omega_b = 0.02207 \pm 0.00015$. In this case, we also obtain very similar results, but with slight indications of improvement in the estimation for the $M_\nu$ best fit\revised{, as can be inferred by looking at the position of the maxima of the 1-dimensional posterior distributions of $M_\nu$ in figure \ref{fig:Allparams} with and without $\omega_b$ fixed}. 
We think this is because $\Omega_\nu \ll \Omega_b$, and hence even a small allowed change in the baryon abundance permits significant variations of the neutrino abundance.

We show our results when we compare against the massive neutrinos case $M_\nu= 0.1\, \text{eV}$ at redshift $z=0.5$ and using a maximum wave-number $k_{\text{max}} = 0.2 \hmpci$. These are shown in the triangular plots of figure \ref{fig:mass01z05} and table \ref{Table:bestfit_m01z05}. \revised{We found upper bounds for the neutrino mass, being $M_\nu< 0.493 \,\text{eV}$ for the effective volume $V_{25}$ and $M_\nu< 0.219 \,\text{eV}$ for $V_{100}$, both at 0.68 c.l. For the rest of the cosmological parameters we} notice that we are not able to recover all of them for $V_{100}$ within the 68\% limits, although they are inside the 95\% limits, while for $V_{25}$ the discrepancies are smaller, and the parameters lie inside the 0.68 c.l. intervals, except the case of the primordial fluctuations amplitude $A_s$, which is overestimated. 
We think the reason for this mismatch is the well-known degeneracy that exists between $A_s$ and $M_\nu$. This leads us to believe of that neutrino mass will not be measured with ongoing and near future surveys using 2-point functions alone, but we should use additional information coming from higher order statistics as the bispectrum \cite{Hahn:2019zob,Kamalinejad:2020izi,Hahn:2020lou} or even other summary statistics as the marked power spectrum and correlation function \cite{White:2016yhs,Massara:2020pli,Philcox:2020fqx,Philcox:2020srd}.

\revised{Finally, we test our modeling and code against the massive neutrinos case $M_\nu= 0.4\, \text{eV}$, but at a higher redshift of $z=1.0$, since these redshifts will be covered by current stage IV experiments as with the DESI-ELGs. We fit up to a maximum wave-number $k_{\text{max}} = 0.2 \hmpci$ and let free the biases $b_{3nl}$ and $b_{s^2}$, and the baryonic matter abundance $\omega_b$. As with the case of $M_\nu= 0.1\, \text{eV}$ at redshift $z=0.5$, we test the cases of $V_{25}$ and $V_{100}$. Our results are shown in the triangular plots of figure \ref{fig:mass04z1} and Table \ref{Table:bestfit_m04z01}. We notice we are able to recover all the parameters for the two effective volumes at 0.68 c.l., except the case of $M_\nu$ for the effective volume $V_{100}$, which shows a small mismatch, but still recovered at 0.95 c.l.}


\section{Code FOLPS$\nu$}
\label{sec:5}

Fast One Loop Power Spectrum in the presence of massive neutrinos (\texttt{FOLPS}$\nu$) is a Python code that computes the redshift space power spectrum multipoles in a fraction of second. It receives as input the linear power spectrum of matter or $cb$ field and the set of cosmological parameters $\{\Omega_m,\, h, \,M_\nu \}$ and nuisance parameters $\{b_1, b_2, b_{s^2}, b_{3nl}, \alpha_0, \alpha_2, \alpha_4, \tilde{c}, \alpha_0^{shot}, \alpha_2^{shot}\}$. (Alternatively, one can use the bias parameters $b_{\mathcal{G}_2}$ and  $b_{\Gamma_3}$, instead of $b_{s^2}$ and $b_{3nl}$.) The code computes the EFT galaxy redshift space power spectrum given in eq.~\eqref{PS_EFT} for both the wiggle and non-wiggle (also computed by the code) linear power spectra. Thereafter, it uses the IR-resummation expression in eq.~\eqref{PsIR}. Finally, the code computes the Legendre multipoles using the integral of eq.~\eqref{Pells}. All the loop integrals are computed using the formulae of  \S\ref{sec:FFTLog} and Appendix \ref{appB}. 
\revised{The code obtains $f_0$ by solving directly the differential equation for the linear growth $D_+$ at very large scales, using as input the matter abundance $\Omega_m$ and neglecting relativistic species.} 
The Hubble dimensionless rate $h$ is used to obtain the non-wiggle linear power spectrum using the fast sine transform recipe of \cite{Hamann:2010pw}. The SPT kernel $G_1(k) = f(k)/f_0$ is obtained with the Hu-Eisenstein approximation of ref.~\cite{Hu:1997vi}.

The code obtains $f_0$ by solving directly the differential equation for $D_+$ at very large scales, as in the massless neutrino case, using as input the matter abundance $\Omega_m$ and neglecting relativistic species. 

\texttt{FOLPS}$\nu$ employs the standard libraries \texttt{NumPy}\footnote{We recommend to use \texttt{NumPy} versions $\geq$ 1.20.0. For older versions, one needs to rescale $c_m$, $c_m^f$ and $c_m^{ff}$  by a factor $1/N$.} and \texttt{SciPy}. 
Other inputs are the FFTLog parameters $k_{\text{min}}$, $k_{\text{max}}$, $N$ and the bias $\nu$ introduced in  \ref{sec:FFTLog}.  
\revised{In addition, the code has the capability to switch between \texttt{fk}- and EdS-kernels, and it can also account for the Alcock-Paczynski effect if required \cite{Alcock:1979mp}.}

\begin{figure}
 	\begin{center}
 	\includegraphics[width=6.05 in]{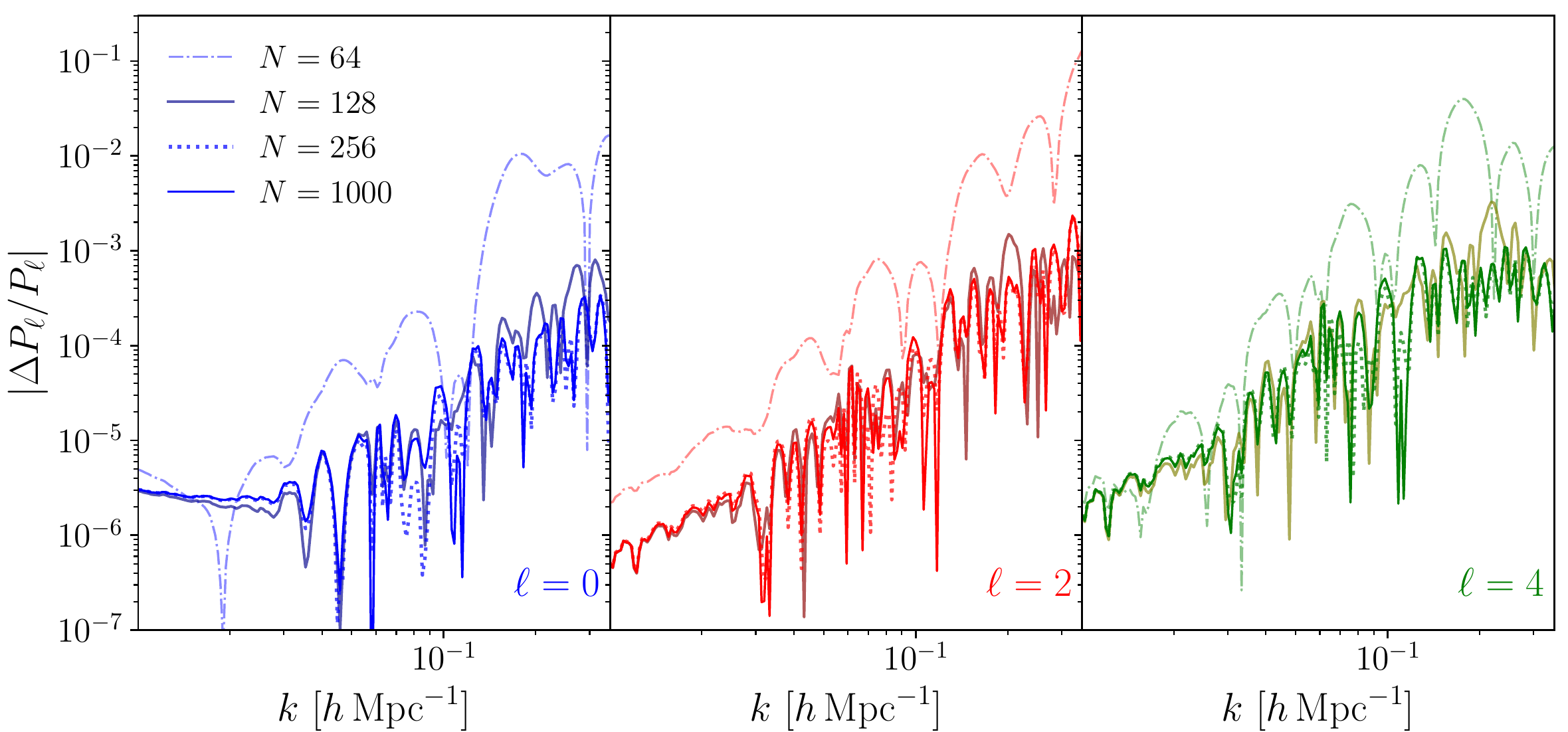}
 	\caption{Accuracy of the code \texttt{FOLPS$\nu$} when using the sampling points $N =  64$, $128$, $256$ and $1000$. The comparisons are against the direct integration, using as much precision as possible.
 	\label{fig:precision}}
 	\end{center}
 \end{figure}

To compute the Fast Fourier Transforms and construct the $M$ matrices, \texttt{FOLPS$\nu$} takes $N$ points logarithmically spaced over the interval $(k_{\text{min}} = 10^{-7} \hmpci, k_{\text{max}} = 100 \hmpci )$. In the following we test the use of different values $N =  64$, $128$, $256$ and $1000$. 
We compute the redshift space power spectrum multipoles for a set of cosmological and nuisance parameters, first using direct integration  (with a very high precision) and then through the \texttt{FOLPS$\nu$} code. Figure \ref{fig:precision} shows the relative difference between these two approaches for the different numbers of sampling points. As expected, by sampling  with $N=64$ we obtain the highest relative difference with respect to the direct integration, while the cases of $N =$ 128, 256 and 1000 provide similar accuracy. In particular, $N = 128$ produces a maximal error of $\sim 0.1 \%$ for the monopole and quadrupole, and $\sim 0.2 \%$ for hexadecapole up to $k = 0.35 \hmpci$, which is even smaller when sampling up to smaller wave-numbers. 
In all the computations, we use the default bias parameter $\nu=-0.1$ for all the functions except those required only for biased tracers, that is for eqs.~(\ref{Eq:Pb_1b_2})\,–\,(\ref{sigma23EdS}), for which we use $\nu = -1.51$. We have tested that these choices yield good convergence.

Let us now discuss the performance of our code. The computational times presented here were obtained with a standard personal computer with the following specifications: ASUS ZenBook with Cpu: 11th Gen Intel(R) Core(TM) i7-1165G7 @ 2.80GHz, RAM: 16 GB, OS: Ubuntu 20.04.4 LTS.

\begin{figure}
 	\begin{center}
 	\includegraphics[width=5.0 in]{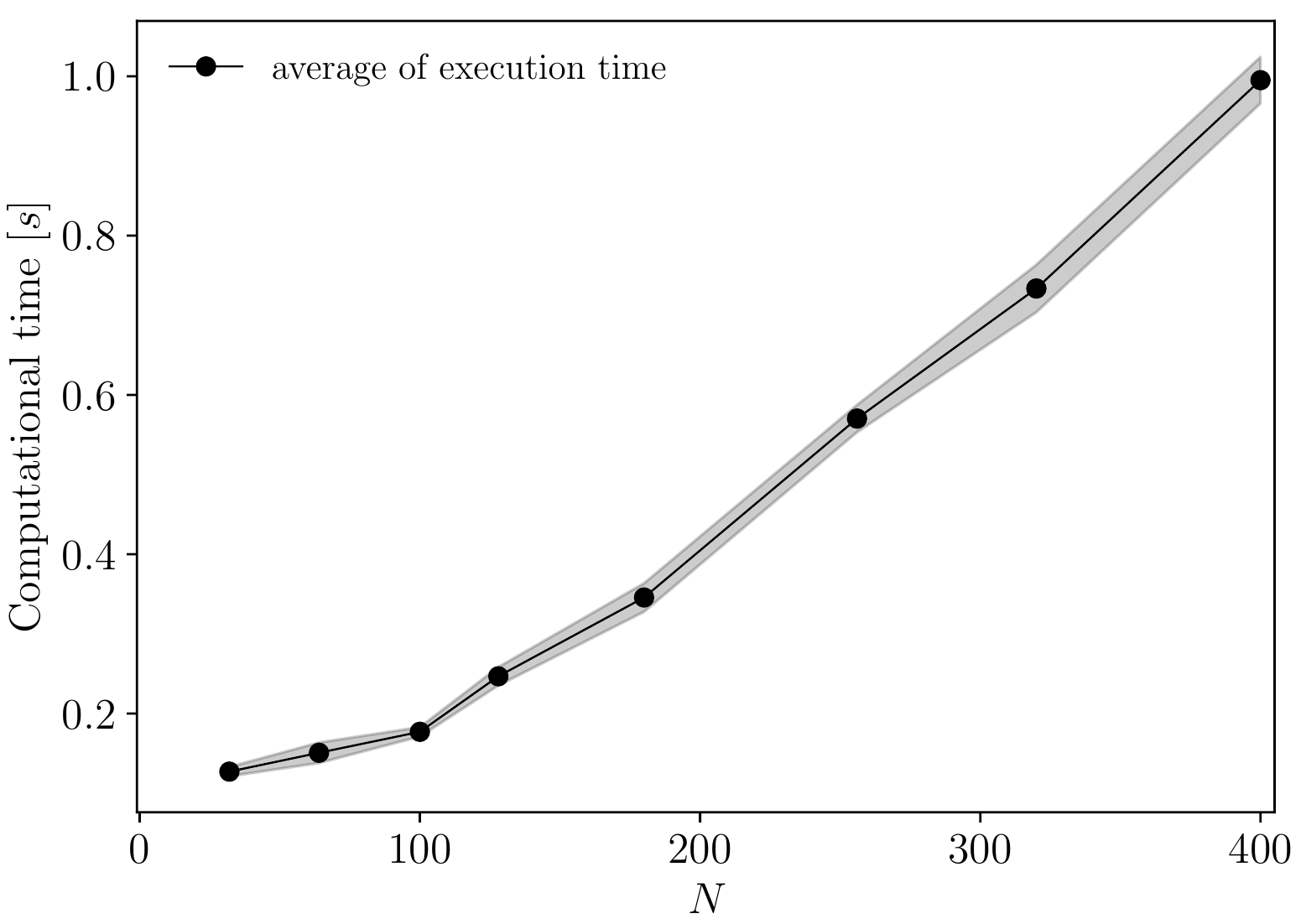}
 	\caption{Performance: Estimated end-to-end time taken by the \texttt{FOLPS$\nu$} code as a function of the number of sampling points $N$.
 	\label{fig:performance}}
 	\end{center}
 \end{figure}

Figure \ref{fig:performance} shows the estimated end-to-end time taken by the \texttt{FOLPS}$\nu$ code as a function of the number of sampling points $N$, here is not included the time taken by the \texttt{classy} module of the code \texttt{CLASS} to compute the linear $cb$ power spectrum. To these computations we have used 191 $k$ points over the interval $[0.01,0.5] \hmpci$. The black dots represent the $N$ values where the code was evaluated, while the black solid line refers to the average execution time, and the gray regions are the standard deviations of 100 runs. In table \ref{Table:times} we decompose some of these times into the different pieces executed during the code evaluation, the symbol “---” in the cells means that the time is unchanged since those pieces do not depend on the FFTLog parameter $N$. From the table we observe that the Hu-Eisenstein approximation is the fastest piece of the code, the next on this list are the inclusion of the IR-resummations and the computation of multipoles from eq.~(\ref{Pells}). Also note that the cosmology-dependent terms $c_m$ ($c^f_m$ and $c^{ff}_m$ for massive neutrinos cosmologies) are computed quickly through the use of the FFT algorithms, while most of the time is spent calculating matrix multiplications, especially the $P_{22}$-{\it type}. 

We notice the computational time of our \texttt{Python} code is competitive to others, as for example, the \texttt{C} language code \texttt{class-pt}, see table 1 in \cite{Chudaykin:2020aoj}. (Other codes, as \texttt{PyBird} \cite{DAmico:2020kxu}, \texttt{FAST-PT} \cite{McEwen:2016fjn,Fang:2016wcf} and \texttt{velocileptors} \cite{Chen:2020fxs} run in a similar amount of time as \texttt{class-pt}.) We think that this is because our expressions entering eq.~\eqref{Pkmu} have been analytically reduced to perform a smaller number of computations. Further, despite using \fk-kernels, the number of $M_{22}$ matrix multiplications is equal to $2\times 26$ against $2\times 24$ when EdS are used (the factor of 2 comes from the fact that we must compute all the matrix multiplications using both wiggle and non-wiggle linear power spectra). Hence our method is expected to run in a similar amount of time than EdS. If one considers that this code is thought to be used inside a pipeline containing a Boltzmann code and an MCMC sampler, the saving time when using EdS kernels is negligible.

Finally, since the matrices $M$ do not depend on the cosmology, they can be pre-computed and stored. However, it may be convenient to allow the user to select other different values of $N$ from those presented here. For that reason, our code computes the matrices $M$, this is the slowest part of the code, and we do not include it in figure \ref{fig:performance} and table \ref{Table:times} because this calculation is performed only one time. That is, the first time the code is called, it computes the $M$ matrices and stores them for the rest of the runs, which can be of the order of thousands in parameter estimations. This is shown in a \texttt{Jupyter} notebook released with the code. For the cases of $N = 64$, $128$ and $256$ the computational time for all the $M$ matrices are 0.118,  0.400, and 1.251 seconds, respectively.

\begin{table}
    \centering
    \begin{tabular}{cccccccc} \toprule
    {$N$} & {$f(k)/f_0$} & {$P_{nw}(k)$} & {$c_m$, $c^f_m$, $c^{ff}_m$} & {$P_{22}$-{\it type}} & {$P_{13}$-{\it type}} & {$P_s^\text{IR}(k,\mu)$, $P_\ell(k)$} & {Total} \\ \midrule
    64  & 0.00136 & 0.0535 & 0.0104 & 0.0477 & 0.0154 & 0.00787 &  0.136 \\\midrule
    128  & --- & --- & 0.0179 & 0.137 & 0.0183 & ---  & 0.236\\\midrule
    256  & --- & --- & 0.0317 & 0.435 & 0.0254 & --- & 0.555\\
    \bottomrule
\end{tabular}
    \caption{\label{Table:times} Computing time in [s] for the different terms for $N=64$, $128$, and $256$. The ``---'' means that the computing time is almost unchanged compared to the case $N=64$. The total number of computed of wave-numbers $k$ is 191.}
\end{table}



\section{Conclusions}\label{sect:conclusions}

During the upcoming years, the increasing constraining power of galaxy surveys will probably yield the tightest bounds on the sum of neutrino masses. The most important cosmological effect is the suppression of the power spectrum and correlation function below the free-streaming scale, which arises because neutrinos cannot be confined at those scales because their large velocity dispersion counteract the gravitational collapse and prevent neutrinos from forming structure. The amplitude and location of these effects \revised{depend} on the neutrino masses in a well understood physical process. However, the free-streaming lies at the onset of non-linearities, and as such the modeling of it requires the use of perturbation theory and non-linear effects as Effective Field Theory and IR-resummations. On the other hand, as surveys become wider and deeper they cover larger volumes and the statistical errors become smaller and systematic errors dominate. As such, improvements on the theoretical description would be very beneficial for data analyses.  In this work we have continued the development of a comprehensive perturbation theory for cosmologies in the presence of massive neutrinos building on the works \cite{Aviles:2020cax,Aviles:2020wme,Aviles:2021que} which compute kernels beyond EdS for the combined cold dark matter and baryonic fluid which is biased to obtain galaxy statistics. The use of more adequate kernels than EdS \revised{has} been studied in several works, and \revised{its  utility was shown}, particularly the Lagrangian approach, when used in COmoving Lagrangian Acceleration (COLA) implementations \cite{Winther:2017jof,Wright:2017dkw} as well as other approximate methods to generate mocks \cite{Moretti:2019bob}, and to initialize full N-body simulations \cite{Elbers:2022tvb}. However, their use for comparing against data has been used very little, if any. There are two important reasons for this: {\emph 1)} the so called full-shape methods that compare directly the consistent Effective field theory for Large Scales Structure to real data have reached sufficient masterization \revised{only} recently \cite{Ivanov:2019pdj,DAmico:2019fhj} in order to be used to constrain cosmological parameters; and, {\emph 2)} the use of complete kernels for cosmologies that contains additional scales, the neutrino mass in our case, requires lengthy computations of 1-loop integrals with kernels that are not known analytically, but should be found through differential equations at each bin of the 2-dimensional integrals.\footnote{\revised{We want to mention that this situation is encountered in some modified gravity models, such as $f(R)$ theories, symmetron fields, or more generally in a large sector of the Hordenski sector; see e.g., \cite{Koyama:2009me,Brax:2013fna,Aviles:2017aor,Aviles:2018qot,Aviles:2020wme} and Appendix B of \cite{Bose:2016qun}. In those cases there is an additional scale given by the associated scalar field (effective) mass $m_{eff}$. Clearly, there is no free-streaming effect, but a universal fifth force of range $1/m_{eff}$.  However, the presence of the additional scale leads to a function $A(k,t)$ different than in $\Lambda$CDM (where $A(t)=3/2 \, \Omega_m H^2 $) and ultimately to a $k$-dependence on the linear growth function and growth rate. For this reason the method developed in this work, particularly the \fk-kernels formalism, is potentially useful also for modified gravity theories. We are currently exploring this possibility and hope to present the results elsewhere in the near future.}}

We want to finish by mentioning that the results found in this work can be applied also to some modified gravity models, such as $f(R)$ theories, symmetron fields, and more generally to a large sector of the Hordenski theory (see e.g., \cite{Koyama:2009me,Brax:2013fna,Aviles:2017aor,Aviles:2018qot,Aviles:2020wme} and Appendix B of \cite{Bose:2016qun}). In those cases, there is an additional scale given by the associated scalar field (effective) mass $m_{eff}$. Clearly, there is no free-streaming effect, but a universal fifth force of range $1/m_{eff}$.  However, the presence of the additional scale leads to a function $A(k,t)$ different than $=3/2 \, \Omega_m H^2 $, and ultimately to a $k$-dependence on the growth rate. For this reason the method developed in this work, particularly the \fk-kernels formalism, is potentially useful also for modified gravity theories. We are currently exploring this possibility and hope to present the results elsewhere in a near future.

 However, the presence of the additional scale leads to a function $A(k,t)$ different than $=3/2 \, \Omega_m H^2 $, and ultimately to a $k$-dependence on the growth rate. For this reason the method developed in this work, particularly the \fk-kernels formalism, is potentially useful also for modified gravity theories. We are currently exploring this possibility and hope to present the results elsewhere in a near future.

In this work, following the results of \cite{Aviles:2021que}, we have developed a method that keeps the scale-dependent growth rates $f(k,t)$ as inherited from the linear theory non-local relation between velocity and density fields given in eq.~\eqref{Eq:linear_theta} [see e.g. eq.~\eqref{intheta2}], while the rest of the pieces are maintaining equal to their EdS counterparts. In this way, we are able to construct kernels with precise analytical expressions (up to the growth rate), but yet  the free-streaming scale is present also at the non-linear level and not only through the input linear power spectrum. The effects of having these more precise kernels are about, or smaller than, the 1 per-cent \revised{level} for neutrino masses around $0.1\,\text{eV}$, but when spanning over uninformative priors one can go up to large masses of $\gtrsim 1\,\text{eV}$, where the use of EdS kernels is at least questionable. Hence, we argue that for a correct sampling of the parameter space, one needs to use a more complete theory. The main objection against it, is not that such a theory is missing, since several PTs have existed from long time \cite{Wong:2008ws,Saito:2009ah}, but that the computational time of loop integrals is prohibitive to be used in MCMC algorithms. However, with the method presented in this work, named by us as \texttt{fk}, we are able to compute the \revised{loop integrals} in a time comparable to that in EdS. Further, since our kernels can now be written as the sum of terms $p^{2 n_1}|\vk-\vp|^{2 n_2} k^{-2(n_1 + n_2)}$ times growth rates, we are able to use FFTLog methods that improve the accuracy and reduce even more the time spent when computing the loop integrals.

We release the code \texttt{FOLPS$\nu$} that is fed with the linear $cb$ power spectrum and a set of cosmological and nuisance parameters, and it computes the galaxy/tracers redshift space power spectrum, including 1-loop corrections, effective field theory counterterms, biasing and IR-resummations. Despite our code \revised{is} written in \texttt{Python} and goes beyond EdS, it is competitive in time with other existing codes. We attribute this to the fact that our final expansion is different, since we are able to write the loop power spectrum in the form of eq.~\eqref{Pkmu}, analytically reducing the number of computations performed by our code. Further, our code does not require any considerably more amount of time than the use of EdS, since actually the number of matrix multiplications is only 10\% larger when using \texttt{fk}-kernels.

In  \S\ref{sec:modelvalidation} we validated our model and code by comparing against the \textsc{Quijote} suite of simulations. We use as our baseline model the most massive case available in these simulations, \revised{$M_\nu=0.4 \, \text{eV}$}, because in this case the use of a more proper theory becomes more relevant. Further, we use halos and a very large volume of $(100\, \hgpc)^3$, corresponding to the 100 realizations in the simulations, in order to have very small statistical errors. As such, any failure in recovering the cosmological parameters of the simulations is likely to be a cause of a wrong modeling in our theory. The most important plot of the model validation is shown in figure \ref{fig:Base_TriangularPlot}, where we fit to four different data sets,  all with the same central points but the corresponding covariance scaling with the inverse of the volume, for effective volumes of  $1,\,5,\,25\,\,\text{and} \,\,100 \, h^{-3}\,\text{Gpc}^3$, named $V_1$, $V_5$, $V_{25}$ and $V_{100}$, respectively. We find that we are able to recover all the cosmological parameters within the 68\% confidence limits, even for the largest volume where the errors are very small. We further compare our constraints with those obtained with the use of EdS, finding modest but noticeable improvements when using our method. Curiously, this happens only for the $M_\nu$ parameter, with EdS failing to recover it at 0.68 c.l. for $V_{100}$.

In our baseline fitting, we fixed the baryon abundance to the simulation's fiducial value. However, we also test the effect of varying it with a Gaussian prior with width given by {\it Planck} 2018. By doing this, being quite small, the neutrino abundance has enough room to move a little bit without affecting  the baryon abundance considerably, and ultimately allowing a slight improvement for the best fit.  

We also compare against the case of $M_\nu = 0.1\,\text{eV}$ finding not as good agreement than the other case. Particularly, we found the mass of the neutrinos is overestimated, \revised{and at the same time, the fluctuations primordial amplitude $A_s$ is overestimated as well, but in an even larger amount.} Since these parameters are highly degenerate, we argue that the galaxy 2-point functions alone will not be sufficient to measure the absolute mass scale of the neutrinos \revised{with current and upcoming surveys}, which is a somewhat believed result; e.g. \cite{Hahn:2019zob}.

We believe that calibrations in our method as the choice of more optimal parameters should be done by using mocks and ultimately comparisons with real data. This is a work in progress that we hope to present soon elsewhere. Particularly with the view of upcoming results for the first year DESI data. 

\acknowledgments

We would like to thank Willem Elbers, Carlos Frenk, Baojiu Li, Martin White and Arka Banerjee for useful discussions and suggestions. AA and HEN are supported by CONACyT Ciencia de Frontera grant No.319359 and CONACyT grant 283151. AA also acknowledges CONACyT grant 102958. 
MV, HEN and SF acknowledges PAPIIT IN108321 and PAPIIT IA103421. MV acknowledges  CONACyT grant A1-S-1351. This research was partially supported through computational and human resources provided by the LAMOD UNAM project through the clusters
Atocatl and Tochtli. LAMOD is a collaborative effort between the IA, ICN and IQ institutes at UNAM.
\\

\appendix
\section{Functions A and D}\label{sect:AandD}


Function $A^\text{TNS}(k,\mu)$ was originally introduced in \cite{Taruya:2010mx} for the TNS (Taruya-Nishimichi-Saito) model. It is given by eq.~\eqref{ATNS} and can be written as 
\begin{align}
 A^\text{TNS}(k,\mu) &= \mu^2 \big[ f_0 I^{1,udd}_1(k)  + f_0^2 I^{2,uud}_{1}(k)\big] +  \mu^4 \big[f_0^2 I^{2,uud}_{2}(k) \nonumber\\
 &\quad + f_0^3 I^{3,uuu}_{2}(k) \big] + \mu^6 f_0^3 I^{3,uuu}_{3}(k),
\end{align}
where 
\begin{align}
I^{m, ijk}_n (k) &= 2 \ip  \va_{nm}(\vk,\vp) P_L(k) P_L(p)  
 + \ip  \tilde{\mathcal{A}}_{nm}(\vk,\vp)  P_L(p) P_L(|\vk-\vp|),
  \label{Eq:Inm_abc} \\ & 
\equiv I^{m,ijk}_{n,a}(k)+I^{m,ijk}_{n,b}(k). \label{Eq:Inm_ijk}
\end{align}
for the following index combination :
\begin{table}[H]
\centering
\setlength{\tabcolsep}{1.10em} 
\renewcommand{\arraystretch}{1.0}
\begin{tabular}{lll}
\hline
\multicolumn{3}{c}{\bf{Index combination}} \\
\hline
\multirow{1}{*}{$m = 1$ } 
& $ n = 1$  & $ijk  = udd$ \\ 
 \hline
\multirow{1}{*}{$m = 2$ } 
& $ n = 1, 2$  & $ijk  = uud$ \\
 \hline
\multirow{1}{*}{$m = 3$ } 
& $ n = 2, 3$  & $ijk  = uuu$ \\
 \hline
\end{tabular}
\caption{\label{Table:Index_} Index combination for $I^{m, ijk}_n$ functions.}
\end{table}
\noindent
That is, the labels $uuu$, $uud$, $\dots$, serve to remind us where each term comes from. For example, the function $I^{2,uud}_{2}(k)$ comes from a bispectrum composed of two velocity fields ($u$) and one density field ($d$). It should not be confused with the function $I^{2,uudd}_{2}(k)$, introduced below, which is obtained from a 4-point correlation $\langle uudd \rangle$. Actually, the total function $I^m_n$ with $m=2, n=2$, is $I^{2}_{2}(k) = P_{\theta\theta}(k) + I^{2,uud}_{2}(k)+I^{2,uudd}_{2}(k)$.

The usual functions in the TNS model are $A_{nm}$, $\tilde{A}_{nm}$ and $a_{nm}$ \cite{Taruya:2010mx}, instead of our bold faced $\va_{nm}$ and $\mathcal{\tilde{A}}_{nm}$. But they are actually the same, as is manifestly under the identifications $\va_{nm}(\vk, \vp) = r^{-2} a_{nm}(\vk, \vp)$ and $\mathcal{\tilde{A}}_{nm}(\vk, \vp) = r^{-2}\tilde{A}_{nm}(\vk, \vp)/(1+r^2-2rx)^2$ (with $r=p/k$ and $x=\hat{\vp}\cdot\hat{\vk}$), and only differ in the used kernels $G_2$ and rates $f(k)/f_0$ \cite{Aviles:2020wme}. Also, the functions $\mathcal{A}_{nm}$ are unnecessary because $\mathcal{A}_{nm}(\vk,\vk - \vp) = \va_{nm}(\vk,\vp)$, and hence the factor ``2'' in eq.~\eqref{Eq:Inm_abc}.

The functions $\va_{nm}(\vk,\vp)$, $\tilde{\mathcal{A}}_{nm}(\vk,\vp)$ are given by
\begin{align}
 \va_{11}(\vk,\vp) & =  2 \frac{\vk \cdot \vp}{p^2}G_1(p)  F_2(-\vk,\vp) + 2  \frac{\vk \cdot (\vk-\vp)}{|\vk-\vp|^2}  G_2(-\vk,\vp),
 \\
\tilde{\mathcal{A}}_{11}(\vk,\vp) & = 4 \frac{\vk \cdot \vp}{p^2} G_1(p) F_2(\vp, \vk - \vp),
\end{align}

\begin{align}
 \va_{12}(\vk,\vp) & = \frac{1}{|\vk-\vp|^2} \left[\frac{(\vk \cdot \vp)^2}{p^2} - k^2 \right] G_1(p) G_2(-\vk, \vp),
 \\
\tilde{\mathcal{A}}_{12}(\vk,\vp) & = \frac{1}{|\vk-\vp|^2} \left[\frac{(\vk \cdot \vp)^2}{p^2} - k^2 \right] G_1(p) G_1(|\vk-\vp|) F_2(\vp, \vk -\vp),
\end{align}

\begin{align}
\va_{33}(\vk, \vp) &= \frac{1}{|\vk-\vp|^2} \left[ k^2 +2 \frac{k^2}{p^2}(\vk \cdot \vp) - 3 \frac{(\vk \cdot \vp)^2}{p^2}\right] G_1(k) G_1(p) G_2(-\vk, \vp),
\\
\tilde{\mathcal{A}}_{33}(\vk, \vp) &= \frac{1}{|\vk-\vp|^2} \left[k^2 + 2 \frac{k^2}{p^2}(\vk \cdot \vp) - 3 \frac{(\vk \cdot \vp)^2}{p^2} \right] G_1(p) G_1(|\vk-\vp|) G_2(\vp, \vk-\vp),
\end{align}

\begin{align}
\va_{22}(\vk, \vp) &= \frac{\va_{33}(\vk, \vp)}{G_1(k)} + \va_{11}(\vk, \vp) G_1(k),   
\\
\tilde{\mathcal{A}}_{22}(\vk, \vp) &= \frac{1}{|\vk-\vp|^2} \left[k^2 + 2 \frac{k^2}{p^2}(\vk \cdot \vp) - 3 \frac{(\vk \cdot \vp)^2}{p^2} \right] G_1(p) G_1(|\vk-\vp|) F_2(\vp, \vk-\vp)
\nonumber \\ & \quad
+ 4 \frac{\vk \cdot \vp}{p^2} G_1(p) G_2(\vp, \vk - \vp),
\end{align}

\begin{align}
\va_{23}(\vk, \vp) & = \va_{12}(\vk, \vp) G_1(k), 
\\
\tilde{\mathcal{A}}_{23}(\vk,\vp) & = \frac{1}{|\vk-\vp|^2} \left[\frac{(\vk \cdot \vp)^2}{p^2} - k^2 \right] G_1(p) G_1(|\vk-\vp|) G_2(\vp, \vk -\vp).
\end{align}
For EdS kernels, all these functions coincide with those reported in TNS paper \cite{Taruya:2010mx}, but here we have reduced the number of different functions $A_{mn}$, $a_{mn}$ and $\tilde{A}_{mn}$ from \revised{being 15 to 8}, plus two more ($\va_{22}$ and $\va_{23}$) that are trivially obtained from the rest: that is, they differ only by factors $G_1(k)$ that can be pulled out of the internal momentum $\vp$ integrals as shown in eqs.~\eqref{Eq:I22a} and \eqref{Eq:I32a}. 


Now, the $D(k, \mu)$ function in eq.~\eqref{DTNS} can be expressed as \cite{Aviles:2021que}
\begin{align}\label{D_function}
D(k,\mu) & = \mu^2  f^2_0 I^{2, uudd}_1(k) 
+ \mu^4  \big[ f^2_0 I^{2, uudd}_2(k) + f^3_0 I^{3, uuud}_2(k)  + f^4_0 I^{4, uuuu}_2 (k)\big]
\nonumber \\ & \quad
+ \mu^6  \big[ f^3_0 I^{3, uuud}_3(k) + f^4_0 I^{4, uuuu}_3(k) \big] + \mu^8 f^4_0 I^{4, uuuu}_4 (k),
\end{align}
with
\begin{align}
I^{2, uudd}_n (k) & = I^{2, uudd}_{n, D} (k) - \delta_{n1} k^2 \sigma^2_{vv} P_{\delta \delta}(k), \hspace{1cm} (n = 1, 2)
\\
I^{3, uuud}_n (k) & = I^{3, uuud}_{n, D} (k) - \delta_{n2} 2k^2 \sigma^2_{vv} P_{\delta \theta}(k), \qquad (n = 2, 3)
\\
I^{4, uuuu}_n (k) & = I^{4, uuuu}_{n, D} (k) - \delta_{n3} k^2 \sigma^2_{vv} P_{\theta \theta}(k), \hspace{0.9cm} (n = 2, 3, 4)
\end{align}
where the label $D$ means the sum of contributions $B$ and $C$. \revised{The velocity dispersion $\sigma^2_{vv}$ is given by
\begin{equation}\label{sigma2vv}
    \sigma^2_{vv} \equiv \frac{1}{6\pi^2} \int^\infty_0 dp \, P^L_{\theta \theta} (p).
\end{equation}
}
Explicitly, the above $I^m_n$ functions are
\begin{align} \label{I2uuddnB}
I^{2, uudd}_{n, B} (k) = \ip \mathcal{B}^n_{1 1}(\vk, \vp) P^L_{\delta \theta}(p) P^L_{\delta \theta}(|\vk - \vp|),
\end{align}
with
\begin{align}
\mathcal{B}^1_{1 1}(\vk, \vp) &= \frac{1}{2|\vk - \vp|^2} \left[ \frac{(\vk \cdot \vp)^2}{p^2} - k^2 \right],
\\
\mathcal{B}^2_{1 1}(\vk, \vp) &= \frac{1}{2|\vk - \vp|^2} \left[k^2 + 2 \frac{k^2}{p^2} (\vk \cdot \vp)  - 3 \frac{(\vk \cdot \vp)^2}{p^2} \right],
\end{align}
and
\begin{align} \label{I2uuddnC}
I^{2, uudd}_{n, C} (k) = 2 \ip \mathcal{C}^n_{1 1}(\vk, \vp)  P_L(p) P^L_{\theta \theta}(|\vk - \vp|),   
\end{align}
with
\begin{align}
 \mathcal{C}^1_{1 1}(\vk, \vp) & = \frac{1}{4 |\vk - \vp|^4} \left[k^2 p^2 - (\vk \cdot \vp)^2 \right],
 \\
\mathcal{C}^2_{1 1}(\vk, \vp) & = \frac{1}{4 |\vk - \vp|^4} \left[2k^4  -k^2p^2 - 4k^2 (\vk \cdot \vp) + 3(\vk \cdot \vp)^2 \right].
\end{align}
Notice these expressions have different power spectrum product combinations, so they can not be written into a single term, as we do in the following for the rest of the $I^{m}_n$ functions for $D(k,\mu)$.

Functions with $m=3$ are 
\begin{align} \label{I3uuudnD}
I^{3, uuud}_{n, D}(k) = -2 \ip \mathcal{D}^n_{21}(\vk, \vp) P^L_{\delta \theta}(p) P^L_{\theta \theta}(|\vk - \vp|),
\end{align}
with
\begin{align}
\mathcal{D}^2_{21}(\vk, \vp) &= \frac{k^4}{2 |\vk - \vp|^4} \left[2-3 \frac{(\vk \cdot \vp)}{k^2} \right] \left[ 1- \frac{(\vk \cdot \vp)^2}{k^2 p^2}\right],
\\
\mathcal{D}^3_{21}(\vk, \vp) & = - \frac{k^4}{2 |\vk - \vp|^4} \left[ 2 + \left(\frac{2}{p^2} - \frac{3}{k^2}\right)(\vk \cdot \vp) - 6\frac{(\vk \cdot \vp)^2}{k^2 p^2} + 5 \frac{(\vk \cdot \vp)^3}{k^4 p^2}\right].
\end{align}

Finally, functions with $m=4$ are given by
\begin{align} \label{I4uuuunD}
I^{4, uuuu}_{n, D} (k) = \ip \mathcal{D}^n_{22}(\vk, \vp) P^L_{\theta \theta}(p)   P^L_{\theta \theta}(|\vk - \vp|), 
\end{align}
with
\begin{align}
\mathcal{D}^2_{22}(\vk, \vp) & = \frac{3}{16 |\vk - \vp|^4} \left[k^2 - \frac{(\vk \cdot \vp)^2}{p^2} \right]^2,
\\
\mathcal{D}^3_{22}(\vk, \vp) & = \frac{k^2}{8p^2 |\vk - \vp|^4} \left[ k^2 - \frac{(\vk \cdot \vp)^2}{p^2}\right] \left[ 2k^2 -3p^2 -12(\vk \cdot \vp) + 15 \frac{(\vk \cdot \vp)^2}{k^2}\right],
\\
\mathcal{D}^4_{22}(\vk, \vp) & = \frac{k^4}{16p^2 |\vk - \vp|^4} \bigg[ 3p^2-4k^2+24(\vk \cdot \vp) + \left( \frac{12}{p^2}-\frac{30}{k^2} \right)(\vk \cdot \vp)^2 - 40 \frac{(\vk \cdot \vp)^3}{k^2 p^2}
\nonumber \\ & \quad
+ 35\frac{(\vk \cdot \vp)^4}{k^4 p^2}
\bigg].
\end{align}
We have reduced the $m=3,4$ contributions to a single term expression. Then, for these cases, the $B$ and $C$ contributions boil down to only one matrix multiplication for each $n$ value when applying the FFTLog formalism.

\bigskip

To complete the full framework, we need to introduce the bias into functions $A^\text{TNS}$ and $D$. These are obtained by substituting $f_0 \rightarrow f_0/b_1$ and weighting the functions with powers of $b_1$, namely
\begin{align}
A^\text{TNS}(k,\mu;f_0) &\, \longrightarrow \, b_1^3 A^\text{TNS}(k,\mu;f_0/b_1), \\
D(k,\mu;f_0) &\, \longrightarrow \, b_1^4 D(k,\mu;f_0/b_1).
\end{align}
The biasing for the $D$ function is exact, because  $D$ is constructed from linear fields only. Instead, the function $A^\text{TNS}$ is also biased by the non-linear $\delta^2_{cb}$ and $s^2$ operators. The biased expression of $A^\text{TNS}(k,\mu)$, includes also $b_2$ and $b_{s^2}$ but the resulting corrections are subdominant, even smaller in amplitude than the other contributions carrying these bias parameters, so we neglect them.  The complete expression including second order linear and tidal biases can be found in Appendix A.1 of \cite{Aviles:2020wme}.

\section{FFTLog contributions}
\label{appB}
In this appendix we present explicit expressions for FFTLog contributions, including their matrices or vectors and UV or IR corrections if necessary. We employ two different values for the bias parameter $\nu$. For the biasing power spectra $P_{b_1b_2},\dots$ we use $\nu = -1.51$, while for the rest of the functions we use $\nu = -0.1$.

In the following, remind that $\nu_1$ and $\nu_2$ should not be confused with the bias $\nu$, but these are defined through eqs.~\eqref{nu1nu2defs}.


\subsection*{Overdensity and velocity power spectra}
We start with the leading non-linear contributions~(\ref{P22dd})\,–\,(\ref{P13vv}), whose approximations are
\begin{align}
&P^{22}_{\delta \delta} (k)   =  k^3 \displaystyle\sum\limits_{m_1, m_2} c_{m_1} k^{-2 \nu_1} M_{22, \delta \delta} (\nu_1, \nu_2) \, c_{m_2} k^{-2 \nu_2},
\\
&M_{22, \delta \delta }(\nu_1, \nu_2 )   = \Big[\nu_1 \nu_2  \left( 98 \nu^2_{12} -14 \nu_{12} +36 \right) - 91\nu^2_{12} + 3\nu_{12}+ 58  \Big] 
 \nonumber\\ &\quad  \qquad \qquad\qquad\times
\frac{\left(\frac{3}{2}-\nu_{12} \right) \left(\frac{1}{2} -\nu_{12} \right)}{196 \nu_1 (1+\nu_1)\left( \frac{1}{2} - \nu_1\right) \nu_2(1+\nu_2)\left( \frac{1}{2} - \nu_2\right)} I(\nu_1, \nu_2).
\end{align}  

\begin{align}
&P^{22}_{\delta \theta}(k) =  2 k^3  \displaystyle\sum\limits_{m_1, m_2} c^f_{m_1} k^{-2\nu_1} M^{f_{\vp}}_{22, \delta \theta }(\nu_1, \nu_2) \, c_{m_2} k^{-2\nu_2},
\\
&M^{f_{\vp}}_{22, \delta \theta }(\nu_1, \nu_2) = \Big[7 (7 \nu_1+3) \nu_2^2+(7 \nu_1 (7 \nu_1-1)-38) \nu_2-21 \nu_1-23 \Big] 
\nonumber\\ & \quad \qquad \qquad\qquad\times 
\frac{(2 \nu_{12}-3) (2 \nu_{12}-1)}{196 \nu_1 (\nu_1+1) \nu_2 (\nu_2+1) (2 \nu_2-1)} I(\nu_1, \nu_2).
\end{align}

\begin{align}
&P^{22}_{\theta \theta} (k) = k^3 \displaystyle\sum\limits_{m_1, m_2} c^{ff}_{m_1} k^{-2\nu_1} M^{f_{\vp}f_{\vp}}_{22, \theta \theta}(\nu_1, \nu_2) \, c_{m_2} k^{-2\nu_2}   
\nonumber \\ & \quad \qquad
+ k^3 \displaystyle\sum\limits_{m_1, m_2} c^{f}_{m_1} k^{-2\nu_1} M^{f_{\vk-\vp} f_{\vp}}_{22, \theta \theta}(\nu_1, \nu_2) \, c^f_{m_2} k^{-2\nu_2},
\\
&M^{f_{\vp}f_{\vp}}_{22, \theta \theta}(\nu_1, \nu_2)  = 
\Big[98 \nu_1^3 \nu_2+7 \nu_1^2 \big(2 \nu_2 (7 \nu_2-8)+1 \big) -\nu_1 \big(2 \nu_2 (7 \nu_2+17)+53 \big)
 \nonumber\\
&\quad  \qquad\qquad\qquad
-12 (1-2 \nu_2)^2\Big]
    \frac{2\nu_{12}-3}{98 \nu_1 (\nu_1+1) \nu_2 (\nu_2+1) (2 \nu_2-1)}
    I(\nu_1, \nu_2),
\\
 &M^{f_{\vk-\vp} f_{\vp}}_{22, \theta \theta}(\nu_1, \nu_2) = \Big[7 \nu_1^2 (7 \nu_2+3)+\nu_1 \big(7 \nu_2 (7 \nu_2-1)-10\big)+\nu_2 (21 \nu_2-10)-37\Big]
     \nonumber\\
&\quad  \qquad\qquad\qquad \times \frac{2\nu_{12} -3 }{98 \nu_1 (\nu_1+1) \nu_2 (\nu_2+1)} I(\nu_1, \nu_2).
\end{align}

\begin{align}
& P^{13}_{\delta \delta}(k)  = k^3 P_L(k) \displaystyle\sum\limits_{m_1} c_{m_1} k^{-2\nu_1} M_{13, \delta \delta}(\nu_1),  
 \\
& M_{13, \delta \delta}(\nu_1) = \frac{1+9\nu_1}{4} \frac{\tan(\nu_1\pi)}{28\pi (\nu_1+1)\nu_1(\nu_1-1)(\nu_1-2)(\nu_1-3)}.
\end{align}

\begin{align}
&P^{13}_{\delta \theta} (k) = \frac{1}{2} k^3 P_L(k)\displaystyle\sum\limits_{m_1} \left( \frac{f(k)}{f_0}c_{m_1} + c^f_{m_1} \right) k^{-2\nu_1} M^{f_{\vk}}_{13, \delta \theta}(\nu_1),
\\
&M^{f_{\vk}}_{13, \delta \theta} (\nu_1) = \frac{9\nu_1-7}{4}
     \frac{\tan(\nu_1\pi)}{28\pi (\nu_1+1)\nu_1(\nu_1-1)(\nu_1-2)(\nu_1-3)}.
\end{align}

\begin{align}
 &    P^{13}_{\theta \theta}(k) = k^{3} \frac{f(k)}{f_0}  P_L(k) \left(  \frac{f(k)}{f_0} \displaystyle\sum\limits_{m_1} c_{m_1} k^{-2 \nu_1}M^{f_{\vk}}_{13, \theta \theta} (\nu_1) +  \displaystyle\sum\limits_{m_1} c^{f}_{m_1}k^{-2\nu_1}M^{f_{\vp}}_{13, \theta \theta} (\nu_1) \right),
    \\
 & M^{f_{\vk}}_{13, \theta \theta} (\nu_1) = -\frac{\tan (\nu_1\pi )}{14 \pi  (\nu_1+1)\nu_1(\nu_1-1)(\nu_1-2)(\nu_1-3)},
    \\
 & M^{f_{\vp}}_{13, \theta \theta} (\nu_1) = M^{f_{\vk}}_{13, \delta \theta} (\nu_1).
\end{align}
Since we use $\nu = -0.1$, the only corrections that must be added are
\begin{align}
P^{13, \text{UV}}_{\delta \delta}(k) &= - \frac{61}{105} P_L(k) k^2 \sigma^2_{\Psi},
\\
P^{13, \text{UV}}_{\delta \theta} (k) &=  - \left(  \frac{23}{21}  \frac{f(k)}{f_0} \sigma^2_\Psi + \frac{2}{21} \sigma^2_v \right) k^2 P_L(k), 
\\
P^{13, \text{UV}}_{\theta \theta} (k) & = -\left( \frac{169}{105} \frac{f(k)}{f_0} \sigma^2_\Psi  +\frac{4}{21} \sigma^2_v \right)  k^2 P^L_{\delta \theta}(k),
\end{align}
where $\sigma^2_\Psi$ and $\sigma^2_v$ are given by eq.~(\ref{Eq:sigmas_psi_v}).

\subsection*{Biasing power spectra}
The integrals of eqs.~(\ref{Eq:Pb_1b_2})\,–\,(\ref{Pb2s2}) can be approximated by
\begin{equation}
    P(k) = k^3 \sum_{m_1, m_2} c_{m_1}k^{-2\nu_1}  M_P(\nu_1, \nu_2) \, c_{m_2}k^{-2\nu_2},
\end{equation}
where ``$P$'' refers to $P_{{b_1} b_2}$, $ P_{b_1 b_{s^2}}$, $P_{b^2_2}$, $ P_{b_2 b_{s^2}}$ and $P_{b^2_{s^2}}$, while the matrices involved are 
\begin{align}
M_{P_{b_1 b_2}}\left(\nu _1,\nu _2\right) & = \frac{(2\nu_{12}-3) (7 \nu_{12}-4)}{28 \nu_1 \nu_2} I(\nu_1,\nu_2), 
\\
M_{P_{b_1 b_{s^2}}}\left(\nu _1,\nu _2\right) & = \Big[14 \nu_1^2 (2\nu_2-1)+\nu_1 \big(4\nu_2 (7 \nu_2-11)-3\big)-\nu_2 (14 \nu_2+3)+2\Big]  
\nonumber\\ &\quad \times \frac{(2\nu_{12}-3) }{168 \nu_1 (\nu_1+1) \nu_2 (\nu_2+1)} I(\nu_1,\nu_2),
\\
M_{P_{b^2_2}}(\nu_1,\nu_2) & = \frac{1}{2} I(\nu_1,\nu_2),
\\
M_{P_{b_2 b_{s^2}}}\left(\nu_1,\nu_2\right) & = \frac{(2 \nu_1-3) (2 \nu_2-3) }{12 \nu_1 \nu_2} I(\nu_1,\nu_2),    
\\
M_{P_{b_{s^2}^2}}\left(\nu _1,\nu _2\right) &= \Big[4 \Big(\big(3+2 (\nu_1-2) \nu_1 \big) \nu_2^2+ (17-4 \nu_1) \nu_1\nu_2 +3 (\nu_1-5) \nu_1\Big)-60 \nu_2+63\Big]
\nonumber\\ &\quad \times
\frac{1}{36 \nu_1 (\nu_1+1) \nu_2 (\nu_2+1)} I(\nu_1,\nu_2).
\end{align}
Notice that for eqs.~(\ref{Pb22})\,–\,(\ref{Pb2s2}), we have to subtract their large scale constant contributions, e.g. $P_{b_2^2}(k) \rightarrow P_{b_2^2}(k) - P_{b_2^2}(k \rightarrow 0)$. 

The integrals~(\ref{Eq:P_b_2_t}) and~(\ref{Eq:P_b_s2_t}) can be approximated by 
\begin{align}
P(k) =  k^3 \sum_{m_1, m_2} c^f_{m_1} k^{-2\nu_1} M_P(\nu_1, \nu_2) \, c_{m_2} k^{-2\nu_2},
\end{align}
here ``$P$'' stands for $P_{b_2,\theta}$, $P_{b_{s^2},\theta}$, while
\begin{align}
M_{P_{b_2,\theta }} \left(\nu _1,\nu _2\right) & =\frac{(7 \nu_1-4) (2 \nu_{12} - 3) }{14 \nu_1 \nu_2} I(\nu_1,\nu_2),
\\
M_{P_{b_{s^2},\theta}} \left(\nu _1,\nu _2\right) &= \frac{(2\nu_{12}-3) \Big[\nu_1 \big(14 \nu_1 (2 \nu_2-1)-30 \nu_2+39 \big)-10 \nu_2-19 \Big] }{84 \nu_1 (\nu_1+1) \nu_2 (\nu_2+1)} I(\nu_1,\nu_2), 
\end{align}
and eq.~(\ref{sigma23EdS}) takes the form
\begin{align}
 &\sigma_3^2 (k) = k^3 \sum_{m_1} c_{m_1} k^{-2\nu_1} M_{\sigma_3^2}(\nu_1), \\
&M_{\sigma_3^2}(\nu_1) = \frac{45 \tan (\nu_1 \pi)}{128 \pi  (\nu_1-3) (\nu_1-2) (\nu_1-1) \nu_1 (\nu_1+1)}.
\end{align}
For $\nu=-1.51$, the biasing power spectra do not require UV or IR corrections.

\subsection*{A function}
We move our attention to eq.~(\ref{Eq:Inm_ijk}), which can be computed through
\begin{align}
I^{1,udd}_{1,a}(k)  &= k^3 P_L(k) \left[\frac{f(k)}{f_0} \sum_{m_1} c_{m_1} k^{-2\nu_1} M^{f_{\vk}}_{\va_{11}}(\nu_1)  + \sum_{m_1} c^f_{m_1} k^{-2\nu_1} M^{f_{\vp}}_{\va_{11}}(\nu_1) \right],
\\
M^{f_{\vk}}_{\va_{11}}(\nu_1) &= \frac{(15-7\nu_1) \tan (\nu_1 \pi)}{56 \pi  (\nu_1-3) (\nu_1-2) (\nu_1-1) \nu_1},
\\
M^{f_{\vp}}_{\va_{11}}(\nu_1) &= \frac{(7\nu_1-6) \tan (\nu_1 \pi)}{56 \pi  (\nu_1-3) (\nu_1-2) (\nu_1-1) \nu_1}.
\end{align}

\begin{align}
I^{2,uud}_{1,a}(k)  &= k^3 P_L(k) \left[\frac{f(k)}{f_0} \sum_{m_1} c^f_{m_1} k^{-2\nu_1} M^{f_{\vk}f_{\vp}}_{\va_{12}}(\nu_1)  + \sum_{m_1} c^{ff}_{m_1} k^{-2\nu_1} M^{f_{\vp} f_{\vp}}_{\va_{12}}(\nu_1) \right],  
\\
M^{f_{\vk}f_{\vp}}_{\va_{12}}(\nu_1) &= \frac{3 (7\nu_1-13) \tan (\nu_1 \pi)}{224 \pi  (\nu_1-3) (\nu_1-2) (\nu_1-1) \nu_1 (\nu_1+1)},
\\
M^{f_{\vp} f_{\vp}}_{\va_{12}}(\nu_1) &= \frac{3 (1-7\nu_1) \tan (\nu_1\pi)}{224 \pi  (\nu_1-3) (\nu_1-2) (\nu_1-1) \nu_1 (\nu_1+1)}.
\end{align}

\begin{align}
I^{3, uuu}_{3, a}(k) &= k^3 \frac{f(k)}{f_0} P_L(k) \Bigg[\frac{f(k)}{f_0} \sum_{m_1} c^f_{m_1} k^{-2\nu_1} M^{f_{\vk} f_{\vk} f_{\vp}}_{\va_{33}}(\nu_1) \nonumber\\ 
&\quad  + \sum_{m_1} c^{ff}_{m_1} k^{-2\nu_1} M^{f_{\vk} f_{\vp} f_{\vp}}_{\va_{33}}(\nu_1) \Bigg], 
\\
M^{f_{\vk} f_{\vk} f_{\vp}}_{\va_{33}}(\nu_1) & =\frac{ \big(\nu_1 (53-28 \nu_1)+21\big) \tan (\nu_1 \pi)}{224 \pi  (\nu_1-3) (\nu_1-2) (\nu_1-1) \nu_1 (\nu_1+1)},
\\
M^{f_{\vk} f_{\vp} f_{\vp}}_{\va_{33}}(\nu_1) &= \frac{\big(\nu_1 (28 \nu_1-17)-21\big) \tan (\nu_1\pi)}{224 \pi  (\nu_1-3) (\nu_1-2) (\nu_1-1) \nu_1 (\nu_1+1)}.
\end{align}
\begin{align}
I^{2, uud}_{2,a}(k) & = \frac{f_0}{f(k)} I^{3, uuu}_{3,a}(k) + \frac{f(k)}{f_0} I^{1, udd}_{1,a}(k), \label{Eq:I22a}
\\
I^{3, uuu}_{2, a}(k) & = \frac{f(k)}{f_0} I^{2, uud}_{1,a}(k), \label{Eq:I32a}
\end{align}
where~(\ref{Eq:I22a}) and~(\ref{Eq:I32a}) reduce to trivial computations. 

The rest of the functions $I^m_n$ are
\begin{align}
&I^{1, udd}_{1, b} (k) = k^3 \sum_{m_1, m_2} c^f_{m_1} k^{-2\nu_1} M^{f_{\vp}}_{\tilde{\mathcal{A}}_{11}}(\nu_1, \nu_2) \, c_{m_2} k^{-2\nu_2},
\\
&M^{f_{\vp}}_{\tilde{\mathcal{A}}_{11}}(\nu_1, \nu_2) = \frac{(2 \nu_{12}-3) (2 \nu_{12}-1) \big( \nu_1 (7 \nu_{12}-4)-5 \big) }{7 \nu_1 (\nu_1+1) (2 \nu_1-1) \nu_2} I(\nu_1, \nu_2).
\end{align}

\begin{align}
&I^{2, uud}_{1,b}(k) = k^3 \sum_{m_1, m_2} c^f_{m_1} k^{-2\nu_1} M^{f_{\vk-\vp} f_{\vp}}_{\tilde{\mathcal{A}}_{12}}(\nu_1, \nu_2)\, c^f_{m_2} k^{-2\nu_2},
\\
&M^{f_{\vk-\vp} f_{\vp}}_{\tilde{\mathcal{A}}_{12}}(\nu_1, \nu_2) = -\frac{(2 \nu_{12}-3) (2 \nu_{12}-1) (7\nu_{12}+6) }{56 \nu_1 (\nu_1+1) \nu_2 (\nu_2+1)} I(\nu_1,\nu_2).
\end{align}

\begin{align}
I^{2, uud}_{2,b} (k) & = k^3 \sum_{m_1, m_2} c^f_{m_1} k^{-2\nu_1} M^{f_{\vk-\vp} f_{\vp}}_{\tilde{\mathcal{A}}_{22}} (\nu_1, \nu_2) \, c^f_{m_2} k^{-2\nu_2}
\nonumber \\ & \quad
+ k^3 \sum_{m_1, m_2} c^{ff}_{m_1} k^{-2\nu_1} M^{f_{\vp} f_{\vp}}_{\tilde{\mathcal{A}}_{22}} (\nu_1, \nu_2) \, c_{m_2} k^{-2\nu_2},
\\
M^{f_{\vk-\vp} f_{\vp}}_{\tilde{\mathcal{A}}_{22}} (\nu_1, \nu_2)  &= \Big[336 \nu_1 \nu_2^4+4 \big( 4 \nu_1 (35 \nu_1 -19)-39 \big) \nu_2^3+48 \nu_1  \big(\nu_1  (7 \nu_1 -3)-9\big) \nu_2^2
\nonumber \\ & \quad
+8 \nu_1  \Big(2 \nu_1  \big(\nu_1  (7 \nu_1 -4)-28 \big)+41\Big) \nu_2+3 \nu_1  \big(4 \nu_1  (10-9 \nu_1 )+1 \big)
\nonumber \\ & \quad
+75 \nu_2-18\Big] \times \frac{2\nu_{12}-3}{56 \nu_1 (\nu_1+1) (2 \nu_1-1) \nu_2 (\nu_2+1) (2 \nu_2-1)} I(\nu_1, \nu_2),
\\
M^{f_{\vp} f_{\vp}}_{\tilde{\mathcal{A}}_{22}} (\nu_1, \nu_2) & = \frac{(2 \nu_{12}-3) \left[ \nu_1 (7 \nu_{12}-4)+3 \nu_2-5 \right] }{7 \nu_1 (\nu_1+1) \nu_2} I(\nu_1,\nu_2).
\end{align}

\begin{align}
&I^{3, uuu}_{2,b} (k) = k^3 \sum_{m_1, m_2} c^{ff}_{m_1} k^{-2\nu_1} M^{f_{\vk-\vp} f_{\vp}f_{\vp}}_{\tilde{\mathcal{A}}_{23}}(\nu_1, \nu_2)\, c^f_{m_2} k^{-2\nu_2},
\\
&M^{f_{\vk-\vp} f_{\vp}f_{\vp}}_{\tilde{\mathcal{A}}_{23}}(\nu_1, \nu_2) = -\frac{(7 \nu_1-1) (2\nu_{12}-3) (2 \nu_{12}-1) }{28 \nu_1 (\nu_1+1) \nu_2 (\nu_2+1)} I(\nu_1, \nu_2).
\end{align}

\begin{align}
&I^{3, uuu}_{3,b} (k) = k^3 \sum_{m_1, m_2} c^{ff}_{m_1} k^{-2\nu_1} M^{f_{\vk-\vp} f_{\vp}f_{\vp}}_{\tilde{\mathcal{A}}_{33}}(\nu_1, \nu_2)\, c^f_{m_2} k^{-2\nu_2},
\\
& M^{f_{\vk-\vp} f_{\vp}f_{\vp}}_{\tilde{\mathcal{A}}_{33}}(\nu_1, \nu_2) = \Big[4 (7 \nu_1+3) \nu_2^2+2 (\nu_1 (14 \nu_1-1)-11) \nu_2-13 (\nu_1+1) \Big]
\nonumber\\ & \quad \qquad\qquad\qquad\qquad\times 
 \frac{(2\nu_{12}-3) (2 \nu_{12}-1)  }{28 \nu_1 (\nu_1+1) \nu_2 (\nu_2+1) (2 \nu_2-1)}   I(\nu_1,\nu_2) .
\end{align}
Finally, for the $A$ functions, we have to add the following corrections
\begin{align}
I^{1, udd, \,\text{UV}}_{1, a} (k) &= \left( \frac{92}{35}  \frac{f(k)}{f_0} \sigma^2_{\Psi} - \frac{18}{7} \sigma^2_v \right) k^2 P_L(k),
\\
I^{2, uud, \,\text{UV}}_{1, a} (k) &= - \left( \frac{38}{35} \frac{f(k)}{f_0} \sigma^2_v + \frac{2}{7} \sigma^2_{vv} \right)k^2 P_L(k), 
\\
I^{3, uuu, \,\text{UV}}_{3, a} (k) &= -\left( \frac{16}{35} \frac{f(k)}{f_0} \sigma^2_v + \frac{6}{7} \sigma^2_{vv} \right)k^2 P^L_{\delta \theta}(k),
\end{align}
with $\sigma^2_\Psi$ and $\sigma^2_v$ given by eq.~(\ref{Eq:sigmas_psi_v}), and $\sigma^2_{vv}$ by eq.~(\ref{sigma2vv}).

\subsection*{D function}

Finally, we display the FFTLog expressions for the $D(k, \mu)$ function of eq.~(\ref{D_function}). 

\bigskip

\noindent For $I^{2, uudd}_{n, D}= I^{2, uudd}_{n, B} + I^{2, uudd}_{n, C}$  with $n=1,2$, corresponding to eqs.~\eqref{I2uuddnB} and \eqref{I2uuddnC}:
\begin{align}
&I^{2, uudd}_{n, B} (k) = k^3 \sum_{m_1, m_2} c^f_{m_1} k^{-2\nu_1} M_{\mathcal{B}^n_{11}}(\nu_1, \nu_2) \, c^f_{m_2} k^{-2\nu_2},
\\
&M_{\mathcal{B}^1_{11}}(\nu_1, \nu_2) = \frac{3-2\nu_{12}}{4\nu_1\nu_2} I(\nu_1, \nu_2),
\\
&M_{\mathcal{B}^2_{11}}(\nu_1, \nu_2) = \frac{(2\nu_{12}-3)(2\nu_{12}-1)}{4\nu_1 \nu_2}I(\nu_1, \nu_2).
\end{align}
and
\begin{align}
&I^{2, uudd}_{n, C}(k)  =  k^3 \sum_{m_1, m_2} c_{m_1}  k^{-2\nu_1}
M_{\mathcal{C}^n_{11}}(\nu_1, \nu_2) \, c^{ff}_{m_2} k^{-2\nu_2},
\\
&M_{\mathcal{C}^1_{11}}(\nu_1, \nu_2)  = \frac{(2\nu_1 -3)(2\nu_{12}-3)}{4\nu_2 (\nu_2 +1)(2\nu_2-1)} I(\nu_1, \nu_2),
\\
&M_{\mathcal{C}^2_{11}}(\nu_1, \nu_2)  = \frac{(2\nu_{12}-3)(2\nu_{12}-1)}{4 \nu_2 (\nu_2 + 1)} I(\nu_1, \nu_2).
\end{align}

\bigskip

\noindent For $I^{3, uuud}_{n, D}(k)$ with $n=2,3$, corresponding to eqs.~\eqref{I3uuudnD}:
\begin{align}
I^{3, uuud}_{n, D}(k) & = k^3 \sum_{m_1, m_2} c^f_{m_1} k^{-2\nu_1} M_{\mathcal{D}^n_{21}}(\nu_1, \nu_2) \, c^{ff}_{m_2} k^{-2\nu_2}, 
\\
M_{\mathcal{D}^2_{21}}(\nu_1, \nu_2) & = \frac{(2\nu_1-4\nu_2-1)(2\nu_{12}-3)(2\nu_{12}-1)}{4 \nu_1 \nu_2 (\nu_2 +1)(2\nu_2-1)} I(\nu_1, \nu_2),
\\
M_{\mathcal{D}^3_{21}}(\nu_1, \nu_2) & = \frac{(3-2\nu_{12})\left(1-4\nu^2_{12} \right)}{4 \nu_1 \nu_2 (\nu_2 + 1)} I(\nu_1, \nu_2).
\end{align}

\bigskip

\noindent Finally for $I^{4, uuuu}_{n, D}(k)$ with $n=2,3,4$, corresponding to eqs.~\eqref{I4uuuunD}:
\begin{align}
&I^{4, uuuu}_{n, D}(k)  = k^3 \sum_{m_1, m_2} c^{ff}_{m_1} k^{-2\nu_1} M_{\mathcal{D}^n_{22}}(\nu_1, \nu_2) \, c^{ff}_{m_2} k^{-2\nu_2}, 
\\
&M_{\mathcal{D}^2_{22}}(\nu_1, \nu_2)  = \frac{3(3-2\nu_{12})(1-2\nu_{12})}{32 \nu_1 (\nu_1 + 1) \nu_2 (\nu_2 + 1)} I(\nu_1, \nu_2),
\\
&M_{\mathcal{D}^3_{22}}(\nu_1, \nu_2)  = \frac{(3-2\nu_{12}) \left(1-4\nu^2_{12} \right) \left[ 1 + 2 \left( \nu^2_1 -4\nu_1 \nu_2 + \nu^2_2 \right) \right]}{16 \nu_1 (\nu_1 + 1) (2\nu_1 - 1) \nu_2 (\nu_2 +1)(2\nu_2 -1)} I(\nu_1, \nu_2),
\\
&M_{\mathcal{D}^4_{22}}(\nu_1, \nu_2)  = \frac{\left(9 -4\nu^2_{12}\right) \left( 1-4\nu^2_{12} \right)}{32 \nu_1 (\nu_1 + 1) \nu_2 (\nu_2 +1)} I(\nu_1, \nu_2).
\end{align}

Functions $D$ do not require IR or UV corrections with our choice of bias $\nu = -0.1$.

\section{Complementary  plots}
\label{app:plots}

In this appendix we show complementary plots to \S \ref{sec:modelvalidation} that we decided to not show in the main text to avoid too much cluttering.

 \begin{figure}
 	\begin{center}
 	\includegraphics[width=6.0 in]{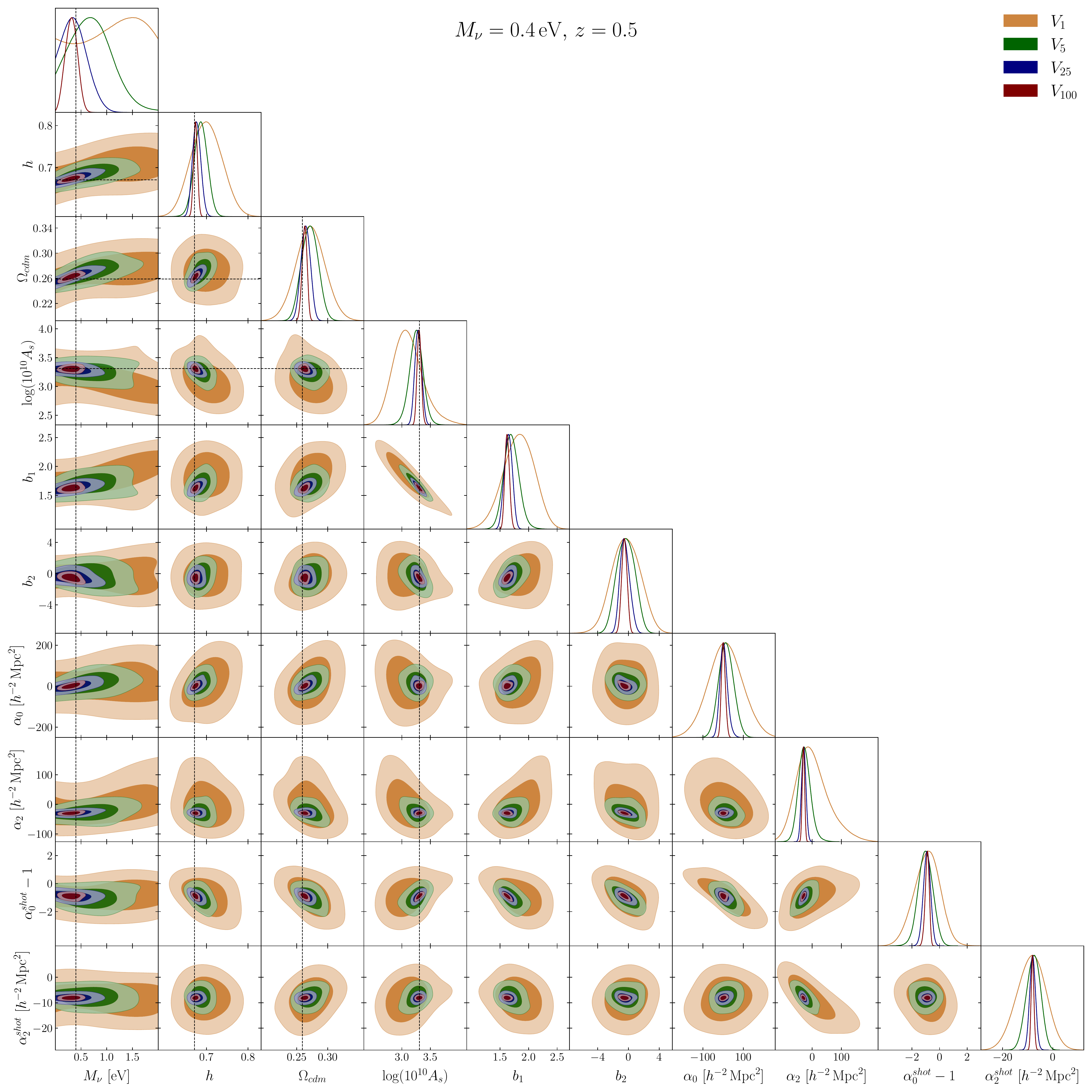}
 	\caption{Contour plots for the posterior distributions at 68 and 95\% confidence level computed with different covariance matrices, corresponding to $V_1$, $V_5$, $V_{25}$ and  $V_{100}$ effective volume data sets. This is the case of mass $M_\nu=0.4 \,\text{eV}$ at redshift $z=0.5$. The fittings are performed using the monopole and quadrupole of the power spectrum up to $k_{\text{max}} = 0.2 \, \hmpci$. Vertical and horizontal dashed lines show the values of the simulations. This is a complementary figure to  \ref{fig:Base_TriangularPlot} and table \ref{Table:bestfit}
 	\label{fig:Base_TriangularPlot_All}}
 	\end{center}
 \end{figure}

 \begin{figure}
 	\begin{center}
 	\includegraphics[width=6.0 in]{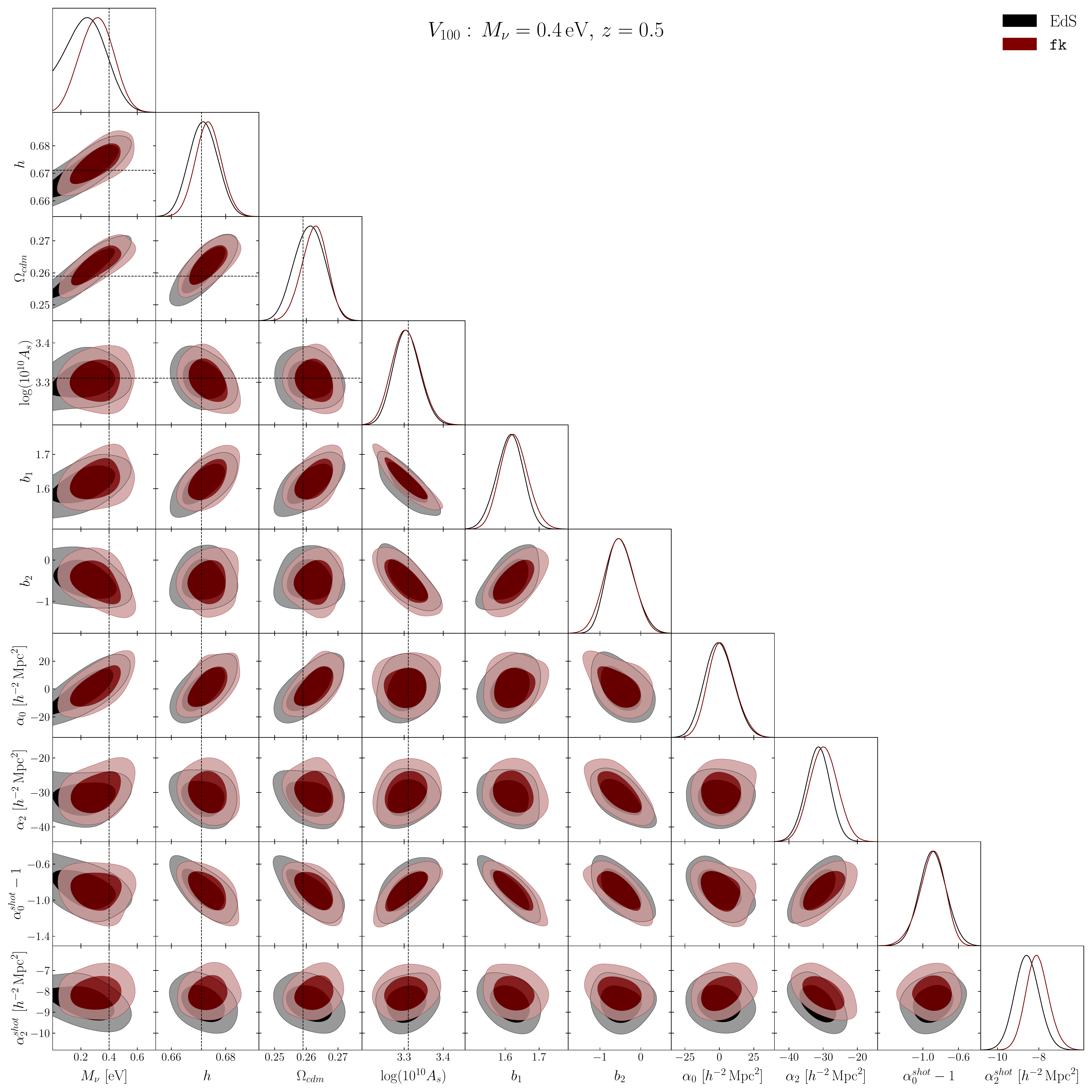}
 	\caption{Comparison on the constraints of the cosmological parameters when using \fk- and EdS-kernels for the volume $V_{100}$. This is the case of $M_\nu=0.4 \,\text{eV}$ at redshift $z=0.5$ when fitting up to $k_{\text{max}} = 0.2 \, \hmpci$
 	\label{fig:fkvsEdS}}
 	\end{center}
 \end{figure}

 \begin{figure}
 	\begin{center}
 	\includegraphics[width=6.0 in]{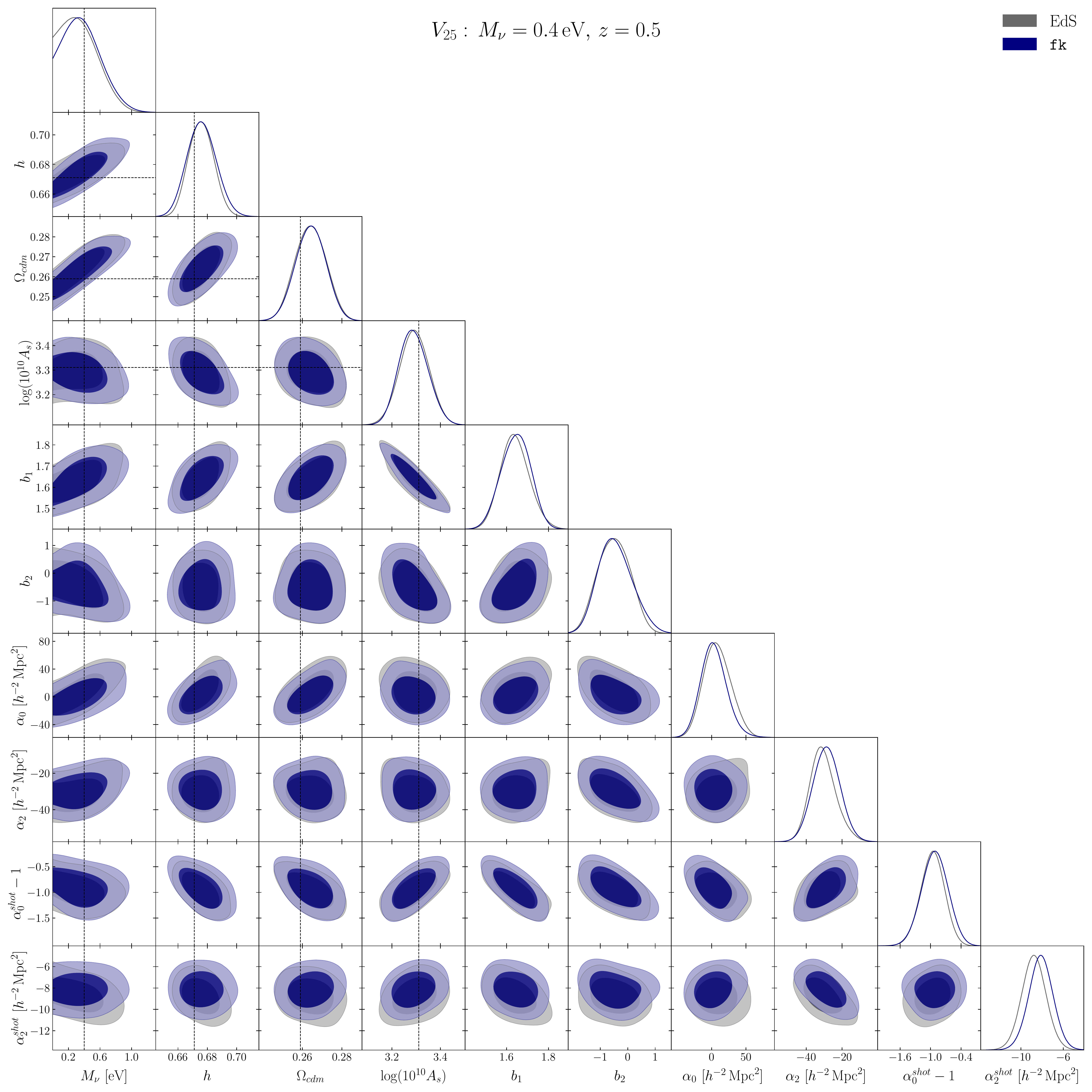}
 	\caption{Comparison on the constraints of the cosmological parameters when using \fk- and EdS-kernels for the volume $V_{25}$. This is the case of $M_\nu=0.4 \,\text{eV}$ at redshift $z=0.5$ when fitting up to $k_{\text{max}} = 0.2 \, \hmpci$
 	\label{fig:fkvsEdS25}}
 	\end{center}
 \end{figure}

\begin{figure}
 	\begin{center}
 	\includegraphics[width=6.0 in]{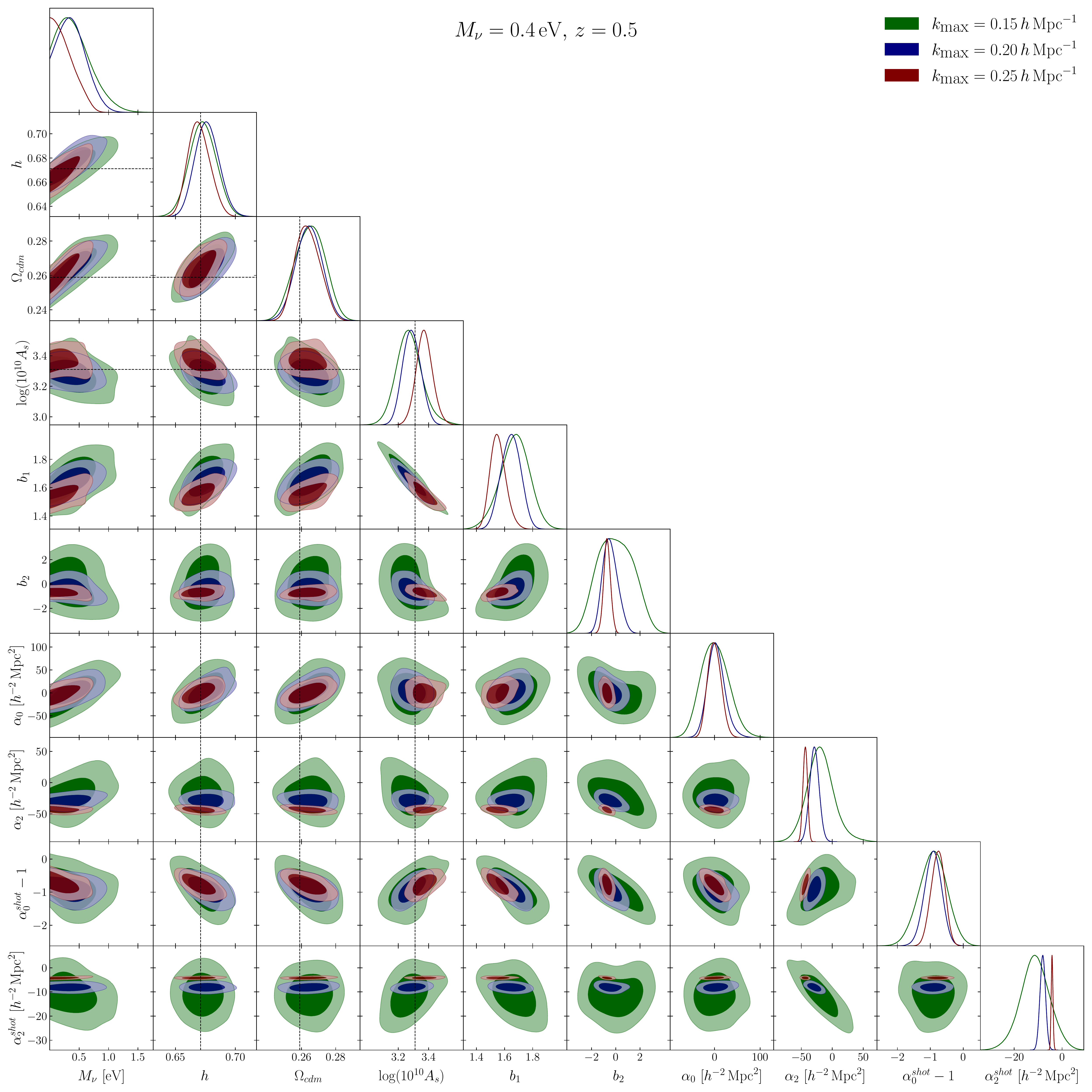}
 	\caption{\revised{Contour plots for the posterior distributions at 68 and 95\% confidence level computed for different values of the maximum wave-number $k_{\text{max}} = 0.15,\,0.2,\,0.25 \hmpci$. This is the case of mass $M_\nu=0.4 \,\text{eV}$ at redshift $z=0.5$ using the DESI-like volume $V_{25}$. Vertical and horizontal dashed lines show the values of the simulations. This is a complementary figure to \ref{fig:kmax}.}
 	\label{fig:kmax_all}}
 	\end{center}
 \end{figure}

 \bibliographystyle{JHEP}  
 \bibliography{bib.bib}

\end{document}